\documentclass[%
amsmath,amssymb,
pra,
]{revtex4-1}

\usepackage{graphicx}
\usepackage{dcolumn}
\usepackage{bm}
\usepackage{hyperref}
\usepackage{subfigure}

\newcommand{\e}{e}
\newcommand{\NA}{\mathrm{NA}}
\newcommand{\amax}{\alpha_{\mathrm{max}}}
\newcommand{\R}{\mathbf{R}}
\newcommand{\C}{\mathbf{C}}

\renewcommand{\vec}[1]{\bm{\mathrm{#1}}}
\renewcommand{\Re}{\operatorname{Re}}
\renewcommand{\Im}{\operatorname{Im}}
\renewcommand{\hat}{\widehat}
\renewcommand{\varphi}{\phi}

\usepackage{bm} 

\newcommand{\Av}{\vec{A}}
\newcommand{\av}{\vec{a}}
\newcommand{\bv}{\vec{b}}
\newcommand{\vv}{\vec{v}}

\newcommand{\Ev}{\vec{E}}
\newcommand{\Hv}{\vec{H}}
\newcommand{\Mv}{\vec{M}}
\newcommand{\kv}{\vec{k}}
\newcommand{\rv}{\vec{r}}
\newcommand{\rp}{\rv_{\perp}}
\newcommand{\kp}{\kv_{\perp}}
\newcommand{\xiv}{\vec{\xi}}
\newcommand{\xip}{\xiv_{\perp}}

\newcommand{\ku}{\hat{\vec{k}}}
\newcommand{\su}{\hat{\vec{s}}}
\newcommand{\pu}{\hat{\vec{p}}}

\newcommand{\MC}{{\mathcal{M}}}
\newcommand{\TC}{{\mathcal{T}}}

\newcommand{\D}{\,d}
\newcommand{\Dkp}{\D\kp}
\newcommand{\Dxip}{\D\xip}
\newcommand{\Drp}{\D\rp}

\newcommand{\FOUR}{\mathcal{F}}

\newcommand{\FII}{\FOUR_{2}}
\newcommand{\FIII}{\FOUR_{3}}

\newcommand{\xu}{\hat{\vec{x}}}
\newcommand{\yu}{\hat{\vec{y}}}
\newcommand{\zu}{\hat{\vec{z}}}

\usepackage{dsfont}
\newcommand{\indicator}{\mathds{1}}

\newcommand{\Ahv}[2]{\begin{pmatrix} \hat{a}_{P, #2}(#1) \\
    \hat{a}_{S, #2}(#1) \end{pmatrix}}

\begin{document}

\preprint{APS/123-QED}

\title{On Maximum Focused Electric  Energy in Bounded Regions}

\author{Jonas Teuwen}
\affiliation{Optics Research Group, Department of Imaging Physics, Delft
  University of Technology, P.O.~Box 5046, 2600 GA Delft, The Netherlands \&
  Radiology and nuclear medicine, Radboudumc, Nijmegen, The Netherlands}
\email{j.j.b.teuwen@tudelft.nl}
\author{H. Paul\ Urbach}%
\email{h.p.urbach@tudelft.nl}
\affiliation{Optics Research Group, Department of Imaging Physics, Delft University of Technology, P.O.~Box 5046, 2600 GA Delft, The Netherlands  \& ITMO University, St. Petersburg, Russia}



\date{\today}

\begin{abstract}
  A general method is presented for determining the maximum  electric energy in a bouded region of optical fields with given time-averaged flux of electromagnetic energy. 
 Time-harmonic fields are considered whose plane wave expansion consists of
 propagating plane waves only, i.e., evanescent waves are excluded. The bounded region can be quite general: it can consist of finitely many points, or be a curve,  a curved surface or a bounded volume. The optimum optical field is eigenfield corresponding to the maximum eigenvalue of a compact linear integral operator which depends on the bounded region. It is explained how these optimum fields can be realized by  focussing  appropriate pupil fields.
The special case that the region is a circular disc perpendicular to the direction of  optical axis is investigated by numerical simulations.
\end{abstract}

\pacs{42}
\maketitle
\tableofcontents

\section{Introduction}
In optics it is often desirable to maximize the electric energy in a
certain bounded region of space. This is for example important 
 to optically excite certain molecules or atoms efficiently,
to trap molecules or small particles using optical tweezers, to
enhance scattering  or absorption of light in some volume and in
numerous other cases \cite{Helseth, Zhan, Xie}. An important method to
realize optimum concentrations of light is by shaping the pupil field
of an objective lens \cite{Janssen}. With spatial light modulators
(SLMs) not only amplitude and phase but also the polarization can be
varied pixel by pixel. In this way pupil fields can be shaped to
achieve optimized focused fields \cite{Neil, Sheppard, Iglesias,
  Sheppard2, Urbach, deBruin}.

In this paper we  present a general mathematical formulation for
achieving optimum concentration of electric energy. With a similar method also
the magnetic energy or the total electromagnetic energy, i.e., the sum of the
electric and magnetic energies, could be maximized, but since at optical
frequencies the main interaction with matter occurs through the electric field,
it is more interesting to maximize the electric energy. To be more precise, we
consider time-harmonic electromagnetic fields which propagate in a given
direction, say the positive $z$-direction, and which have numerical aperture
$\NA$ smaller than the refractive index $n$ of the medium in which the
propagation takes place. This means that the wave vectors of the plane wave
expansion of the field make an angle with the positive $z$-axis which  does not
exceed  the angle $\amax$ where $\sin \amax=\NA/n$. The waves in the angular
spectrum are thus all propagating and there are no evanescent waves.

In Section~\ref{section.Lagrange} we formulate the optimisation problem
in terms of the plane wave amplitudes. The problem is to
determine the amplitudes for which the   electric energy 
in a given region is maximum for the given values of the
$\NA$ and for given  mean power flow. The region can be quite
general: it can for example be a bounded 3D volume,  a bounded curved surface, a
bounded curve or it can consist of one or several points. Furthermore, by a
slight generalisation of the formulation of the optimisation problem, we include
the case of maximizing the squared modulus of only a specific electric field
component, instead of the electric energy. Because our formulation is general,
it includes many previously studied optimisation problems such as
\cite{Sheppard,Urbach2009} as special cases. 

We remark that when evanescent waves would be allowed in the plane wave
expansion, the maximum electric energy in any bounded region can, for every
prescribed value of the mean flow of power, be made infinite. The reason is that
the evanescent waves do not contribute to the mean power flow and therefore
their amplitude is not constrained by it. The evanescent waves do however
contribute to the electric energy density, therefore the energy density can be
made arbitrarily large if evanescent waves would be taken into account.
Excluding evanescent waves from the optimisation problem means that in this
paper we study only fields that are radiated by sources which are many
wavelengths from the region where the energy is maximized. We assume in
particular that there are no structures and objects close to the region of
interest which could generate evanescent waves by scattering.

Many different groups have contributed to the shaping and optimisation
of optical fields in or near the  focal point of a lens. This has led
to important applications and to improved optical sensitivities and
resolution. In contrast to most previous work where field enhancements are
studied,  we aim at determining the maximum possible energy in a given region
and for a given $\NA$ and power flow. The optimisation problem has infinitely many variables (i.e., all amplitudes of the plane waves inside the given $\NA$) and  hence it is a problem in an infinitely dimensional function space. This means that the  optimum fields obtained by our method are fundamental and are not only interesting from the point of view of applications but are also of theoretical interest.

In Section~\ref{section.optprobl} and Section~\ref{subsection.expra} the
optimisation problem is formulated mathematically and expressed in terms of the
plane wave amplitudes.
By applying the Lagrange multiplier rule to the optimisation problem,
it is shown in Section~\ref{section.Lagrange} that the optimum plane
wave amplitudes are eigenfield of a linear integral operator
corresponding to the maximum eigenvalue. This linear integral operator
is compact and Hermitian when the proper scalar product is chosen. Since such an
operator has a maximum eigenvalue, existence of an optimum field is garanteed.
It should be remarked that the optimum field is not always unique: it can happen
that there are several distinct optimum fields and as a matter of fact an
example is  discussed in Section~\ref{section.results}.
In Section~\ref{subsection.scaling} a scaling property is derived which shows
that if the region over which the electric energy is optimized is scaled by
multiplying with a parameter $\sigma>0$, the optimum fields remain unchanged
when the total power, the  $\NA$  and the ratio $\lambda/\sigma$, where
$\lambda$ is the wavelength in vacuum, are kept constant.
  In
Section~\ref{sec:pupil-field} we explain how the optimum electromagnetic fields can
be obtained in practice  in the focal region of a positive lens of
numerical aperture $\NA$, by shaping the pupil field appropriately
using e.g., SLMs.

In Section~\ref{section.disc} we study in  detail the special
case of maximizing the electric energy in a disc perpendicular to the
$z$-axis. By using cylindrical coordinates and applying a Fourier
series to expand the functions with respect to polar angle,  the  2D integral
equation becomes equivalent to a set of 1D integral equations, with
the radial variable as integration variable. In
Section~\ref{section.results} we first discuss the case that the disc has
vanishing radius, which means that the average of the electric energy in the
focal point of the lens is optimized. For this case the solutions can be
obtained in closed form  and we retrieve previously published results. Then we
consider discs with positive radius. In this case the solutions can only be
obtained by numerical computations. It is found that when the radius of the disc
is varied, only two types of solutions occur, namely one for which the optimum
field in the pupil is predominantly linearly polarised in some direction,
whereas the second type has azimuthal polarised pupil field.  As the radius and
the $\NA$ are varied, the numerically solutions are alternating between these
two cases. For certain values of the $\NA$ and radius of the disc, both type of solutions occur, i.e., both give the same maximum electric energy.

\section{The Optimisation problem}
\label{section.optprobl}
Consider a time-harmonic electromagnetic field in an unbounded 
 homogeneous  nonmagnetic  lossless medium with refractive
index $n$. The electromagnetic field is written as
\begin{align}
  \label{eq:electric-field-real}
  \vec{\mathcal{E}}(\rv, t) &= \Re[\Ev(\rv) \e^{-i \omega t}],  \\
  \label{eq:magnetic-field-real}
  \vec{\mathcal{H}}(\rv, t) &= \Re[\Hv(\rv) \e^{-i \omega t}],
\end{align}
where the frequency $\omega > 0$ and  $\Ev(\rv)$ and $\Hv(\rv)$
are the complex time-independent electric and magnetic fields.
We will assume that with respect to the cartesian coordinate system $(x,
y, z)$ with unit vectors $\xu, \yu, \zu$, the electromagnetic
field (\ref{eq:electric-field-real}, \ref{eq:magnetic-field-real}) has
numerical aperture $\NA \leq n$ and is propagating in the positive
$z$-direction. This means that the plane wave vectors of the angular
spectrum of the fields have angles with the positive $z$-axis that are
smaller than $\amax$, with $\NA = n\sin\amax$. The complex electric
field can be expanded into plane waves
\begin{equation}
  \label{eq:electric-field}
  \Ev(\rv) = \frac{1}{4\pi^2} \iint_{\Omega} \Av(\kp) \,e^{i \kv \cdot
    \rv} \Dkp,
\end{equation}
where $\Omega$ is the disk in two-dimensional reciprocal space with radius $k_0 \NA$:
\begin{equation}
  \label{eq:Omega-kx-ky}
  \Omega = \bigl\{(k_x, k_y) : \sqrt{k_x^2 + k_y^2} \leq k_0 \NA \bigr\},
\end{equation}
where $k_0 = \omega \sqrt{\epsilon_0 \mu_0}$ is the wave number in
vacuum and
the vectors $\kv$ and $\kp$ are defined by
\begin{equation*}
  \kp = k_x \xu + k_y \yu,  \text{ and } \kv = \kp + k_z \zu,
\end{equation*}
where $k_z = \sqrt{k^2 - |\kp|^2}$ and $k=k_0 n$ is the wave number
inside the medium. We choose the usual branch of the square root so that the cut is along the negative real axis and the square root of a positive real number is positive. Hence,  the plane waves of (\ref{eq:electric-field}) are
 propagating in the positive $z$-direction. Faraday's law
$\nabla \times \Ev = i\omega \mu_0  \Hv$ implies that the complex magnetic
field $\Hv$ can be written as
\begin{equation}
  \label{eq:magnetic-field}
  \Hv(\rv) = \frac{1}{\omega \mu_0} \frac{1}{4\pi^2}
  \iint_{\Omega} \kv \times
  \Av(\kp)\e^{i \kv \cdot \rv} \Dkp.
\end{equation}
Apart from the fact that the fields $\Ev$ and $\Hv$ consist of a
superposition of plane waves that propagate in the positive $z$-direction and whose wave vectors have angle with the $z$-axis which 
does not exceed $\amax$, the  field is completely general. 
For the time being we will not consider
how such a field can be realized in practice. This issue will be
addressed in Section~\ref{sec:pupil-field} where the focussing of an appropriate pupil field is  described.

Let   $S$  be  a bounded set. $S$ can be quite general:
  it can for example consist of finitely many 
points, be a curve, a (curved) surface or a bounded
volume.  It will be convenient
in what follows to associate with $S$ a distribution $T_S$ in $\R^3$ 
defined 
such that for every  smooth test function $\phi(\rv): \R^3\mapsto \R$:
\begin{equation}
  \langle T_S, \phi \rangle_{\R^3} = \frac{1}{|S|} \int_S \phi \D{S}.
  \label{eq.defT_S}
\end{equation}
If $S$ is a set  of finitely many points, then $|S|$ is the number of points and the integral should be interpreted as the sum of the  values of $\phi$ in those points. In other words, if $S$ is a set of points, $T_S$ is a sum of delta-functions at these points, divided by the number of points in $S$.
If $S$ is a curve, surface or volume, $|S|$ is the length, surface area or volume, respectively. 
Hence,  
$\langle T_S,\phi\rangle_{\R^3}$ is simply the average of $\phi$ over $S$. 
The subscript $\R^3$ at the bracket  emphasizes that $T_S$ is a distribution on $\R^3$.
We further elaborate on these examples in Section~\ref{sec:examplesop}.

Because the electric field is free of divergence  it follows from
\eqref{eq:electric-field} that $\Av(\kp) \cdot \kv = 0$, i.e., $\Av(\kp)$ is  perpendicular to the wave vector.
To incorporate this property we will write the plane wave amplitudes on the 
 positively oriented orthonormal basis in reciprocal space defined by:
\begin{equation}
  \label{eq:basis-sp}
  \su(\kp) = \frac{\ku \times \zu}{|\ku \times \zu|} =\frac1{|\kp|}
  \begin{pmatrix}
    k_y\\
    -k_x\\
    0
  \end{pmatrix},
  \;\;\;\;
  \pu(\kp) = \frac{\su \times \ku}{|\su \times \ku|} =\frac1k\frac1{|\kp|}
  \begin{pmatrix}
    -k_x k_z\\
    -k_y k_z\\
    |\kp|^2
  \end{pmatrix},
\end{equation}
where for a vector $\mathbf{v}$:  $|\mathbf{v}|=\sqrt{|v_x|^2+|v_y|^2+|v_z|^2}$.
Note that $\ku \cdot \pu = 0$ and $\ku \cdot \su = 0$. 
We  write $\Av: \Omega \to \C^3$ as 
\begin{equation}
  \label{eq:Amplitudes-image-basis}
  \Av(\kp) = a_p(\kp) \pu(\kp) + a_s(\kp) \su(\kp),
\end{equation}
where $a_p$ is the component parallel to the plane through the wave vector $\kv$ and the $z$-axis and $a_s$ is the component  perpendicular to this plane. To  prevent confusion with $\Av$, which is a vector with three components, the  vector field with two components: $(a_p, a_s): \Omega \mapsto \C^2$ will be denoted as $\av$, i.e., we write
\begin{equation}
  \label{eq.defav}
  \av(\kp) = 
  \begin{pmatrix}
    a_p(\kp) \\
    a_s(\kp)
  \end{pmatrix}
\end{equation}
The electromagnetic field written in terms of $a_p$ and $a_s$ becomes
\begin{equation}
  \label{eq:electric-field2}
  \Ev(\rv) = \frac{1}{4\pi^2} \iint_{\Omega} [ a_p(\kp) \pu(\kp) + a_s(\kp) \su(\kp)] \e^{i \kv \cdot
    \rv} \Dkp,
\end{equation}
\begin{equation}
  \label{eq:magnetic-field2}
  \Hv(\rv) = n \sqrt{\frac{\epsilon_0}{\mu_0}} \frac{1}{4\pi^2} \iint_{\Omega} [ a_p(\kp) \su(\kp) - a_s(\kp)\pu(\kp)] \e^{i \kv \cdot
    \rv} \Dkp.
\end{equation}
The time-averaged power flow in the positive $z$-direction is given by
the integral over a plane  $z = \text{constant}$ of  the $z$-component of half the real part of the complex Poynting vector $\vec{S} = \Ev \times \Hv^\ast$:
\begin{equation}
  \label{eq:Power-flow-Poynting}
  P = \iint_{\R^2} \frac{1}{2} \Re\{\vec{S}(\rv) \} \cdot
  \zu \D{x}\D{y}.
\end{equation}
Note that, since there is no absorption, the integral \eqref{eq:Power-flow-Poynting} does not dependent on the  chosen  plane $z = \text{constant}$. Using Plancherel's theorem together with $\Av \cdot \kv = 0$ , we get as in \cite[Equation
25]{Urbach2009} that the power flow \eqref{eq:Power-flow-Poynting} can
be expressed in the amplitudes of the plane waves as
\begin{equation}
 P(\Ev) =
  \frac{1}{\omega \mu_0} \frac{1}{8\pi^2} \iint_{\Omega}
  |\av(\kp)|^2 k_z \Dkp = \frac{1}{\omega \mu_0} \frac1{8 \pi^2} \iint_\Omega [|a_p(\kp)|^2 + |a_s(\kp)|^2] k_z \Dkp.
  \label{eq:Power-flow-Poynting-amplitudes}
\end{equation}
To formulate the optimisation problem we define the functional $G_{S,\Pi}$ as follows:
\begin{equation}
  G_{S,\Pi}(\Ev) = \frac1{|S|} \iint_{S} |\Pi(\Ev)|^2 \D S = \langle T_S,
  |\Pi(\Ev)|^2 \rangle_{\R^3},
  \label{eq.defGS}
\end{equation}
where $\Pi: \C^3 \mapsto \C^3$ is a projection on some linear subspace of $\C^3$. 
The goal is to determine the electric field $\Ev$ for which
$G_{S,\Pi}(\Ev)$ is maximal for given power $P(\Ev) = P_0$. 

We give a number of examples.
\begin{enumerate}
\item Let $\hat{\vv}$ be a real unit vector and let $\Pi(\Ev)=\Ev \cdot \hat{\vv}$, i.e., $\Pi$ is the projection on the direction defined by $\hat{\vv}$.
  Then 
  \begin{equation}
    G_{S,\Pi}(\Ev)= \frac1{|S|} \iint_{S}  |\Ev\cdot \hat{\vv}|^2 \D{S},
    \label{eq.GSPex1D}
  \end{equation}
  is the average over the region $S$ of the squared modulus of the projection
  of $\Ev$ along $\hat{\vv}$. The optimisation problem then amounts to
  maximizing the average over the region $S$ of the squared modulus  of the
  component of $\Ev$ along the unit vector $\hat{\vv}$.
\item $\Pi = \mathcal{I}$, the identity, i.e.\ $\Pi(\Ev) = \Ev$. In this case 
  \begin{equation}
    G_{S,\Pi}(\Ev) = \frac1{|S|} \iint_{S} |\Ev|^2 \D S,
    \label{eq.GSPex3D}
  \end{equation}
  is the averaged electric energy in the region S and the optimisation
  problem amounts to maximizing the electric energy averaged over the region $S$.
\item $\Pi(\Ev)=E_x  \xu + E_y \yu$, i.e., $\Pi$ is the projection on
  the $z = 0$ plane and
  \begin{equation}
    G_{S,\Pi}(\Ev))= \frac1{|S|} \iint_{S} |E_x|^2 + |E_y|^2  \D{S}.
    \label{eq.GSPex2D}
  \end{equation}
  Hence in this case the squared modulus of the electric field perpendicular to the $z$-axis is maximized over the
  region $S$.
\end{enumerate}

\subsection{Expression of the optimisation problem in terms of  plane wave amplitudes}
\label{subsection.expra}
We will express the optimisation problem in terms of plane wave amplitudes $\av$. First we 
 express  functional $G_{S,\Pi}$ in terms of  $\av$. 
We remark that \eqref{eq:electric-field} implies that for every $z$:
\begin{equation}
  \label{eq.FE}
  \mathcal{F}_2(\Pi(\Ev))(\kp, z) = \Pi(\Av)(\kp) \e^{i k_z z}, 
\end{equation}
where $\mathcal{F}_2$ is the 2D  Fourier transform defined by
\begin{equation}
  \label{eq:def_Fourier-transform}
  \FOUR_2(f)(\kp) = \iint_{\R^2} f(\rv_\perp) \e^{-i \rv_\perp \cdot \kp} \D\rv_\perp.
\end{equation}
Its inverse is given by
\begin{equation}
  \label{eq:inv-Fourier-transform}
  \FOUR^{-1}_2(g)(\rv_\perp) = \frac1{(2\pi)^2} \iint_{\R^2} g(\kp) \e^{i \kp \cdot \rv_\perp} \D\kp,
\end{equation}
where $\rv_\perp=(x, y)$.
Furthermore, let $\mathcal{F}_{3}$ be the 3D Fourier transform defined by
\begin{equation}
  \FIII(f)(\xip,\xi_z) = \iiint_{\R^3} f(\rp,z) \e^{-i(\xip\cdot \rv + \xi_z z)} \Drp \D{z},
  \label{eq.defFIII}
\end{equation}
where $\xip=(\xi_x, \xi_y)$. It may seem more natural to use $\kp,
k_z$ as Fourier variables, but in this paper the combination of $\kp$,
$k_z$ always implies that $k_z=\sqrt{k_0^2n^2 - |\kp|^2}$ whereas in
the 3D Fourier transform the three variables are 
independent and to prevent confusion we use therefore $\xip$, $\xi_z$ as
 variables of the 3D Fourier transform. We shall often write
\begin{equation}
  k_z(\kp)=\sqrt{k_0^2 n^2 - |\kp|^2},
  \label{eq.defkzkp}
\end{equation}
to emphasize the dependence of $k_z$ on $\kp$.
The following result is derived in Appendix~\ref{appendix.A}:
\begin{equation}
  G_{S,\Pi}(\Ev) = \frac1{(2\pi)^4} \iint_{\Omega} \iint_{\Omega} \FIII(T_S)(\kp - \kp', k_z(\kp) - k_z(\kp'))  \Pi(\Av)(\kp') \cdot
  \Pi( \Av)(\kp)^\ast  \Dkp \Dkp',
  \label{eq.GSAv}
\end{equation}
This is the expression of $G_{S,\Pi}$  in terms of the plane wave amplitudes $\Av$.
By substituting $\Av(\kp) = a_p(\kp) \pu(\kp) + a_s(\kp) \su(\kp)$ we find 
\begin{equation}
  \Pi(\Av)(\kp') \cdot \Pi( \Av)(\kp)^\ast   =  
  \MC_\Pi(\kp, \kp') \av(\kp') \cdot \av(\kp)^\ast,
  \label{eq.A*A*}
\end{equation}
where $\MC_\Pi$ is the real matrix defined by
\begin{equation}
  \label{eq:matrix-first-def}
  \MC_\Pi(\kp', \kp) = 
  \begin{pmatrix}
    \Pi( \pu)(\kp)\cdot \Pi(\pu)(\kp') &\;\; \Pi(\pu)(\kp)\cdot \Pi(\su)(\kp')\\
    \Pi(\su)(\kp)\cdot \Pi(\pu)(\kp') & \; \; \Pi(\su)(\kp)\cdot \Pi(\su)(\kp')
  \end{pmatrix}.
\end{equation}
This matrix is real because the vectors $\su$ and $\pu$ are real.
We remark that 
\begin{equation}
  \MC_\Pi(\kp, \kp')= \MC_\Pi(\kp,\kp')^{\text{T}},
  \label{eq.MvMvT}
\end{equation}
where the right-hand side is the transpose matrix. 
By substituting \eqref{eq.A*A*} in \eqref{eq.GSAv} we obtain the desired expression of $G_{S,\Pi}$ in terms of $\av$:
\begin{equation}
  G_{S,\Pi}(\av) = \frac{1}{(2\pi)^4} \iint_{\Omega} \iint_{\Omega} \FIII(T_S)(\kp
  - \kp', k_z(\kp) - k_z(\kp'))\, \MC_\Pi(\kp, \kp') \av(\kp') \cdot \av(\kp)^\ast \Dkp \Dkp'.
  \label{eq:GSav}
\end{equation}
\noindent\textbf{Remark:} Because $T_S$ is a real distribution on $\R^3$, its 3D Fourier transform satisfies:
\begin{equation}
  \FIII(T_S)(\xip, \xi_z)^\ast = \FIII(T_S)(-\xip, -\xi_z).
  \label{eq.FIIIsym}
\end{equation}
With this property and  \eqref{eq.MvMvT} one can easily verify that
the expression in the right-hand side of \eqref{eq:GSav} is real, as
should be.

The optimisation problem can now be formulated as a problem for the
vector function $\av: L^2(\Omega)^2 \mapsto \C^2$:
\\ \\ \noindent
\textbf{Optimisation Problem 1}: $\operatorname{max\,arg} G_{S, \Pi}(\av)$,  for $\av \in L^2(\Omega)^2$ with $P(\av) = P_0$, 
\\ \\
\noindent
where the power is written as function of $\av$ and $P_0$ is the  total  power. It is easy to see that
the equality constraint on the power can be replaced by the inequality
constraint $P(\av)\leq P_0$. In fact, if $P(\av) < P_0$, then
$G_{S,\Pi}(\av)$  is increased by multiplying $\av$ by a
number larger than 1.
So optimisation problem 1 is equivalent to:
 \\ \\ \noindent
\textbf{Optimisation Problem 2}:  $\operatorname{max\,arg}
G_{S,\Pi}(\av)$,  for $\av\in L^2(\Omega)^2$ with $P(\av)\leq P_0$.\\
\vspace*{0.0em}\\

\section{Lagrange multiplier rule for the optimum plane wave amplitudes}
\label{section.Lagrange}
If $\av$ is a solution of Problem 2, it will satisfy the
Lagrange multiplier rule \cite{Luenberger1969}. To formulate this we need to compute the Gateaux derivatives of the functionals $G_{S,\Pi}$ and $P$. 
For the Gateaux derivative of $G_{S,\Pi}$ (see \eqref{eq:GSav}) we get
\begin{equation}
  \begin{split}
    \delta G_{S,\Pi}(\av)(\bv) &= \lim_{t \to 0} \frac1t [G_{S,\Pi}(\av + t\bv) - G_{S,\Pi}(\av)]\\
    & = \frac{1}{8\pi^4} \Re \iint_{\Omega}
    \iint_{\Omega}  \FIII(T_S)(\kp-\kp',
    k_z(\kp)-k_z(\kp'))\,\MC_\Pi(\kp,\kp')\av(\kp')\cdot
    \bv(\kp)^* \Dkp \Dkp',
  \end{split}
  \label{eq:Gateaux-derivative-GS}
\end{equation}
where in the last step we have used \eqref{eq.MvMvT} and
\eqref{eq.FIIIsym}. We can similarly compute the Gateaux derivative  of $P$ (see \eqref{eq:Power-flow-Poynting-amplitudes}):
\begin{equation}
  \label{eq:Gateaux-derivative-P}
  \delta P(\av)(\bv) = \frac{1}{\omega \mu_0} \frac{1}{4\pi^2} \Re \iint_\Omega \av(\kp) \cdot \bv(\kp)^* k_z(\kp) \Dkp.
\end{equation}
Let $\av$ be a  solution of Problem 2. According  to the Lagrange multiplier rule there exists a number  $\Lambda'>0$ such that
\begin{equation}
  \delta G_{S,\Pi}(\av)(\bv) -\Lambda'  \delta P(\av)(\bv)=0, \;\; \text{for
    all } \bv \in L^2(\Omega)^2.
  \label{eq.LMR1}
\end{equation}
By substituting \eqref{eq:Gateaux-derivative-GS} and \eqref{eq:Gateaux-derivative-P}, and by choosing subsequently  $\bv$ real-valued and purely imaginary-valued, one can derive that 
\begin{equation}
  \iint_\Omega \FIII(T_S)(\kp-\kp',k_z(\kp)-k_z(\kp'))
  \MC_\Pi(\kp,\kp')\av(\kp') \Dkp' - \frac{2\pi^2 \Lambda'}{ \omega \mu_0}
  k_z(\kp) \av(\kp)=  0,
  \text{ for all } \kp \in \Omega.
  \label{eq.LMR2}
\end{equation}
If we define
\begin{equation}
  \Lambda = \frac{2\pi^2 \Lambda'}{  \omega \mu_0},
  \label{eq.defLambda}
\end{equation}
and the operator $\TC_{S,\Pi}: L^2(\Omega)^2 \mapsto L^2(\Omega)^2$ by
\begin{equation}
  \TC_{S,\Pi}(\av)(\kp) = \frac{1}{k_z(\kp)} \iint \FIII(T_S)(\kp-
  \kp', k_z(\kp) - k_z(\kp')) \MC_\Pi(\kp, \kp') \av(\kp') \Dkp',
  \label{eq.defoperatorTS_1}
\end{equation}
then \eqref{eq.LMR2} implies that $\av$ is eigenvector of operator $\TC_{S,\Pi}$ with eigenvalue $\Lambda$:
\begin{equation}
  \TC_{S,\Pi}(\av) - \Lambda \av = 0.
  \label{eq.LMR3}
\end{equation}
Note that, since $G_{S,\Pi}$ and $P$ are quadratic functionals
\begin{equation}
  \delta G_{S,\Pi}(\av)(\av) = 2G_{S,\Pi}(\av), \text{ and } \delta P(\av)(\av) = 2 P(\av).
  \label{eq.special}
\end{equation}
Then  \eqref{eq.LMR1} implies for the eigenvector satisfying $P(\av) = P_0$:
\begin{equation}
  G_{S,\Pi}(\av) = \Lambda' P_0.
  \label{eq.GSopt}
\end{equation}
We conclude that  the eigenfield with the largest eigenvalue is the solution of the optimisation problem.

Summarizing, we have found that for any bounded set $S$, (e.g., a set
 of finitely many points, a curve, a (curved) surface or a
volume)  the plane wave amplitudes of the field of which the average value of $|\Pi(\mathbf{E}))|^2$ over $S$ is maixum for a given power and numerical aperture,
is given by the eigenfield met maximum eigenvalue of  operator
$\TC_{S,\Pi}$ whose kernel  depends on the set $S$ and the projection $\Pi$. The function $\FIII(T_S)$ which occurs in the kernel of $\TC_{S,\Pi}$
is the 3D Fourier transform of the distribution $T_S$ defined by 
\eqref{eq.defT_S}, evaluated at spatial frequencies $\kp, k_z(\kp)$. The numerical aperture determines the domain $\Omega$ of the space $L^2(\Omega)^2$ for the operator and the eigenfields. 

\subsection{Examples}\label{sec:examplesop}
We  give some examples of the operator $\mathcal{T}_S$. 
\begin{enumerate}
\item If $S$ consists of one point: $S=\{({\rv_\perp}_0, z_0)\}$, with ${\rv_\perp}_0=(x_0,y_0)$, the optimisation problem amounts  to maximizing $|\Pi(\Ev)({\rv_\perp}_0,z_0)|^2$, i.e., the squared modulus of the projection $\Pi(\Ev)$ in point $({\rv_\perp}_0,z_0)$, for the given power. In particular, if $\Pi=\mathcal{I}$ (the identity), then the electric energy density in point $({\rv_\perp}_0,z_0)$ is maximized, whereas if $\Pi(\Ev)=\Ev\cdot\hat{\vv}$, the optimisation problem amounts to maximizing  the modulus of the component of the electric field along the direction $\hat{\vv}$  in  point $({\rv_\perp}_0,z_0)$.   We have
  \begin{equation}
    T_S(\rv_\perp,z) = \delta(\rv_\perp -{\rv_\perp}_0, z - z_0),
    \label{eq.TSex1}
  \end{equation}
   and hence
  \begin{equation}
    \FIII(T_S)(\xip, \xi_z) = \e^{-i\xip \cdot {\rv_\perp}_0 -i\xi_z z_0}.
    \label{eq.FTex1}
  \end{equation}
  Therefore operator \eqref{eq.defoperatorTS_1} becomes
  \begin{equation}
    \TC_{S,\Pi}(\av)(\kp) =
    \frac{\e^{-i {\rv_\perp}_0\cdot \kp} \e^{-i z_0 k_z(\kp)}}{k_z(\kp)} 
    \iint e^{i {\rv_\perp}_0\cdot \kp'} \e^{i z_0 k_z(\kp')}  \MC_\Pi(\kp, \kp') \av(\kp') \Dkp'.
    \label{eq.TPSex1}
  \end{equation}
\item Let $S$ be the part of the $z$-axis given by $-\ell/2< z <
  \ell/2$. Then the optimisation problem is to maximize the average
  value of $|\Pi(\Ev)|^2$ over the part of the $z$-axis given by $-\ell/2 \leq z \leq \ell/2$. We have
  \begin{equation}
    T_{S}(\rv_\perp,z) = \delta(\rv_\perp) \frac{1}{\ell} \indicator_{[-\ell/2, \ell/2]}(z),
    \label{eq.TSex2}
  \end{equation}
  where $\indicator_D(x) = 1$ if $x$ is in $D$ and $0$ elsewhere. $T_{S}$ has Fourier transform,
  \begin{equation*}
    \FIII(T_S)(\xip, \xi_z) = \frac{1}{\ell} \int_{-\ell/2}^{\ell/2} \e^{-i z \xi_z}
    \D{z} = \operatorname{sinc}\Bigl(\frac{\ell \xi_z}{2} \Bigr).
  \end{equation*}
  Hence \eqref{eq.defoperatorTS_1} becomes
  \begin{equation*}
    \TC_{S,\Pi}(\av)(\kp) = \frac{1}{k_z(\kp)} \iint_{\Omega} \operatorname{sinc}\biggl(\ell \frac{k_z(\kp) - k_z(\kp')}{2} \biggr) \MC_\Pi(\kp,\kp') \av(\kp') \Dkp',
    \label{eq.TSPex3}
  \end{equation*}

\item If $S = B_R$ is the sphere of radius $R > 0$ and centre the origin, then the optimisation problem is to maximize for the given power  the average value over this sphere of $|\Pi(\Ev)|^2$. There holds for $\rv=(x,y,z)$:
  \begin{equation}
    T_S(\rv) = \frac{\indicator_{B_R}(\rv)}{|B_R|}=\frac{ \indicator_{B_R}(\rv)}{\frac{4 }{3} \pi R^3},
    \label{eq.TSex3}
  \end{equation}
  with $ \indicator_{B_R}(\rv)=1$ if $r<R$ and $=0$ otherwise. We have
  \begin{equation}
    \label{eq:Fourier-transform-disc-R}
    \FIII(T_S)(\xip, \xi_z) = 2 \frac{J_{3/2}(R\sqrt{|\xip|^2 + \xi_z^2})}{(R\sqrt{|\xip|^2 + \xi_z^2})^{3/2}}.
  \end{equation}
  Hence,
  \begin{equation}
    \TC_{S,\Pi}(\av)(\kp) = \frac{2}{k_z(\kp)} \iint_{\Omega}
    \frac{J_{3/2}(R\sqrt{|\kp - \kp'|^2 + |k_z(\kp) - k_z(\kp')|^2})}{(R\sqrt{|\kp - \kp'|^2 + |k_z(\kp) - k_z(\kp')|^2})^{3/2}} \MC_{\Pi}(\kp,\kp') \av(\kp') \Dkp'.
    \label{eq.TSPex3}
  \end{equation}
\item If $S$ is the circular disc $D_R$ of radius $R > 0$ in the plane $z =
  0$ with centre the origin, then
  \begin{equation}
    T_S(\rv_\perp, z) = \frac{\indicator_{D_R}(\rv_\perp)}{|D_R|} \delta(z) = \frac{\indicator_{D_R}(\rv_\perp)}{\pi R^2} \delta(z) ,
    \label{eq.TSex4}
  \end{equation}
  where $\indicator_{D_R}(\rv_\perp)=1$ if $r_\perp<R$ and $=0$ otherwise. We have
  \begin{equation}
    \label{eq:Fourier-transform-disc-R}
    \FIII(T_S)(\xip, \xi_z) = 
    2 \frac{J_1(R|\kp|)}{R|\kp|}.
  \end{equation}
  Hence,
  \begin{equation}
    \TC_{S,\Pi}(\av)(\kp) = \frac{2}{k_z(\kp)} \iint_{\Omega}
    \frac{ J_1(R|\kp - \kp'|)}{R|\kp-\kp'|}  \MC_{\Pi}(\kp,\kp') \av(\kp') \Dkp'.
    \label{eq.TSPex4}
  \end{equation}
  We will study the optimisation problem for the disc in more detail in Section~\ref{section.disc} and 
  following sections.
\end{enumerate}

\subsection{Mathematical properties of the eigenvalue problem}
\label{subsection.mathprop}
We equip the  space $L^2(\Omega)^2$ of square integrable vector fields $\av:\Omega \mapsto C^2$ (where $\Omega$ is, as before, the circle of
finite numerical aperture \eqref{eq:Omega-kx-ky}) with the scalar product:
\begin{equation}
(\av, \bv) = \iint_\Omega \av(\kp) \cdot \bv(\kp)^* k_z(\kp) \Dkp.
\label{eq.scalarpr}
\end{equation}
This scalar product differs from the usual one by the factor $k_z(\kp)$ in the integrand, but the corresponding norm is equivalent to the usual $L^2$-norm. Hence also with respect to this scalar product, $L^2(\Omega)^2$ is a Hilbert space. Moreover, the power $P(\av)$ is proportional to $(\av, \av)$. However this is not the motivation for introducing this scalar product:  the reason is that with respect to this scalar product, operator $\TC_{S,\Pi}$ is symmetric:
\begin{equation}
(\TC_{S,\Pi}(\av), \bv)= (\av, \TC_{S,\Pi}(\bv)).
\label{eq.TSPisymm}
\end{equation}
It is can furthermore be verified that the kernel of operator $\TC_{S,\Pi}$ is square integrable with respect to 
the measure $kz(\kp)k_z(\kp') \Dkp, \Dkp'$:
\begin{equation}
  \iint_\Omega \iint_\Omega  \frac{1}{k_z(\kp)^2} \bigl|  \FIII(T_S)(\kp -
  \kp',k_z(\kp) - k_z(\kp')) \MC_\Pi^{ij}(\kp, \kp') \bigr|^2  k_z(\kp) k_z(\kp') \Dkp \Dkp' < \infty,
  \label{eq.HS}
\end{equation}
for $i, j = 1,2$.
This property implies that operator $\TC_{S,\Pi}$ is a Hilbert-Schmidt operator,
hence it is a self-adjoint compact operator $L^2(\Omega)^2\mapsto L^2(\Omega)^2$. Therefore
the spectrum of $\TC_{S,\Pi}$ is  real and discrete with all eigenvalues having
a finite number of linear independent eigenvectors. Furthermore, there
exists a basis of $L^2(\Omega)^2$ of eigenvectors of
$\TC_{S,\Pi}$ which is orthonormal with respect to the scalar product \eqref{eq.scalarpr}. The eigenvectors corresponding to the largest eigenvalue are,
after being properly normalized to give the maximum allowed power, the
solution of the optimisation problem. If the largest eigenvalue is not
degenerate, the optimum field is unique. However in general it can
happen that the largest eigenvalue is degenerate and then a finite
number of linear independent solutions of the optimisation problem
exist.

\subsection{Scaling law}
\label{subsection.scaling}
The optimisation problem depends on the chosen set $S$, the projection $\Pi$, the numerical aperture $\NA$, the wavenumber $k=k_0 n=2\pi n/\lambda$ and the power $P_0$. Suppose that $S$, $\Pi$ and  $\NA$ have been chosen  and suppose that we change the size of the set $S$ by multiplying it by a number $\sigma>0$:
$S \rightarrow \sigma S$. We have
\begin{align}
  \FIII(T_{\sigma S})(\xip,\xi_z) & = \frac{1}{\sigma S}  \iiint_{\sigma S} \e^{-i(\xip\cdot \rv + \xi_z z)} \Drp \D{z} \nonumber \\
  & = \frac{1}{S} \iint_{S} e^{-i \sigma \xip \cdot \rv' + \sigma \xi_z z'} \Drp' \D{z'} \nonumber \\
  & =  \FIII(T_{ S})(\sigma\xip,\sigma\xi_z). 
  \label{eq.FIIIscale}
\end{align}
Then
\begin{align}
 \FIII(T_{\sigma S})(\kp-\kp',k_z(\kp)-k_z(\kp')) &=  \FIII(T_{\sigma S})\left( k\frac{\kp}{k}-k \frac{\kp'}{k}, k \sqrt{1-k_\perp^2/k^2}-
 k \sqrt{1-k_\perp^2/k^2}\right)  \nonumber \\
 &= \FIII(T_{\sigma k S}) \left( \frac{\kp}{k}-\frac{\kp'}{k},  \sqrt{1-k_\perp^2/k^2}-
  \sqrt{1-k_\perp^2/k^2}\right).
  \label{eq.FIIIscale2}
  \end{align}
  Since $\MC_\Pi$ actually is a function of $\kp/k, \kp'/k$ we  write in this section
  \[
  \MC_\Pi\left(\frac{\kp}{k},\frac{\kp'}{k}\right)
  \]
  instead of $\MC_\Pi(\kp,\kp')$.
  Substitution into  (\ref{eq.defoperatorTS_1}) then gives
  \begin{align}
  \TC_{\sigma S,\Pi}(\av)(\kp)  =   \frac{1}{k_z(\kp)} \iint_\Omega \FIII(T_{\sigma S})(\kp-
  \kp', k_z(\kp) - k_z(\kp')) \MC_\Pi \left(\frac{\kp}{k}, \frac{\kp'}{k}\right) \av(\kp') \Dkp'
   \nonumber \\
   = \frac{k}{\sqrt{1-\frac{k_\perp^2}{k^2}}} \iint_{k_\perp'/k\leq \sin \amax} 
   \FIII(T_{\sigma k S}) \left( \frac{\kp}{k}-\frac{\kp'}{k},  \sqrt{1-k_\perp^2/k^2}-
  \sqrt{1-k_\perp^2/k^2}\right) \MC_\Pi\left(\frac{\kp}{k}, \frac{\kp'}{k}\right) \av(\kp')\,  \mbox{d}\left(\frac{\kp'}{k}\right) \nonumber \\
  \label{eq.T_S_rescaled}
\end{align}
After dividing by $k$ this expression  only depends  on the product of $\sigma$ and $k$ and not on $\sigma$ and $k$ separately.  By dividing eigenvalue problem
(\ref{eq.LMR3}) for $\sigma S$  by $k$ we obtain the eigenvalue problem
\begin{equation}
  \frac{1}{k}\TC_{\sigma S,\;\Pi}(\av) -\frac{ \Lambda}{k} \av = 0,
  \label{eq.LMR3Ssigma}
\end{equation}
  which  depends on $\sigma$ and $k$ only through the product $\sigma k$. We therefore conclude that
  the eigenvectors $\av$ are the same if $\sigma k$ is  kept constant while the eigenvalues $\Lambda$ are proportional to $k$, i.e., inversely proportional to the wavelength. Then (\ref{eq.defLambda}) implies that
  \begin{equation}
  \Lambda' = \frac{\omega \mu_0 \Lambda}{  2\pi^2 } =  \sqrt{\frac{\mu_0}{\epsilon_0}} \frac{1}{2\pi^2 n} k \Lambda
  \propto k^2.
  \label{eq.Lambdaprime}
  \end{equation}
  and hence with  (\ref{eq.GSopt}) it follows that for fixed $\sigma k$ and fixed power $P_0$ the maximum value of the object function
  is proportional to $k^2$.
  
  Summarizing we conclude that if $\amax$,  $P_0$ and the product $\sigma k$ are fixed, where  $\sigma$ is a scaling parameter of the set $S$ and $k$ is the wavenumber, the optimum fields are the same, while the maximum of the object function depends quadratically on the wavenumber.

\section{Realisation of the optimum fields}\label{sec:pupil-field}
An obvious way to realize the optmum field is in the focal region of a
lens using  spatial light modulators (SLMs) to shape the field in the entrance pupil.
The numerical aperture of the lens should be at least as large as that
of the optimum field. Since the plane wave amplitude of the electric field 
 in the focal region  corresponds 1-to-1 to the electric field
in the entrance pupil, the desired amplitude, phase and
polarization of these plane waves  can be obtained by programming a number 
 SLMs in series \cite{Neil, Sheppard, Iglesias,
  Sheppard2, Urbach, deBruin}.
Let $\{\xu, \yu,\zu\}$ be the standard Euclidean basis in the focal
region, with $\zu$ in the direction of the optical axis and pointing
away from the lens. Let  $\hat{\vec{x_\mathnormal{e}}}$,
$\hat{\vec{y_\mathnormal{e}}} $  be unit vectors of the Euclidean coordinate
system in the entrance pupil of the lens that are parallel to $\xu$ and $\yu$
respectively. We will use polar coordinates $\rho_e$ and $\varphi_e$ in the lens
pupil:
\begin{equation}
  \label{eq:polar-coordinate-pupil}
  x_e = \rho_e \cos \varphi_e, \quad y_e = \rho_e \sin \varphi_e.
\end{equation}
The unit vectors $\hat{\vec{\rho_\mathnormal{e}}}$ and
$\hat{\vec{\phi_\mathnormal{e}}}$ are then given by
\begin{align}
  \label{eq:pupil-rho}
  \hat{\vec{\rho_\mathnormal{e}}} &= \cos \varphi_e \hat{\vec{x_\mathnormal{e}}} + \sin \varphi_e \hat{\vec{y_\mathnormal{e}}},\\
  \hat{\vec{\phi_\mathnormal{e}}} &= -\sin \varphi_e \hat{\vec{x_\mathnormal{e}}} + \cos \varphi_e \hat{\vec{y_\mathnormal{e}}}.
\end{align}
Note that $\{\hat{\vec{\rho_\mathnormal{e}}}, \hat{\vec{\phi_\mathnormal{e}}}, \zu\}$ is a
positively oriented basis. Any beam incident on the lens is
predominantly propagating parallel to the optical axis and therefore the
$\zu$-component of its field is neglected. Using the polar basis,  the electric field at a point
$(\rho_e, \varphi_e)$ in the entrance pupil is written as
\begin{equation}
  \label{eq:pupil-basis-expansion}
  \Ev^e(\rho_e, \varphi_e) = E_\rho^e(\rho_e, \varphi_e)
  \hat{\vec{\rho_\mathnormal{e}}} +  E_\varphi^e(\rho_e, \varphi_e) \hat{\vec{\phi_\mathnormal{e}}}.
\end{equation}
We write the vector amplitude $\av$ of the plane wave on the $(\ku,
\pu, \su)$ basis as before as
\begin{equation*}
  \av(\kp) = a_p(\kp) \pu(\kp) + a_s(\kp) \su(\kp).
\end{equation*}
The  point in the pupil  and the  corresponding wave vector $\kv =
k_x \xu + k_y \yu + k_z \zu$, of the angular spectrum
of the field in the focal region  are related by
\begin{align}
  k_x &= -k \frac{k x_e}{f} = -k \frac{\rho_e}{f} \cos\varphi_e,   \label{k_xpupil} \\
  k_y &= -k \frac{k y_e}{f} = -k \frac{\rho_e}{f} \sin\varphi_e,  \label{k_ypupil}
\end{align}
where $f$ is the focal distance. According to the theory of Ignatowski
\cite{Ignatowski1, Ignatowski2}, and Richards and Wolf \cite{Richards1959}
the radial and azimuthal 
components of the pupil field are proportional to  $a_p$ and  $a_s$, respectively:
\begin{align}
  \label{eq:Erho}
  E_\rho^e(\rho_e, \varphi_e) &= \frac{\sqrt{k k_z}}{2 \pi i f}
                                a_p\Bigl(-k \frac{\rho_e}{f} \cos\varphi_e, -k
                                \frac{\rho_e}{f} \sin\varphi_e \Bigr),\\
  \label{eq:Ephi}
  E_\varphi^e(\rho_e, \varphi_e) &= \frac{\sqrt{k k_z}}{2 \pi i f}
                                   a_s\Bigl(-k \frac{\rho_e}{f} \cos\varphi_e, -k
                                   \frac{\rho_e}{f} \sin\varphi_e \Bigr).
\end{align}
where the factor $\sqrt{k k_z}/(2\pi f)$ is included to account for energy
conservation and where
\begin{equation}
  \label{eq:Ekz}
  k_z = k \sqrt{1 - \frac{\rho_e^2}{f^2}}.
\end{equation}
Hence, written on the $\{\xu, \yu\}$ basis, $\Ev^e$ becomes:
\begin{equation}
  \begin{split}
    \Ev^e(\rho_e, \varphi_e) &= \frac{\sqrt{k k_z}}{2 \pi i f}
    [ a_p(-k_x, -k_y) \cos\phi_e - 
    a_s(-k_x, -k_y) \sin\phi_e] \xu\\
    &\quad + \frac{\sqrt{k k_z}}{2 \pi i f}
    [a_p(-k_x, -k_y) \sin\phi_e + 
    a_s(-k_x, -k_y) \cos\phi_e] \yu.
  \end{split}
  \label{eq.Epupil}
\end{equation}
The pupil field can be quite general as every point of the pupil can have its own elliptical state of polarization and the phase difference between the fields in different points of the pupil can be arbitrary. 

\section{Optimising the electric energy in a disc}
\label{section.disc}
In the remainder of this paper we will study the example of Section~\ref{sec:examplesop}, where the region $S$ is the
disc $S=D_R=\{ (\rv,z); r< R, z=0\}$ and the projection is the identity: $\Pi=\mathcal{I}$. Hence
\begin{equation}
  G_{S,\Pi}(\av) =
  \frac{1}{\pi R^2} \iint_{D_R} |\Ev(x,y,0)|^2  \D{x} \D{y},
  \label{eq.defGS}
\end{equation}
and the optimisation problem amounts to finding the  field of which the electric energy averaged over the disc $D_R$  is maximum for given power $P_0$. The optimum plane wave amplitude $\av$ is the  eigenvector: 
\begin{equation}
  \TC_{S,\Pi}(\av) - \Lambda \av = 0.
  \label{eq.LMR3b}
\end{equation}
correspnding to  the largest eigenvalue $\Lambda$ of operator $\TC_{S, \Pi}$ defined by \eqref{eq.TSPex4}:
\begin{equation}
  \TC_{S,\Pi}(\av)(\kp) = \frac{2}{k_z(\kp)} \iint_{\Omega}
  \frac{ J_1(R|\kp - \kp'|)}{R|\kp-\kp'|}  \MC_{\Pi}(\kp,\kp') \av(\kp') \Dkp'.
  \label{eq.TSPex4b}
\end{equation}

\subsection{ Expressions  in terms of azimuthal and polar angles}
It is convenient to change the integration variables from $\kp$ to azimuthal and polar angles $0< \alpha < \amax$ and $0 < \beta < 2\pi$, where $\amax= \arcsin(\NA/n)$. We have
\begin{equation}
    k_x  = k \sin\alpha\cos\beta, \;\;\;\; 
    k_y = k \sin\alpha \sin\beta, \label{eq.kxkyalphabeta}
\end{equation}
so that the normalised wavevector $\ku$ is 
\begin{equation}
  \label{eq:Image-space-basis}
    \ku(\kp)= \ku(\alpha, \beta) =
                                   \begin{pmatrix}
                                     \sin \alpha \cos\beta \\ \sin\alpha \sin\beta \\ \cos\alpha
                                   \end{pmatrix},
                                   \end{equation}
    and  $\pu$ and $\su$ are given by
    \begin{equation}
    \pu(\alpha, \beta) =
                         \begin{pmatrix}
                           -\cos \alpha \cos \beta \\ -\cos \alpha \sin \beta \\ \sin \alpha\\
                         \end{pmatrix},
    \;\;\;\; 
    \su(\beta) =
                 \begin{pmatrix}
                   \sin \beta \\ -\cos \beta \\ 0
                 \end{pmatrix}.
                 \label{eq.pusu}
\end{equation}
Writing
\begin{equation}
  k_x' = k \sin\alpha'\cos\beta', \;\;\;\;
  k_y' = k \sin\alpha'\sin\beta',
  \label{eq.kxkyprime}
\end{equation}
we get
\begin{equation}
  \label{eq:dependence-kpmkpp}
  |\kp - \kp'|^2 = k^2 [\sin^2 \alpha + \sin^2 \alpha' - 2 \sin\alpha
  \sin\alpha' \cos(\beta - \beta')].
\end{equation}
and therefore
\begin{equation}
  \frac2{k_z(\kp)} \frac{J_1(R|\kp - \kp'|)}{R|\kp - \kp'|}  = 
  \frac{2}{k \cos \alpha} \frac{ J_1\left(k R \sqrt{ \sin^2\alpha + \sin^2 \alpha' - 2 \sin\alpha \sin\alpha' \cos(\beta-\beta')}\right)}{ k R \sqrt{ \sin^2\alpha + \sin^2 \alpha' - 2 \sin\alpha \sin\alpha' \cos(\beta-\beta')}}.
  \label{eq.J1}
\end{equation}
Furthermore, using (\ref{eq:matrix-first-def}) with $\Pi=\mathcal{I}$, 
\begin{equation}
  \MC_{\Pi}(\kp,\kp')= \MC_\Pi(\alpha,\alpha',\beta-\beta'),
  \label{eq.Malpha}
\end{equation}
where
\begin{equation}
  \MC_\Pi(\alpha,\alpha',\beta)=
  \begin{pmatrix}
    \cos \alpha \cos \alpha' \cos \beta + \sin\alpha \sin\alpha' \;\; & \cos\alpha \sin\beta  \\
    -\cos\alpha' \sin\beta\;\; &   \cos\beta
  \end{pmatrix}.
  \label{eq.MCalphabeta}
\end{equation}
Using
\begin{equation}
  \Dkp'= \!\D k_x' \D k_y' = k^2 \sin\alpha' \cos\alpha' \D\alpha' \D\beta',
  \label{eq.Dkp}
\end{equation}
we conclude that  \eqref{eq.TSPex4b} becomes
\begin{equation}
  \TC_{S,\Pi}(\av)(\alpha,\beta) = \int_0^{\amax} \int_0^{2\pi} C_R(\alpha,\alpha',\beta-\beta')
  \MC_\Pi(\alpha,\alpha',\beta-\beta') \av(\alpha',\beta') \D\alpha' \D\beta'.
  \label{eq.TSP_conv}
\end{equation}
where
\begin{equation}
  C_R(\alpha,\alpha',\beta) = \frac{2k \cos \alpha' \sin \alpha'}{\cos\alpha}  \frac{ J_1(k R \sqrt{ \sin^2\alpha + \sin^2 \alpha' - 2 \sin\alpha \sin\alpha' \cos\beta})}{ k R \sqrt{ \sin^2\alpha + \sin^2 \alpha' - 2 \sin\alpha \sin\alpha' \cos\beta}}.
  \label{eq.defCR}
\end{equation}
Note that the integral with respect to $\beta'$ is a convolution.
\subsection{Fourier series}
We shall use a Fourier series for $\beta \mapsto \av(\alpha,\beta)$:
\begin{equation}
  \av(\alpha,\beta) = \sum_\ell \hat{\av}(\alpha,\ell) \e^{i \ell \beta},
  \label{eq.FSav}
\end{equation}
Let  $\MC_R$  be the matrix
\begin{equation}
  \MC_R(\alpha,\alpha,\beta) = C_R(\alpha,\alpha'\beta) \MC_\Pi(\alpha,\alpha',\beta).
  \label{eq.defMCR}
\end{equation}
Writing
\begin{equation}
  \MC_R(\alpha,\alpha',\beta) = \sum_\ell \widehat{\MC_R}(\alpha,\alpha',\ell) \e^{i \ell \beta},
  \label{eq.FSMCR}
\end{equation}
it follows that
\begin{equation}
  \widehat{\MC_R}(\alpha,\alpha',\ell) =\sum_{\ell'}
  \widehat{C_R}(\alpha,\alpha',\ell-\ell') \widehat{\MC_\Pi}(\alpha,\alpha',\ell').
  \label{eq.FSMCR2}
\end{equation}
where $\widehat{C_R}(\alpha,\alpha',\ell)$ are the Fourier coefficients of 
$\beta \mapsto C_R(\alpha,\alpha',\beta)$ and 
\begin{equation}
  \label{eq:kernel-matrix-ell}
  \widehat{\MC_\Pi}(\alpha, \alpha', \ell) = \delta_{\ell, 0} \begin{pmatrix}
    \sin\alpha  \sin\alpha' & 0\\
    0 & 0
  \end{pmatrix} + \frac{\delta_{\ell, \pm 1}}2 \begin{pmatrix}
    \cos\alpha \cos\alpha' & \mp i \cos\alpha\\
    \pm  i \cos\alpha'& 1
  \end{pmatrix},
\end{equation}
where $\delta_{\ell, \ell'} = 1$ when $\ell = \ell'$ and $=0$ otherwise.
Hence,
\begin{equation}
  \label{eq:convolution-theorem}
  \widehat{\MC_R}(\alpha, \alpha',\ell) =  
  \widehat{C_R}(\alpha, \alpha',\ell+1) \widehat{\MC_\Pi}(\alpha, \alpha',-1)
  +
  \widehat{C_R}(\alpha, \alpha',\ell) \widehat{\MC_\Pi}(\alpha, \alpha',0)
  +
  \widehat{C_R}(\alpha, \alpha',\ell-1) \widehat{\MC_\Pi}(\alpha, \alpha',1).
\end{equation}
TheFourier  coefficients $\widehat{C_R}(\alpha,\alpha', \ell)$ are computed in Appendix~\ref{sec:fourier-coefficients}. 
Operator \eqref{eq.TSP_conv} can now be written as:
\begin{equation}
  \TC_{S,\Pi}(\av)(\alpha,\beta) =2\pi  \sum_\ell \int_0^{\amax} \widehat{\MC_R}(\alpha,\alpha',\ell)
  \hat{\av}(\alpha',\ell ) \D\alpha' \e^{i \ell \beta}.
  \label{eq.FSTSP}
\end{equation}
By computing the Fourier coefficients of \eqref{eq.LMR3b} it follows that the 
eigenvalue problem is equivalent to the following set of eigenvalue problems
\begin{equation}
  2\pi \int_0^{\amax} \widehat{\MC_R}(\alpha,\alpha',\ell)
  \hat{\av}(\alpha', \ell ) \D\alpha'- \Lambda \hat{\av}(\alpha,\ell)=0, \;\; \mbox{ for all integers }
  \ell,
  \label{eq.eigenell}
\end{equation}
(where  eigenvalue $\Lambda$ depends on $\ell$).
Hence we have obtained an eigenvalue problem for every Fourier component $\hat{\av}(\alpha,\ell)$.
Because $\C_R$  and $\MC_\Pi$ are real-valued, we have
\begin{align}
  \widehat{C_R}(\alpha,\alpha', -\ell) &= \widehat{C_R}(\alpha, \alpha', \ell)^\ast, \label{eq.hatCRprop}\\
  \widehat{\MC_\Pi}(\alpha,\alpha', -\ell) &= \widehat{\MC_\Pi}(\alpha, \alpha', \ell)^\ast,  \label{eq.widehatMRpropr}
\end{align}
and hence also
\begin{equation}
  \widehat{\MC_R}(\alpha,\alpha',-\ell)= \widehat{\MC_R}(\alpha, \alpha', \ell)^\ast.
  \label{eq.widehatMRporpr}
\end{equation}
This implies that if $\hat{\av}(\alpha, \ell)$ is a solution of the eigenvalue problem for $\ell$,
$\hat{\av}(\alpha, -\ell)^*$ is solution of the eigenvalue problem for $-\ell$. Furthermore the eigenvalues for $\ell$ and $-\ell$ are the same. We may therefore assume that the eigenfields
$\av(\alpha,\beta)$ are real and harmonic in $\beta$:
\begin{equation}
  \av(\alpha,\beta) = \hat{\av}(\alpha,\ell)^\ast\e^{-i\ell \beta}+
  \hat{\av}(\alpha,\ell) \e^{i\ell \beta} = 2 \Re[
  \hat{\av}(\alpha,\ell) \e^{i \ell\beta}].
  \label{eq.sol_av}
\end{equation}

It is clear that when $\hat{\av}$ is a solution for given $\ell$, so
is $\hat{\av} \e^{i \ell\psi}$, for arbitrary $\psi$.
This implies that for every eigenvector $\av(\alpha,\beta)$,  
$\av(\alpha, \beta + \psi)$ is also eigenvector.  This reflects the rotational symmetry of the problem.

The optimum field we are looking for is eigenvector for the value of
$\ell$ for which the eigenvalue of \eqref{eq.eigenell} is largest. 
Because $C_R$ is a an analytic function of $\beta$, we have for $\ell $ large that $\widehat{C_R}(\alpha,\alpha',\beta) \rightarrow 0$ faster than any power $\ell^{-m}$, $m=1,2,\ldots$ and uniformly for $0 < \alpha, \alpha' < \amax$. Hence also 
\begin{equation}
  |\widehat{M_R}(\alpha,\alpha',\ell) | \leq C \frac{1}{\ell^m}, \mbox{  for } \ell \rightarrow \infty
  \label{eq.estimate}
\end{equation}
for some constant $C$ (depending on $m$) and uniformly in $\alpha$, $\alpha'$. This shows that the eigenvalues of the operator \eqref{eq.eigenell} become arbitrary small in the limit $\ell \rightarrow \infty$. Therefore, the maximum eigenvalue occurs for some finite $\ell$.
As
discussed in the section with numerical results, it can happen that
the eigenvalues for different $\ell$ are the same and  both maximum. In
that case there are two fields with  different $\ell$  which both
are solutions of the optimisation problem. 

FInally, we express also the  power flux \eqref{eq:Power-flow-Poynting-amplitudes} of the solution  in terms of the Fourier coefficients of the optimum  plane wave amplitudes:
\begin{eqnarray}
  P(\av) &=& \frac{1}{\omega \mu_0} \frac{1}{8\pi^2} \iint_{\Omega} |\av(\kp)|^2 k_z \Dkp \nonumber \\
         &=& \frac{1}{\omega \mu_0} \frac1{8 \pi^2} \iint_\Omega [|a_p(\kp)|^2 + |a_s(\kp)|^2] k_z \Dkp \nonumber\\
         &=& n \sqrt{\frac{\epsilon_0}{\mu_0}} \frac{k^2}{8\pi^2} \int_0^{\amax} \! \int_0^{2\pi} [a_p(\alpha,\beta)|^2 + |a_s(\alpha,\beta)|^2 ] \cos^2\alpha \sin\alpha  \D\beta \D\alpha \nonumber \\
         &= & n \sqrt{\frac{\epsilon_0}{\mu_0}} \frac{k^2}{4\pi} \int_0^{\amax} [ |\widehat{a}_p(\alpha,\ell)|^2 + |\widehat{a}_s(\alpha,\ell)|^2 ] \cos^2 \alpha \sin \alpha \D\alpha,
          \label{eq.PowerFourier}
\end{eqnarray}
for the optimum $\ell$.


\subsection{Optimum pupil fields}
From \eqref{k_xpupil}, \eqref{k_ypupil} and (\ref{eq.kxkyalphabeta})
 it follows that the pupil coordinates $\rho_e$, $\phi_e$ are related to $\alpha$, $\beta$ by
\begin{equation}
  \beta = \phi_e+ \pi \text{ and } \sin \alpha = \frac{\rho_e}{f}. \label{eq.alpharho_e}
\end{equation}
Let 
\begin{equation}
  \av(\alpha,\beta)=2\Re[\hat{\av}(\alpha,\ell)e^{i \ell \beta}] =2 \Re \Biggl[ 
  \begin{pmatrix}
    \widehat{a_p}(\alpha,\ell) \\
    \widehat{a_s}(\alpha,\ell)
  \end{pmatrix} \e^{i \ell \beta} \Biggr],
  \label{eq.avop}
\end{equation}
be a solution of eigenvalue problem \eqref{eq.eigenell} for the value of $\ell$ for which the eigenvalue is maximum.  If we  normalize $\av$  such that  the power satisfies $P(\av)=P_0$,  $\av$ is a solution of the optimisation problem. According to  \eqref{eq:Erho} and \eqref{eq:Ephi} the radial and azimuthal components of the corresponding pupil field are
\begin{align}
  E_\rho^e(\rho_e, \phi_e) &= 2\frac{k (1-\rho_\e^2/f^2)^{1/4}}{2\pi i f} \Re [\widehat{a_p}(\alpha, \ell) \e^{i \ell \phi_e}],   \label{eq.Erho_opt}\\ 
  E_\phi^e(\rho_e,\phi_e) &=2 \frac{k (1-\rho_e^2/f^2)^{1/4}}{2\pi i f} \Re [\widehat{a_s}(\alpha, \ell) \e^{i \ell \phi_e}], 
                            \label{eq.Ephi_opt}
\end{align}
where the irrelevant factor $\e^{i \ell \pi}=(-1)^{\ell}$ has been omitted.
On the cartesian basis we have (see \eqref{eq.Epupil}):
\begin{align}
  \Ev^e(\rho_e, \varphi_e) &=  2\frac{k ( 1 -\rho_e^2/f^2)^{1/4}}{2 \pi i f}
                             \Bigl\{ \Re[ \widehat{a_p}(\alpha, \ell) \e^{i \ell \phi_e}]  \cos\phi_e - 
                             \Re[\widehat{a_s}(\alpha, \ell) \e^{i\ell \phi_e}]  \sin\phi_e \Bigr\} \xu \nonumber \\
                           &  \quad +2 \frac{k ( 1 -\rho_e^2/f^2)^{1/4}}{2 \pi i f} 
                             \Bigl\{ \Re[ \widehat{a_p}(\alpha, \ell)\e^{i \ell \phi_e}] \sin\phi_e + 
                             \Re[\widehat{a_s}(\alpha, \ell) \e^{i \ell \phi_e}]  \cos\phi_e \Bigr\} \yu.
                             \label{eq.Epupil_opt2}
\end{align}
It is seen  that  the optimum pupil field is linear polarized, but that the direction of the polarisation  strongly varies throughout the pupil. By multiplying \eqref{eq.Erho_opt} and
\eqref{eq.Ephi_opt} by $\e^{-i\omega t}$ and taking the real part, it follows that the azimuthal and polar components of the time dependent electric field are in phase throughout the pupil, i.e., they all have value zero at the same time during a period of the field oscillation.

\subsection{Optimum field in the focal region}
The optimum field in the focal region is the (rescaled) Fourier transform of the optimum pupil field. 
We rewrite the Foruier transforms in terms of integrals over polar and azimuthal angles.
We have, for some $\ell$:
\begin{align}
  a_p(\alpha,\beta) &= 2\Re[ \widehat{a_p}(\alpha,\ell) \e^{i\ell \beta}], \label{eq.aPopt} \\
  a_s(\alpha,\beta) &=2 \Re[ \widehat{a_s}(\alpha,\ell) \e^{i\ell \beta}], \label{eq.aSopt}
\end{align}
where $\alpha$ and $\beta$ are related to $\kp$ by (\ref{eq.kxkyalphabeta}).
By applying the change of integration variables $\kp \mapsto (\alpha,\beta)$ to \eqref{eq:electric-field2}, using \eqref{eq.Dkp}, we find that the optimum electric field in the focal region is given by
\begin{align}  
  \Ev(\rv) &=  \frac{1}{4\pi^2}  \iint_\Omega [a_p(\kp \pu(\kp) + a_s(\kp)\su(\kp) ] \e^{ i \kv\cdot \rv} \Dkp \nonumber \\
           &=  \frac{k^2}{2\pi^2} \int_0^{\amax}\! \!\int_0^{2\pi}  \Re [
                \widehat{a_p}(\alpha,\ell) \e^{i \ell \beta}   \pu(\alpha, \beta)  + \widehat{a_s}(\alpha,\ell)
                \e^{i \ell \beta} \su(\beta)] \e^{i k ( x \sin\alpha \cos \beta + y \sin\alpha \sin \beta +z\cos \alpha)}\, \cos \alpha \sin \alpha \, \D\alpha \D\beta.  \nonumber \\
  \label{eq.Efocal}
\end{align}
Expressed in cylindrical coordinates
\begin{equation}
  x=\rho \cos \phi,  \; y=\rho \sin \phi.
  \label{eq.cyl}
\end{equation}
this becomes
\begin{align}  
  \Ev(\rho,\phi,z) =  \frac{k^2}{2\pi^2} \int_0^{\amax}\!\!\int_0^{2\pi} \Re [
  \widehat{a_p}(\alpha,\ell) \e^{i \ell \beta}   \pu(\alpha, \beta)  + \widehat{a_s}(\alpha,\ell)
  \e^{i \ell \beta} \su(\beta)] \e^{i k ( \rho
  \sin\alpha \cos (\beta-\phi) +z \cos \alpha)}\,\cos\alpha \sin \alpha\, 
  \D\alpha \D\beta.
  \label{eq.Evfocal}
\end{align}
For the magnetic field we have similarly from \eqref{eq:magnetic-field}:
\begin{align}
  \Hv(\rho,\phi,z) & = \frac{1}{\omega \mu_0} \frac{1}{4\pi^2}
             \iint_{\Omega} \bigl[ a_p(\kp) \su(\kp) - a_s(\kp)\pu(\kp)\bigr]
             \e^{i \kv \cdot \rv} \Dkp \nonumber \\
           & = n \sqrt{\frac{\epsilon_0 }{\mu_0}}  \frac{k^2}{2\pi^2}\int_0^{\amax}\! \!\int_0^{2\pi} \Re \Bigl[
             \widehat{a_p}(\alpha,\ell)\e^{i \ell \beta} \su(\beta) - \widehat{a_s}(\alpha,\ell)
             \e^{i \ell \beta} \pu(\alpha, \beta) \Bigr] \e^{i k ( \rho
             \sin\alpha \cos (\beta-\phi) +z\cos \alpha)} \cos\alpha \sin \alpha \D\alpha \D\beta.
             \label{eq.Hvfocal}
\end{align}
The integrals over $\beta$ in (\ref{eq.Evfocal}) and (\ref{eq.Hvfocal}) can be computed analytically.
The derivation and results are given in Appendix \ref{appendix_C}.

\section{Results for the maximum energy in a disc}
\label{section.results}
We start with a special case for which the solution can be computed in closed form.
\subsection{The solution for a disc with radius $R = 0$}
This means that we are maximizing the electric energy density in the
origin, i.e., \eqref{eq.defGS} becomes
\begin{equation}
  G_{S,\Pi}(\av) =
  \frac{1}{\pi R^2} \iint_{D_R} |\Ev(x,y,0)|^2  \D{x} \D{y} \rightarrow |\Ev(\mathbf{0})|^2.
  \label{eq.defGSR0}
\end{equation}
We have
\begin{equation}
C_{R=0}(\alpha,\alpha',\beta)= \frac{k \cos\alpha'\sin\alpha'}{\cos\alpha},
\label{eq.CR0}
\end{equation}
so that
\begin{eqnarray}
\hat{C_{R=0}}(\alpha,\alpha',\ell)   =    \frac{k \cos\alpha'\sin\alpha'}{\cos\alpha} \delta_{\ell,0}. \label{eq.hatCR0} 
\label{eq.CR0ell}
\end{eqnarray}
Then, (\ref{eq:kernel-matrix-ell}) implies:
\begin{eqnarray}
 \widehat{\MC_{R=0}}(\alpha, \alpha',0) & = & 
 \frac{k \cos\alpha'\sin\alpha'}{\cos\alpha}
 \begin{pmatrix}
     \sin\alpha \sin \alpha'  \, \, & 0 \\
    0  \,\, &  0  
  \end{pmatrix},
  \label{eq.hatMC_R00} \\
 \widehat{\MC_{R=0}}(\alpha, \alpha',1) & = & 
 \frac{k \cos\alpha'\sin\alpha'}{2\cos\alpha}
 \begin{pmatrix}
     \cos\alpha \cos \alpha'  \, \, &- i \cos \alpha \\
    i \cos \alpha'  \,\, &  1  
  \end{pmatrix},
  \label{eq.hatMC_R01}  \\
 \widehat{\MC_{R=0}}(\alpha, \alpha',\ell) & = & 
 \begin{pmatrix}
    0  \, \, & 0 \\
    0  \,\, &  0  
  \end{pmatrix},  \;\; \mbox{ if } \ell > 1.
  \label{eq.hatMC_R0ell}
\end{eqnarray}
Hence the optimum solution   either has $\ell=0$ or $\ell=1$. 

We consider first $\ell=0$.
Substitution of  (\ref{eq.hatCR0})  into  (\ref{eq.eigenell}) with $\ell=0$, implies:
\begin{eqnarray}
2\pi k \tan \alpha   \int_0^{\amax}  \cos\alpha' \sin^2\alpha'  \hat{a}_p(\alpha',0) d\alpha' = \Lambda \hat{a}_p(\alpha,0),
\label{eq.integral0}
\end{eqnarray}
and
\begin{equation}
\hat{a}_s(\alpha,0)=0.
\label{eq.as0}
\end{equation}
Hence, $\hat{a}_p(\alpha,0)$ is proportional to $\tan \alpha$ and using this fact it follows from (\ref{eq.integral0}) and
(\ref{eq.PowerFourier}):
\begin{equation}
\Lambda = 2\pi k \left( \frac{2}{3}- \cos \amax + \frac{1}{3}\cos^3 \amax \right),
\label{eq.Lambda0}
\end{equation}
and
\begin{equation}
\hat{a}_p(\alpha,0)= 2\pi \left(\frac{2}{k n} \frac{P_0}{\Lambda}\right)^{1/2} \left( \frac{\mu_0}{\epsilon_0}\right)^{1/4}  \tan \alpha.
\label{eq.solap0}
\end{equation}

Next we consider the case $\ell=1$. By substituting (\ref{eq.hatMC_R01}) into  (\ref{eq.eigenell}) with $\ell=1$, one finds
\begin{eqnarray}
\int_0^{\amax} \cos^2\alpha' \sin\alpha' \; \hat{a}_p(\alpha',1) d\alpha' - i \int_0^{\amax} \cos\alpha'\sin\alpha'\; \hat{a}_s(\alpha',1) d\alpha' = \frac{\Lambda }{\pi k} \; \hat{a}_p(\alpha,1), \label{eq.ap1} \\
\frac{i}{\cos\alpha} \int_0^{\amax} \cos^2\alpha'\sin\alpha'\; \hat{a}_p(\alpha',1)+ \frac{1}{\cos\alpha} \int_0^{\amax} \cos\alpha'\sin\alpha' \; \hat{a}_s(\alpha',1)d\alpha' = \frac{\Lambda}{\pi k} \hat{a}_s(\alpha,1).
\label{eq.as1}
\end{eqnarray}
Hence,
\begin{equation}
\hat{a}_p(\alpha,1)= C_p, \;\; \;\; \hat{a}_s(\alpha,1)=\frac{C_s}{\cos\alpha},
\label{eq.conclus}
\end{equation}
where $C_p$ and $C_s$ are constants. Substituting (\ref{eq.conclus}) into (\ref{eq.ap1}) and (\ref{eq.as1}) implies
\begin{equation}
\begin{pmatrix}
     \frac{1}{3}(1-\cos^3\amax) \;\; & -i(1-\cos\amax) \\
     \frac{i}{3}(1-\cos^3 \amax) \;\; & 1 - \cos \amax 
  \end{pmatrix} \begin{pmatrix}
      C_p \\
      C_s
      \end{pmatrix} = 
      \frac{\Lambda}{\pi k} \begin{pmatrix}
          C_p \\
      C_s
      \end{pmatrix}
  \label{eq.matCpCs} 
  \end{equation}
The largest eigenvalue  is given by
\begin{equation}
    \Lambda = \pi k\left( \frac{4}{3} - \cos \amax  -\frac{1}{3} \cos^3 \amax \right),
    \label{eq.Lambda1}
    \end{equation}
    with eigenvector
    \begin{equation}
    \begin{pmatrix}
      C_p \\
      C_s
      \end{pmatrix}=\begin{pmatrix}
        1 \\
        i 
        \end{pmatrix}
        \label{eq.eigenv}
        \end{equation}
        Eigenvalue (\ref{eq.Lambda1}) is for all $\amax>0$ strictly larger than eigenvalue (\ref{eq.Lambda0}), hence the optimum solution corresponds to $\ell=1$.  We have
        \begin{eqnarray}
        \Re\left[ \hat{a}_p(\alpha,1)e^{i\beta}\right] & = & C_p \cos \beta, \label{eq.ap3} \\
          \Re\left[ \hat{a}_s(\alpha,1)e^{i\beta}\right] & = & -C_p \frac{\sin \beta}{\cos \alpha}, \label{eq.as4}
          \end{eqnarray}         
        The corresponding pupil field that gives the optimum field in the focal region follows from 
        (\ref{eq.Epupil_opt2}):
        \begin{eqnarray}
  \Ev^e(\rho_e, \varphi_e) 
                             &=&  C_p \frac{k}{ \pi i f}
                            ( 1 -\rho_e^2/f^2)^{1/4}  
                             \Bigl\{ \left[\cos(2 \varphi_e)+\frac{\sin^2\varphi_e}{(1 -\rho_e^2/f^2)^{1/2}} \right]\xu
                             + \frac{1}{2}\sin(2\varphi_e) \left[ 1- \frac{1}{(1 -\rho_e^2/f^2)^{1/2}}\right] \yu 
                              \Bigr\}.
                             \label{eq.Epupil_opt3}
\end{eqnarray}
The constant $C_p$ can be determined by substituting $\hat{a}_p(\alpha,1)=C_p$ and  $\hat{a}_s(\alpha,1)=C_p/\tan \alpha$ into  (\ref{eq.PowerFourier}) and requiring that the power equals $P_0$.
We remark that the result (\ref{eq.sol_av}) agrees with the solution obtained by different methods in \cite{Sheppard} and \cite{deBruin}.

\begin{figure}
  \centering
 	\includegraphics[width=0.5\textwidth]{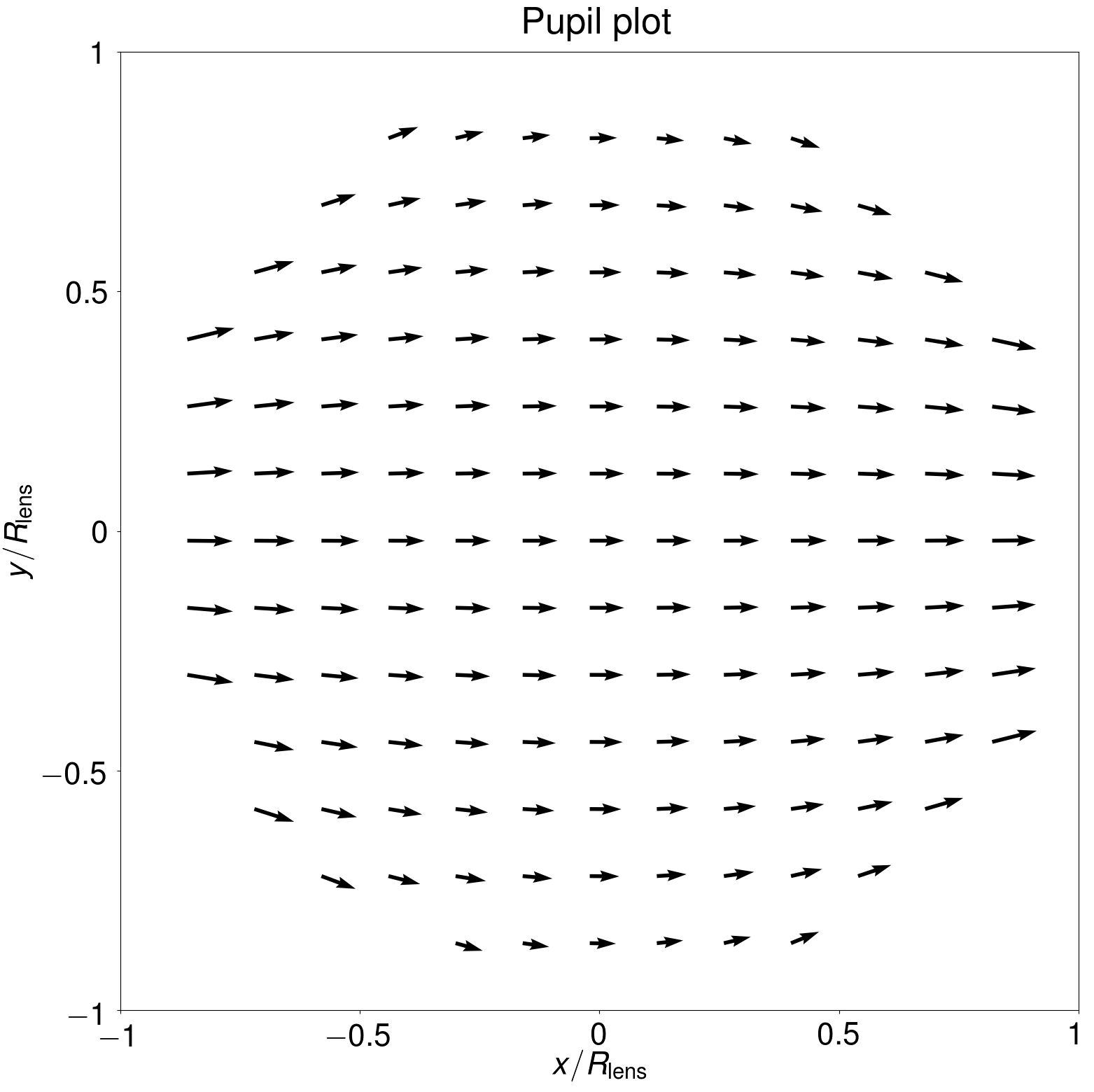}
 	 \caption{Snapshot of the optimum pupil field for $R = 0$ and $NA=0.95 $. For $R=0$ the optimum solution always has $\ell=1$.}
 	\label{Fig.pupilR0NA95}
\end{figure}

\begin{figure} 
 \centering
   	 \includegraphics[width=0.4\textwidth]{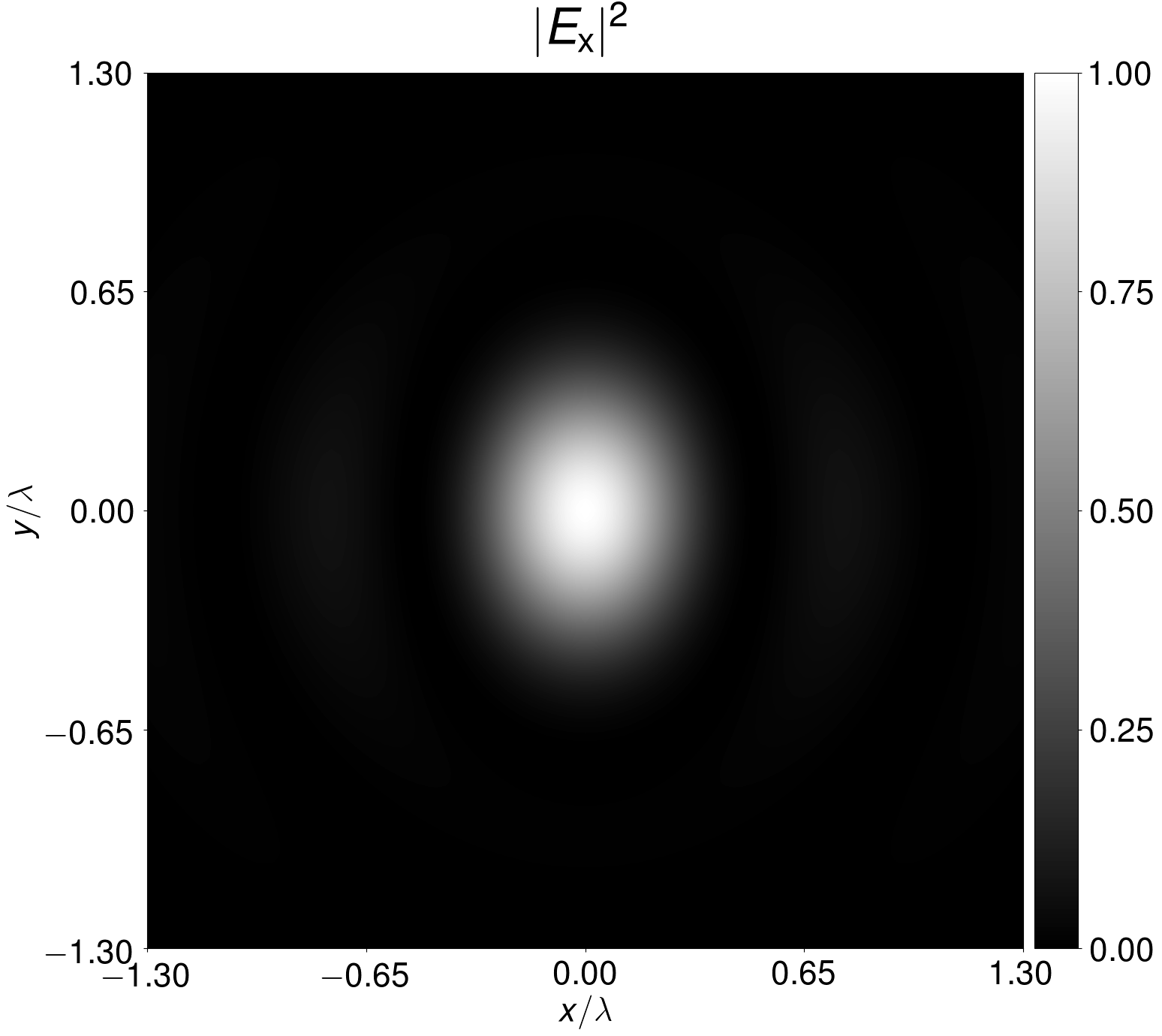}
   	    \includegraphics[width=0.4\textwidth]{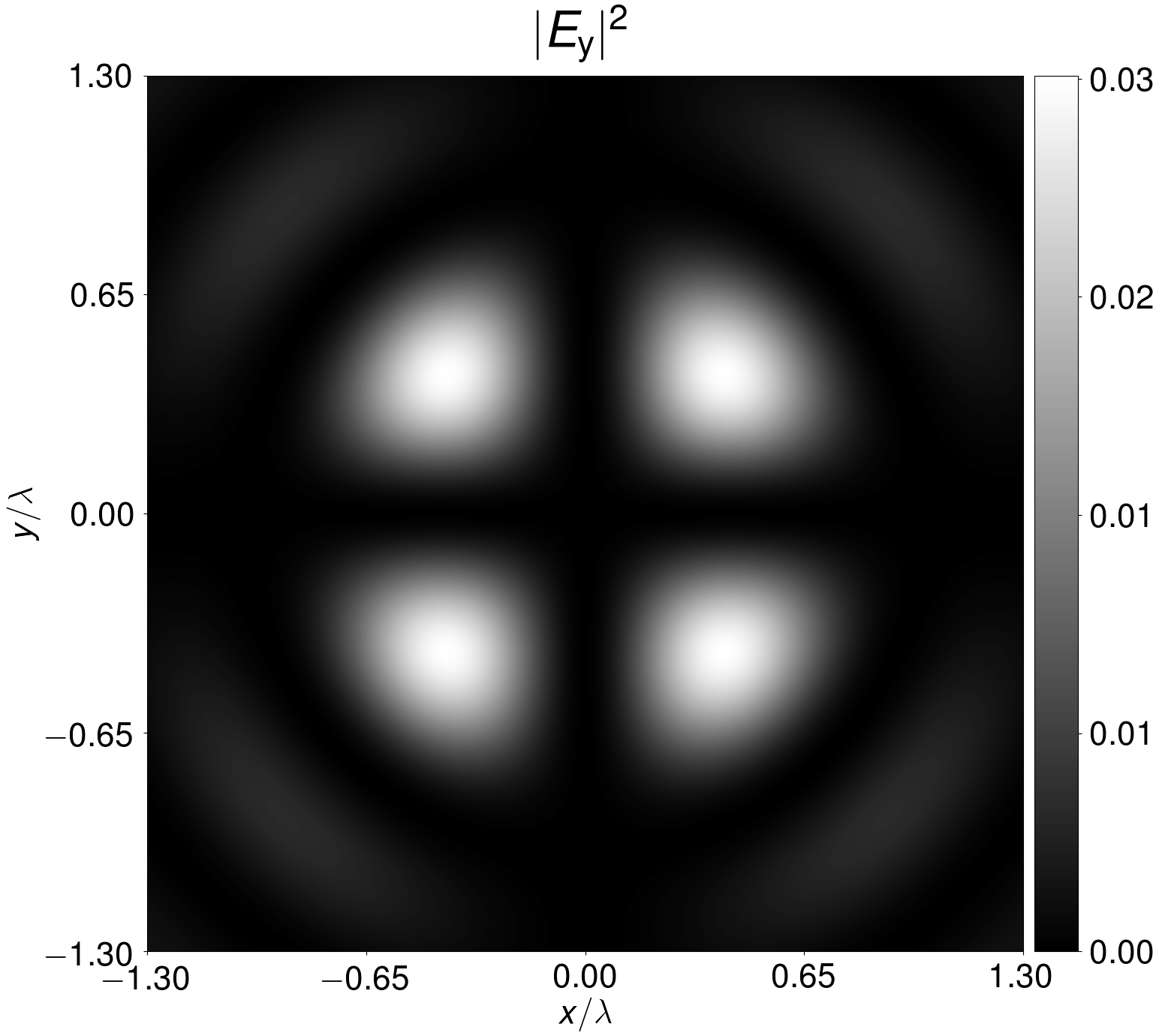} 
   	     \\
   \includegraphics[width=0.4\textwidth]{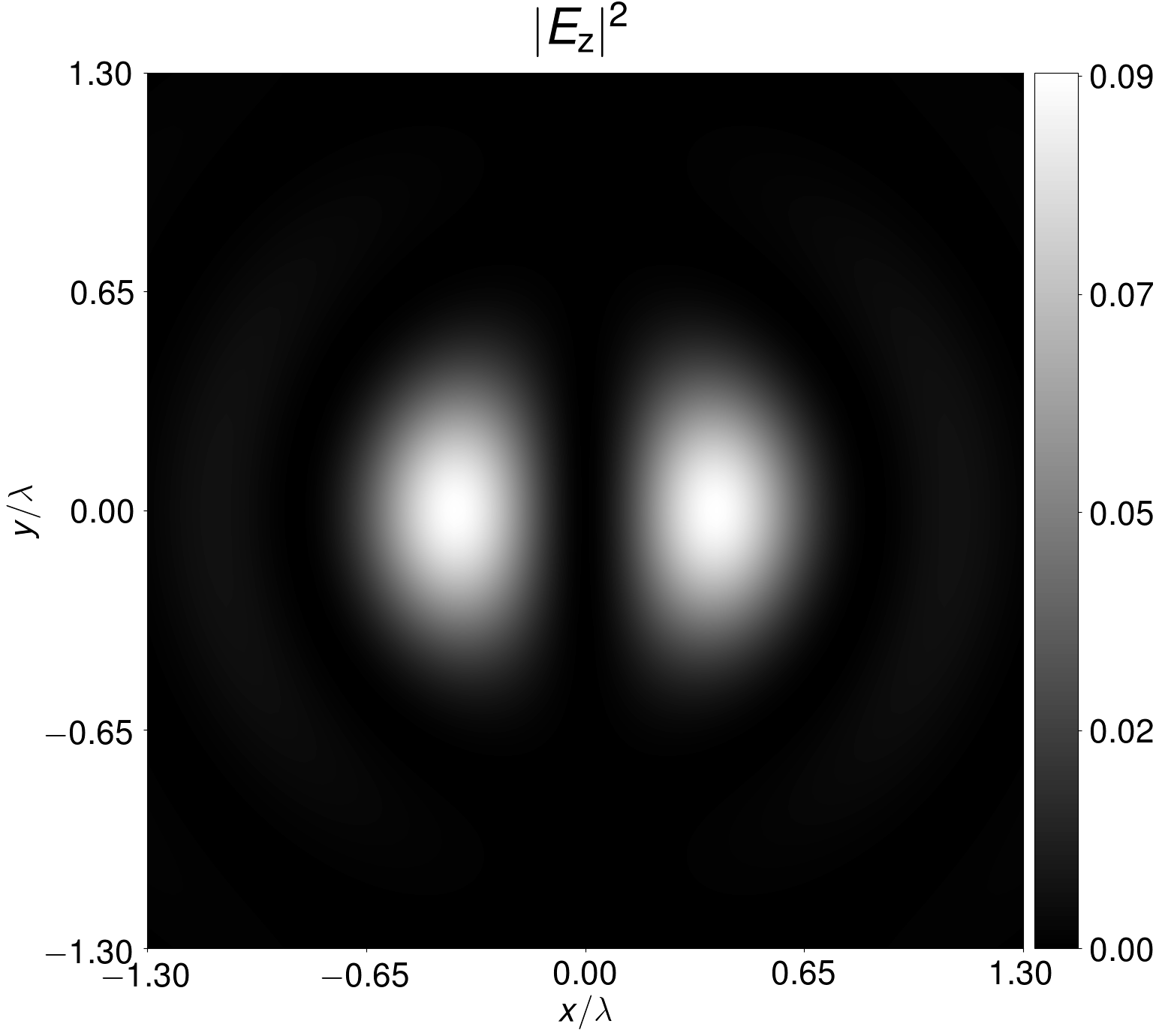}
  \includegraphics[width=0.4\textwidth]{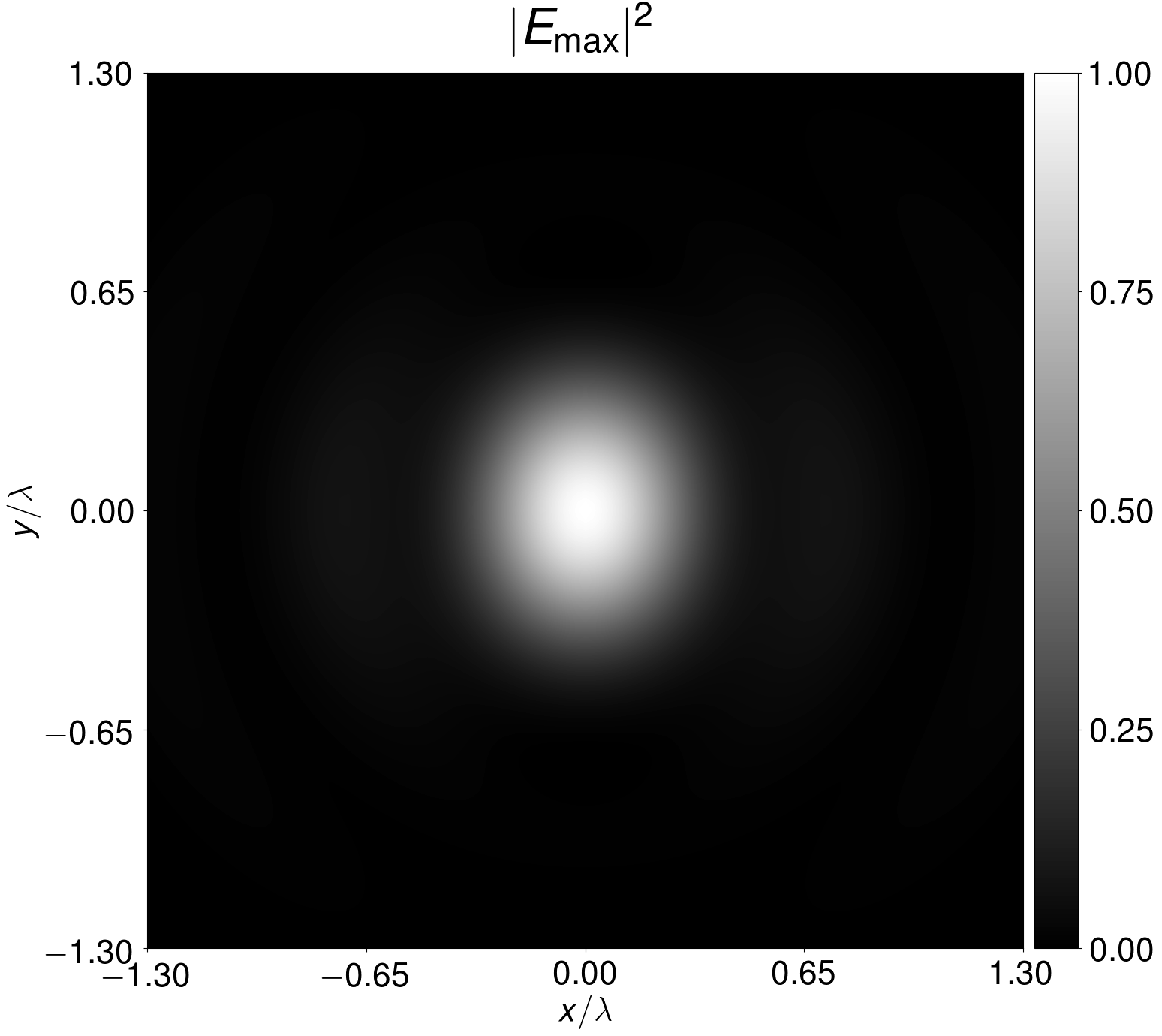} 
   \caption{Optimum focused squared electric field amplitudes and electric energy density in the focal plane 
   for $R = 0$ and $\NA=0.95$. The solution has $\ell=1$. Top left: $|E_x|^2$, top right: $|E_y|^2$, bottom left: $|E_z|^2$ and bottom right:
    $|\bm{E}|^2$. The amplitude and energy density are normalised such that the maximum of the energy density is unity.  }
   \label{Fig.ExyzefocalR0NA95}
\end{figure}

As has been mentioned after (\ref{eq.Epupil_opt2}), the pupil field is linearly polarised in all pupil points. 
It follows from (\ref{eq.Epupil_opt3}) that the pupil field is predominantly linearly polarised parallel to the $x$-axis with more or less constant amplitude. This is confirmed by Fig.~\ref{Fig.pupilR0NA95} where a snapshot of the optimum  pupil field is shown when $NA=0.95$. As function of time the electric field vectors in all pupil points  oscillate harmonically parallel to the direction of the arrows. The amplitudes of the $x$, $y$, and $z$-components of the optimum  electric field in the focal plane and the optimum electric energy density in the focal plane  are shown in 
Figs. \ref{Fig.ExyzefocalR0NA95}. This focal field indeed resembles  that of the vectorial Airy spot, i.e., the focused field of a linearly polarised plane wave.

As was stated after   (\ref{eq.sol_av}), if $\av(\alpha,\beta)$ is eigenvector, so is $\av(\alpha,\beta+\psi)$ for arbitrary $\psi$. The latter solution is predominantly polarised parallel to the direction which makes an angle $\psi$ with the $x$-axis. Hence there is nothing special about the $x$-axis and it is therefore more appropriate to state that the optimum pupil fields for the case $R=0$ are similar to that of a  linearly polarised plane wave. When the numerical aperture is increased, the difference between the optimum pupil field and that of a linear polarised plane wave becomes bigger.


\subsection{Optimum fields for general $R$}
For general $R>0$ the optimisation problem can not be  solved in closed form but  instead numerical computations are necessary. We explain how this can be done in Appendix \ref{appendix:discretization}.
In Fig.~\ref{Fig.MaxNAR} the maximum of the electric energy density is shown as function of $R/\lambda$ and $\NA$ for power $P_0=1$. According to the scaling law discussed in Section~\ref{subsection.scaling}, for given $\NA$, the eigenfields are the same  if $R/\lambda$ is kept constant and are independent of the power $P_0$. The maximum of the object functional $G_ {S, \Pi}$, i.e., the maximum of the average electric energy density over the disc with radius $R$, increases as $1/R^2$ when $R/\lambda$ is kept constant and is proportional to $P_0$. Hence, Fig.~\ref{Fig.MaxNAR} contains information of the solutions of the optimisation problem for all $0.40< \NA <0.95 $ and for the values of $R$ and $\lambda$ for which $0< R/\lambda< 2$.

\begin{figure}
  \centering
  \includegraphics[width=0.45\textwidth]{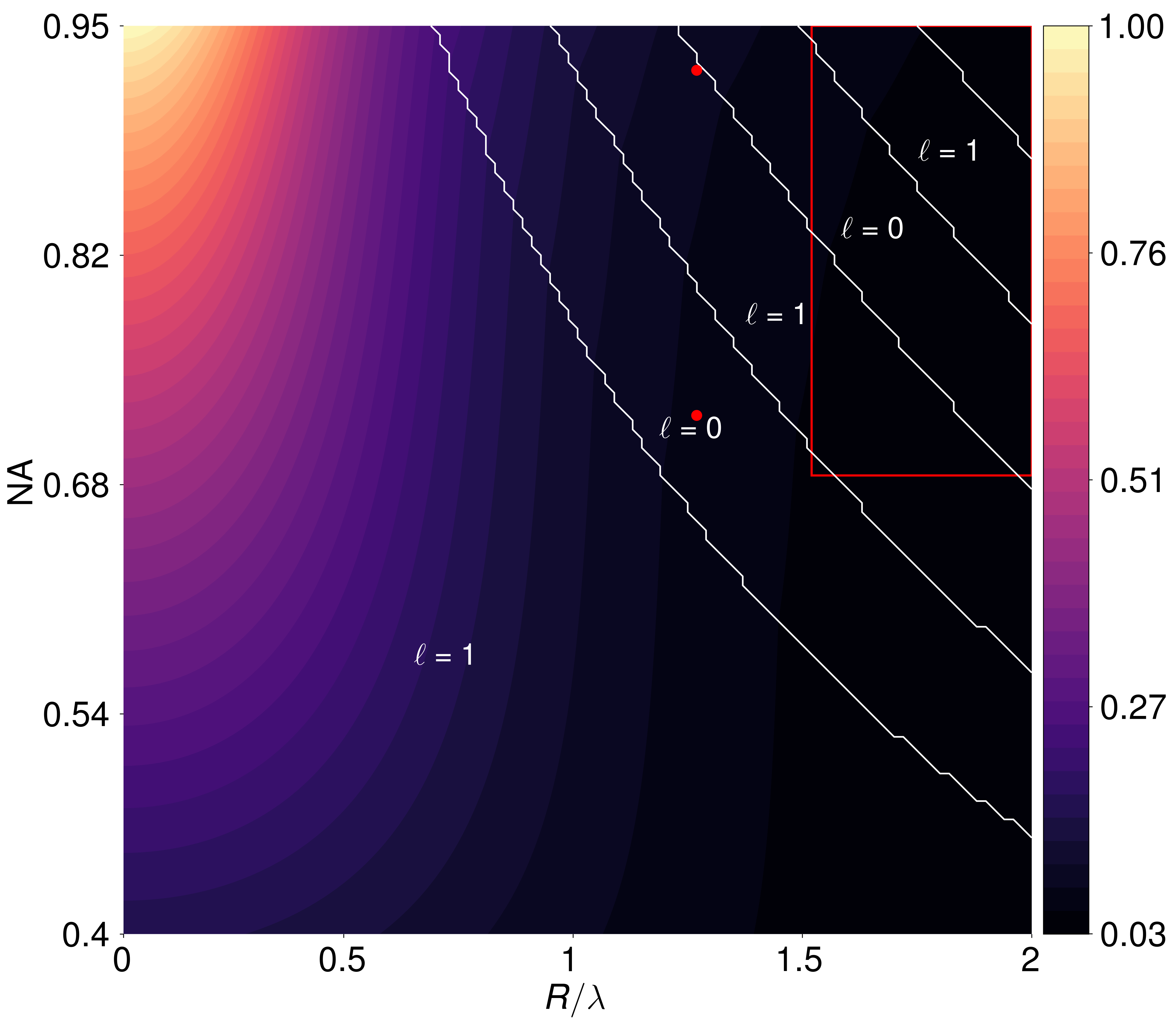}
 \includegraphics[width=0.45\textwidth]{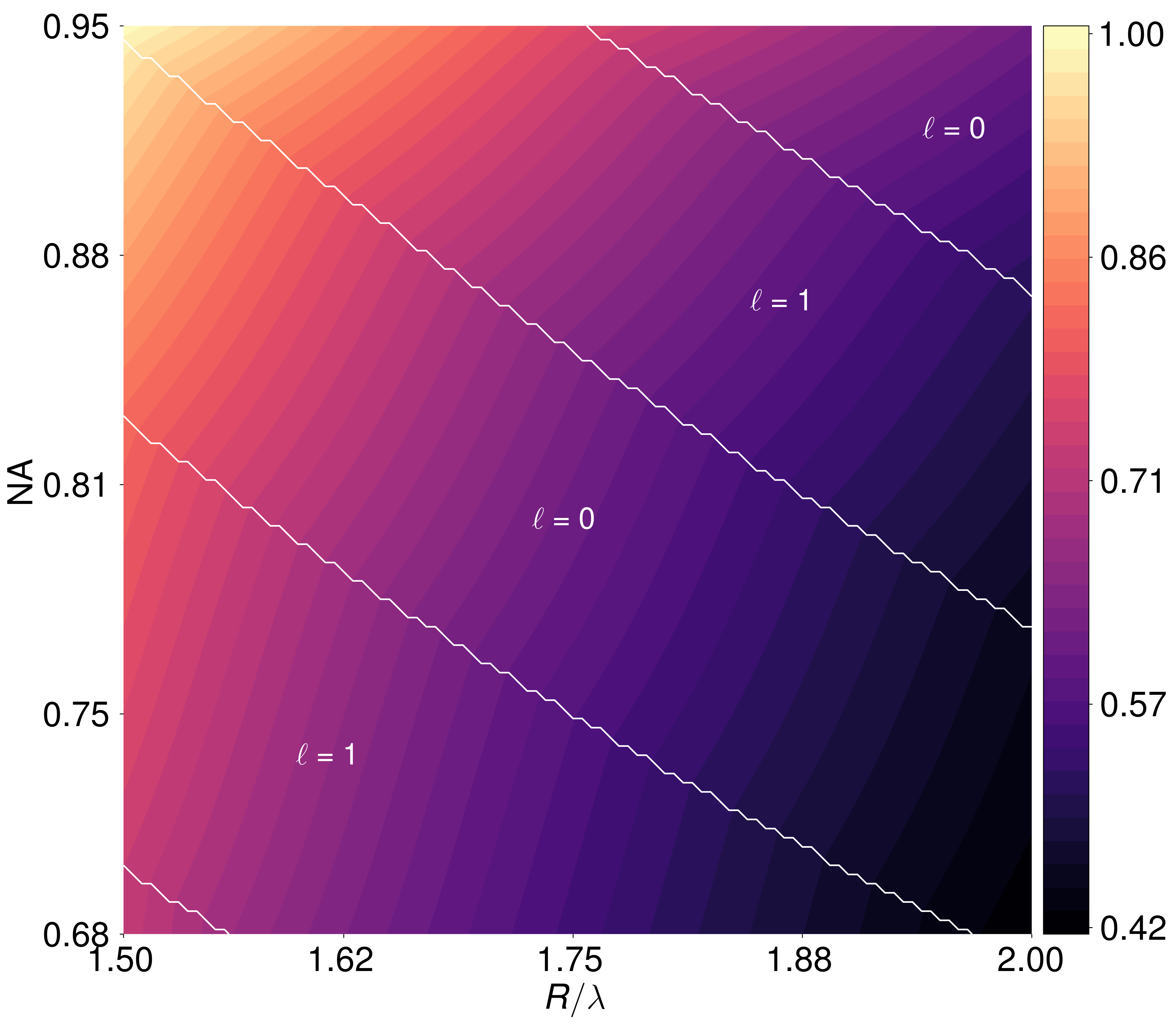}
 \caption{Contour plot of the maximum energy density averaged over a disc of radius $R$, as function of $\NA$  and  $R/\lambda$. The values in the left figure are normalized with respect to the maximum which occurs in this plot for $R/\lambda=0$ and $\NA=0.95$. The right plot is an elargement of the part inside the red rectangle in the left figure and is normalized to the maximum occuring in this rectangle. depending on the values of $R$ and $\NA$ there holds either $\ell=0$ or $\ell=1$. }
 \label{Fig.MaxNAR}
\end{figure}

It is seen in Fig.~\ref{Fig.MaxNAR} that the maximum average electric energy density  monotonically increases  with $\NA$ for fixed $R/\lambda$ and that it monotonically decreases for increasing $R/\lambda$ when $\NA$ is fixed. 
Furthermore, for all optimisation problems for which we have computed the solution, we found that either $\ell=0$ or $\ell=1$, i.e., no value $\ell>1$ was found to be optimal. The regions in Fig.~\ref{Fig.MaxNAR} for different values of $\ell$ are separated by curves where solutions for both $\ell=0$ and $\ell=1$ occur. These curves seem to 
satisfy  $ \NA . R/\lambda =C$, with $R/\lambda \geq C$ and the constant $C$ depends on the curve. 
 
\begin{figure}
  \centering
   \includegraphics[width=0.5\textwidth]{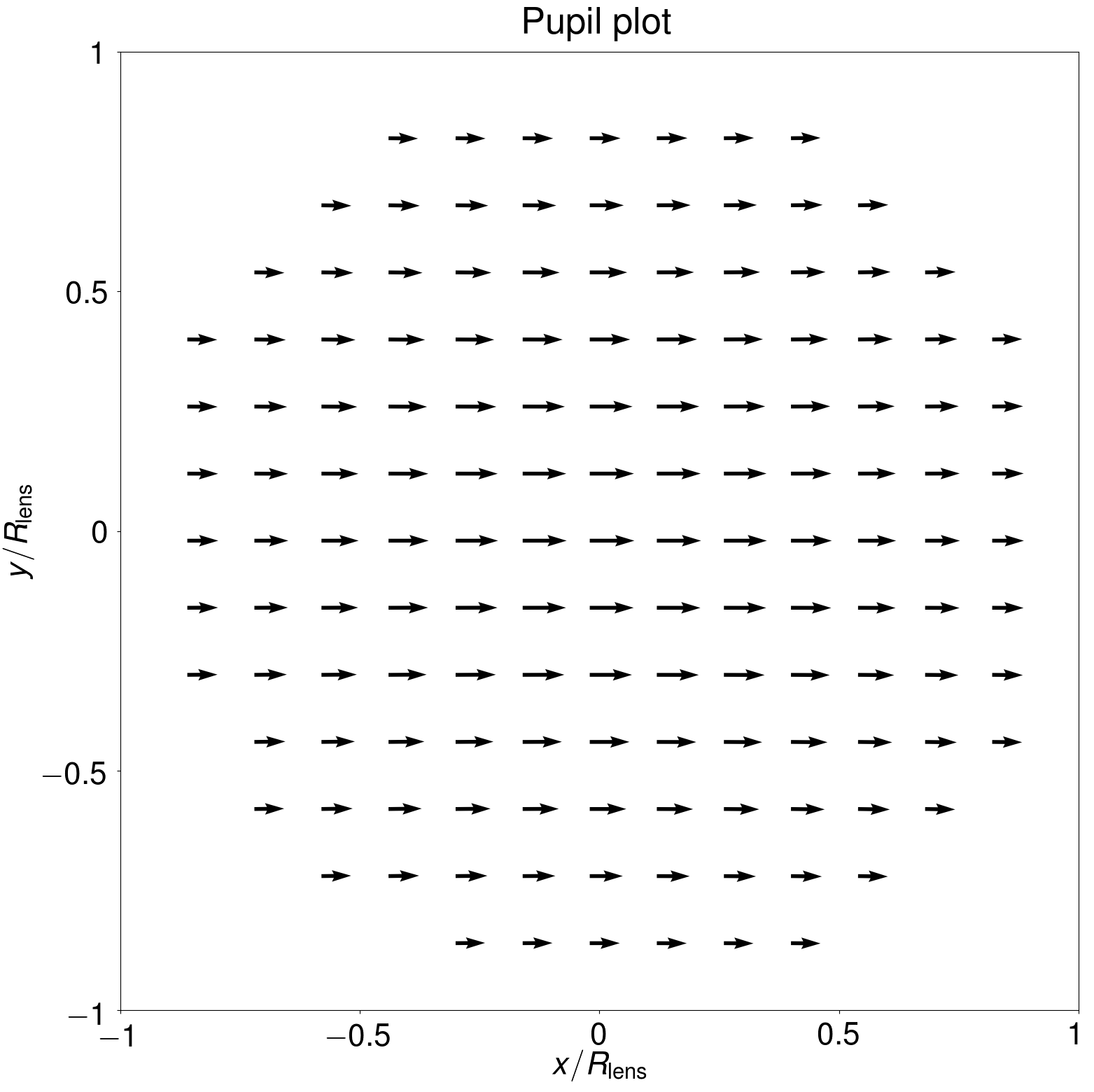}
    \caption{Pupil field for $R = 0.5\lambda$ and $\NA=0.95$. In this case  $\ell = 1$.}
  \label{Fig.pupilR0.5NA95}
\end{figure}

\begin{figure}
  \centering
    \includegraphics[width=0.4\textwidth]{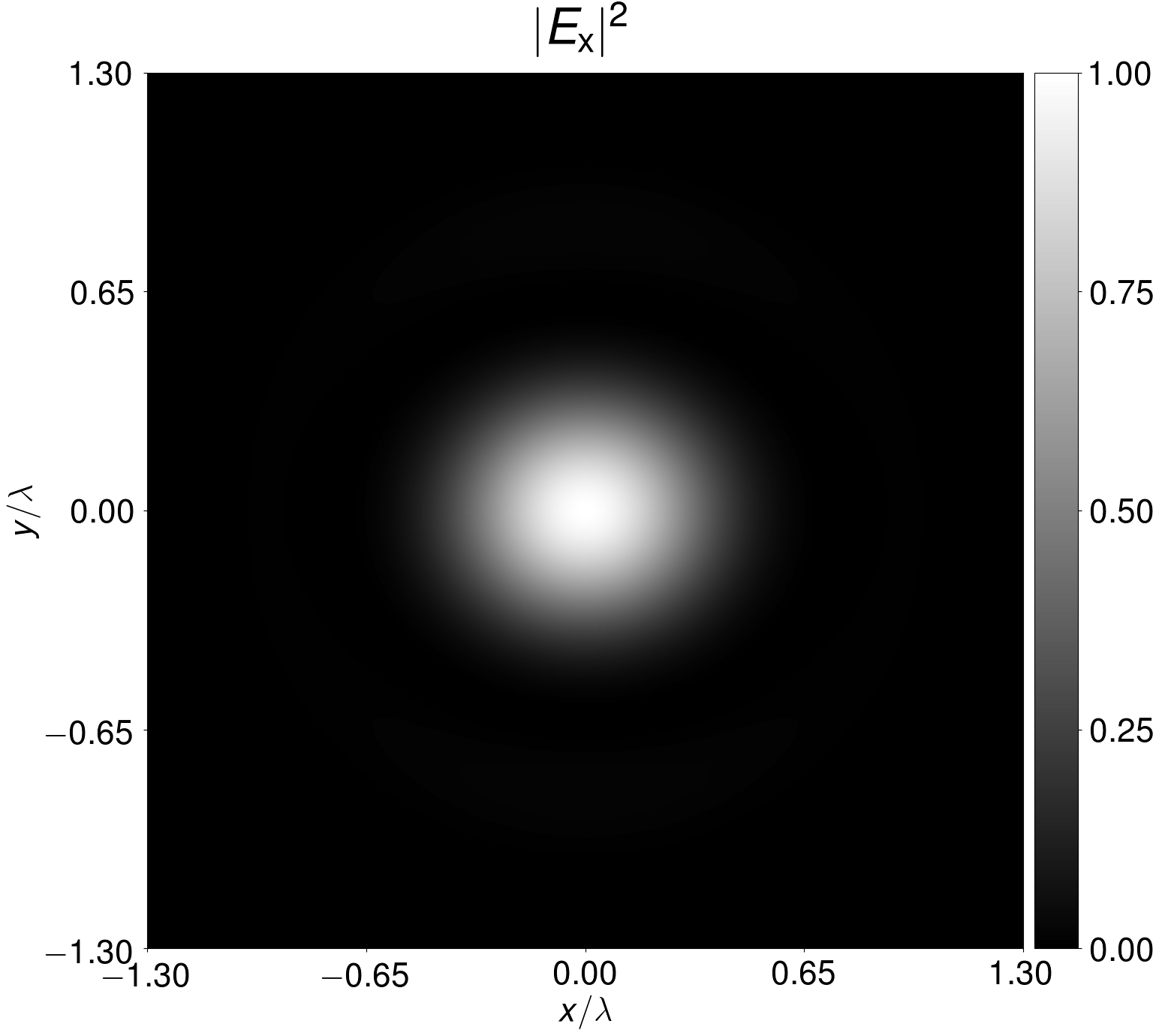}
    \includegraphics[width=0.4\textwidth]{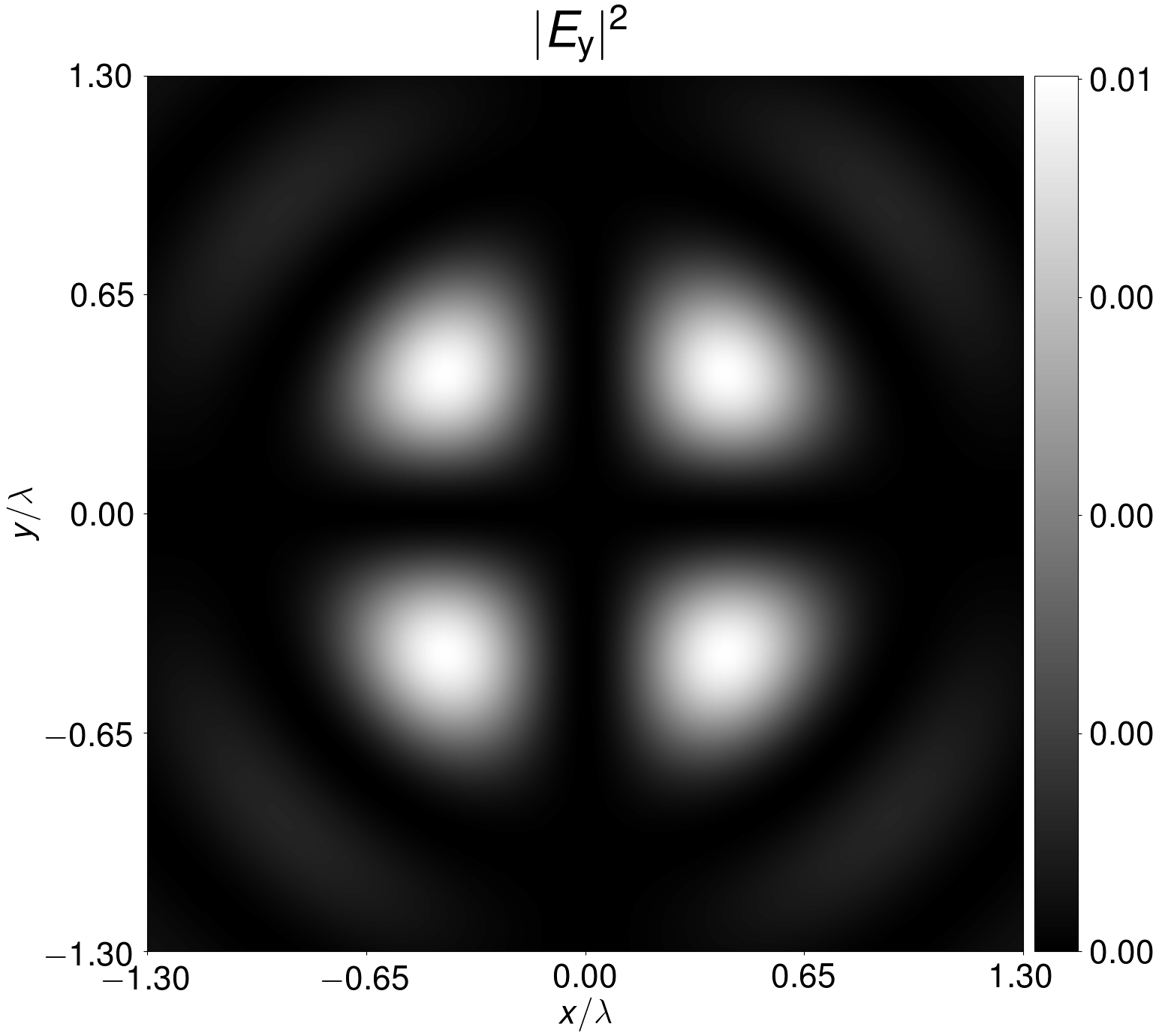}
 \\
    \includegraphics[width=0.4\textwidth]{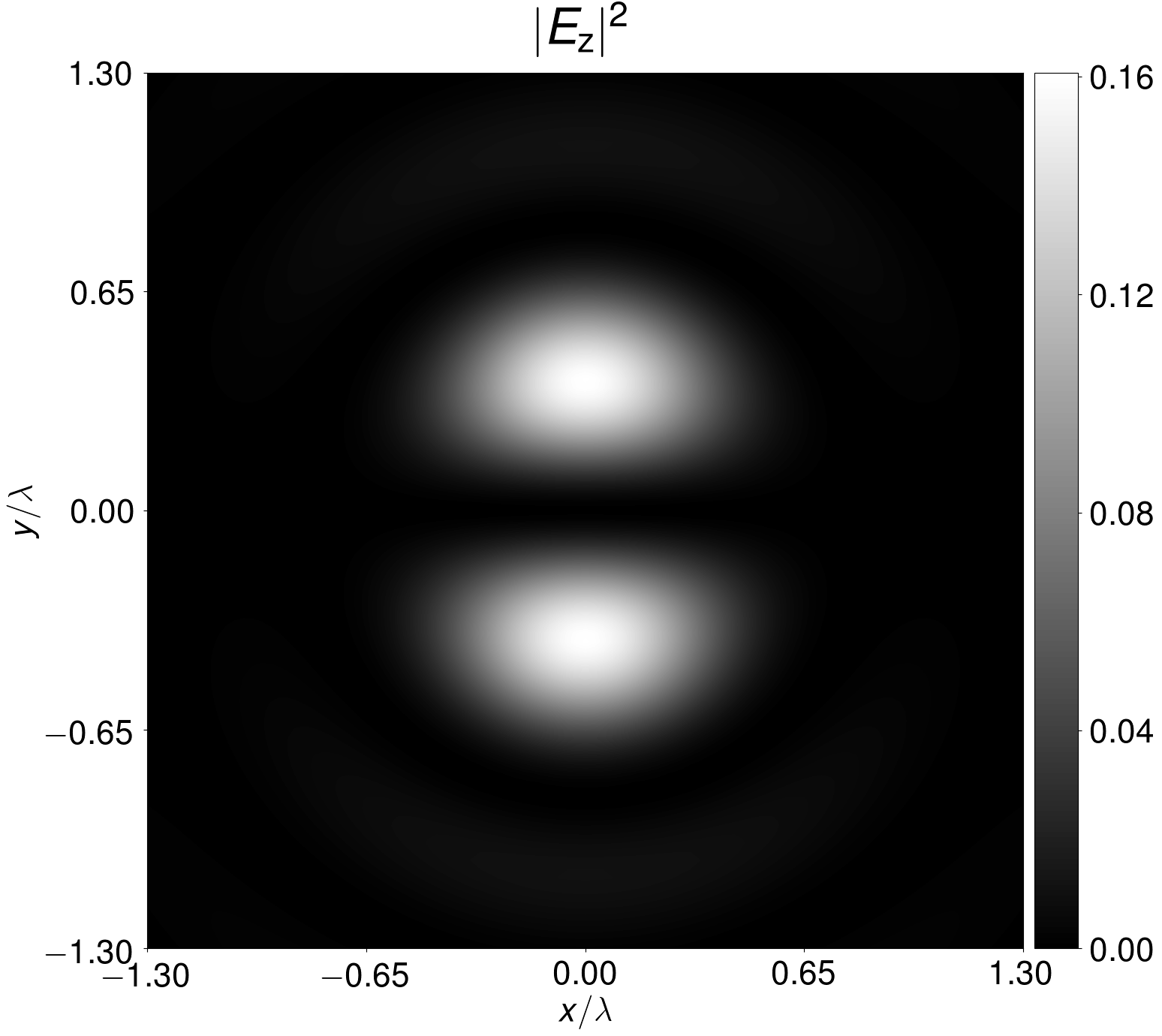}
    \includegraphics[width=0.4\textwidth]{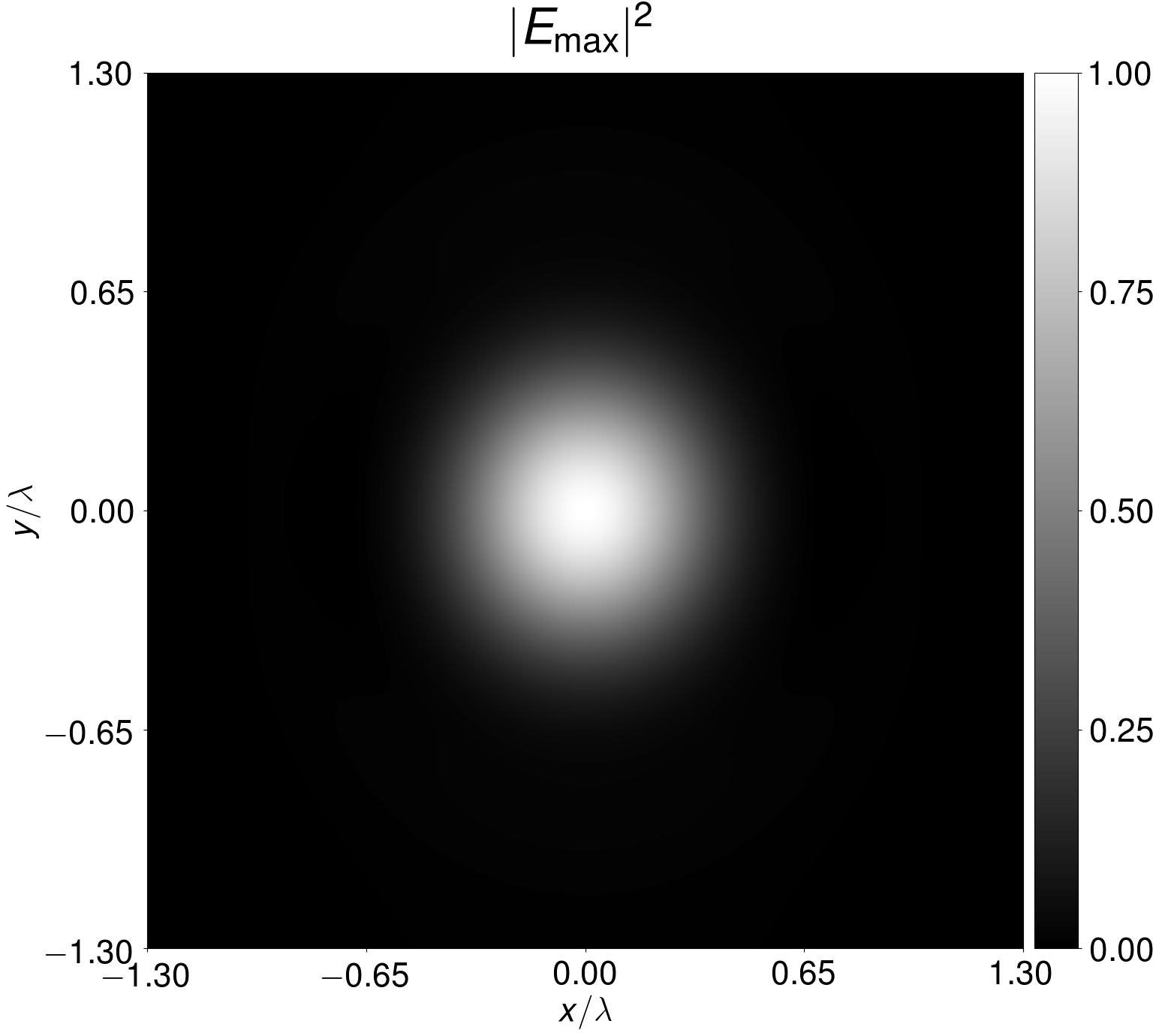}
    \caption{Optimum focused squared electric field amplitudes and electric energy density in the focal plane 
   for $R = 0.5\lambda$ and $\NA=0.95$. The solution has $\ell=1$. Top left: $|E_x|^2$, top right: $|E_y|^2$, bottom left: $|E_z|^2$ and bottom right:
    $|\bm{E}|^2$. The amplitude and energy density are normalised such that the maximum of the energy density is unity. }
  \label{Fig.ExyzefocalR0.5NA95}
\end{figure}

 When $R=0.5\lambda$ and $\NA=0.95$, the solution is in the large region where
 $\ell=1$  which also contains  $R=0$. The optimum pupil field is shown in Fig.
 \ref{Fig.pupilR0.5NA95} and the corresponding focal field is shown in Figs.
 \ref{Fig.ExyzefocalR0.5NA95}. The pupil field is  similar to that of a linear
 polarised plane wave although the amplitude decreases towards the rim of the
 pupil.

\begin{figure}
  \centering
  \includegraphics[width=0.5\textwidth]{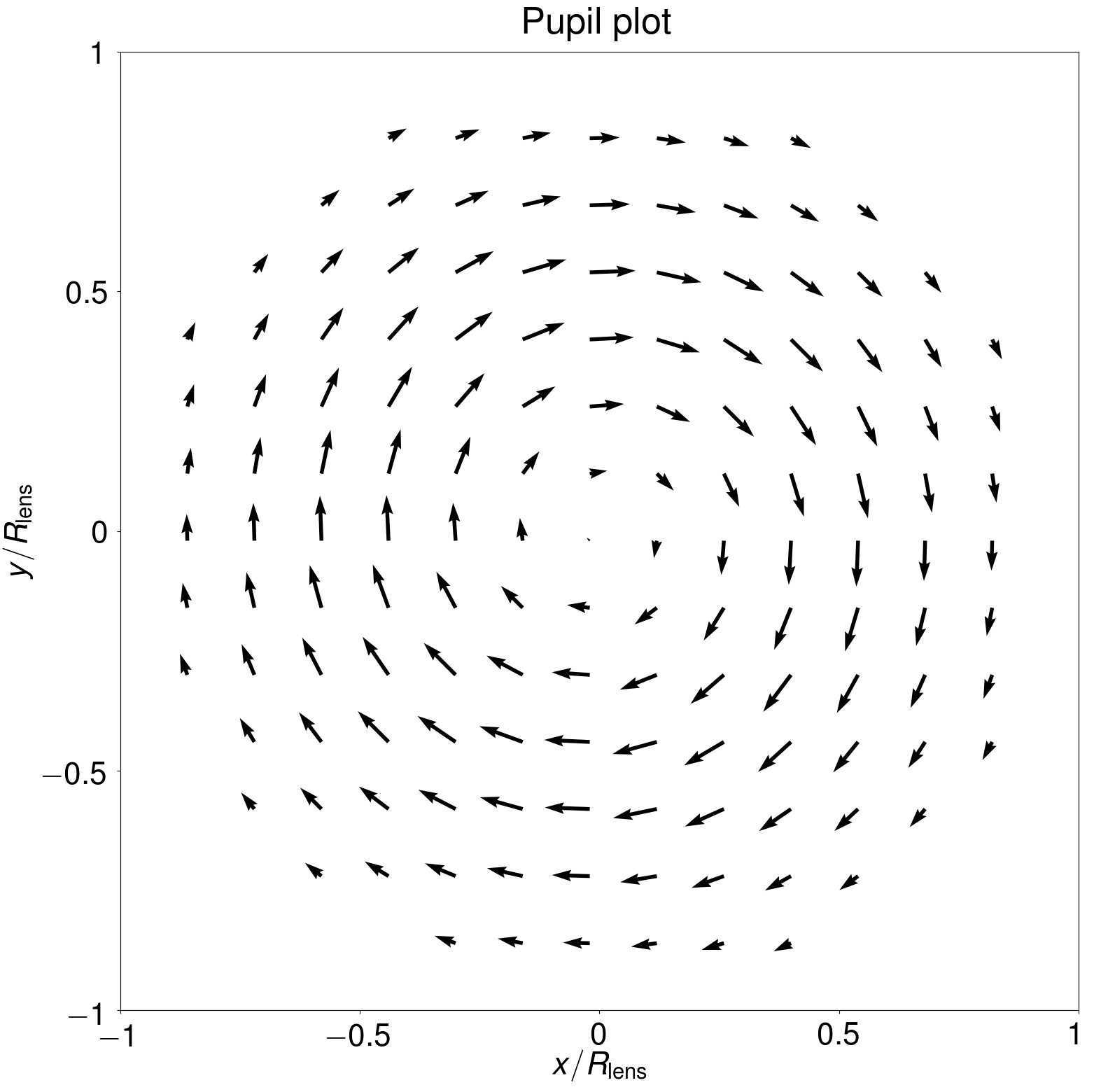}
   \caption{Pupil field for $R = 1.25\lambda$ and $\NA=0.75$. In this case  $\ell = 0$.}
  \label{Fig.pupilR1.25NA75}
\end{figure}

\begin{figure}
  \centering
   	 \includegraphics[width=0.4\textwidth]{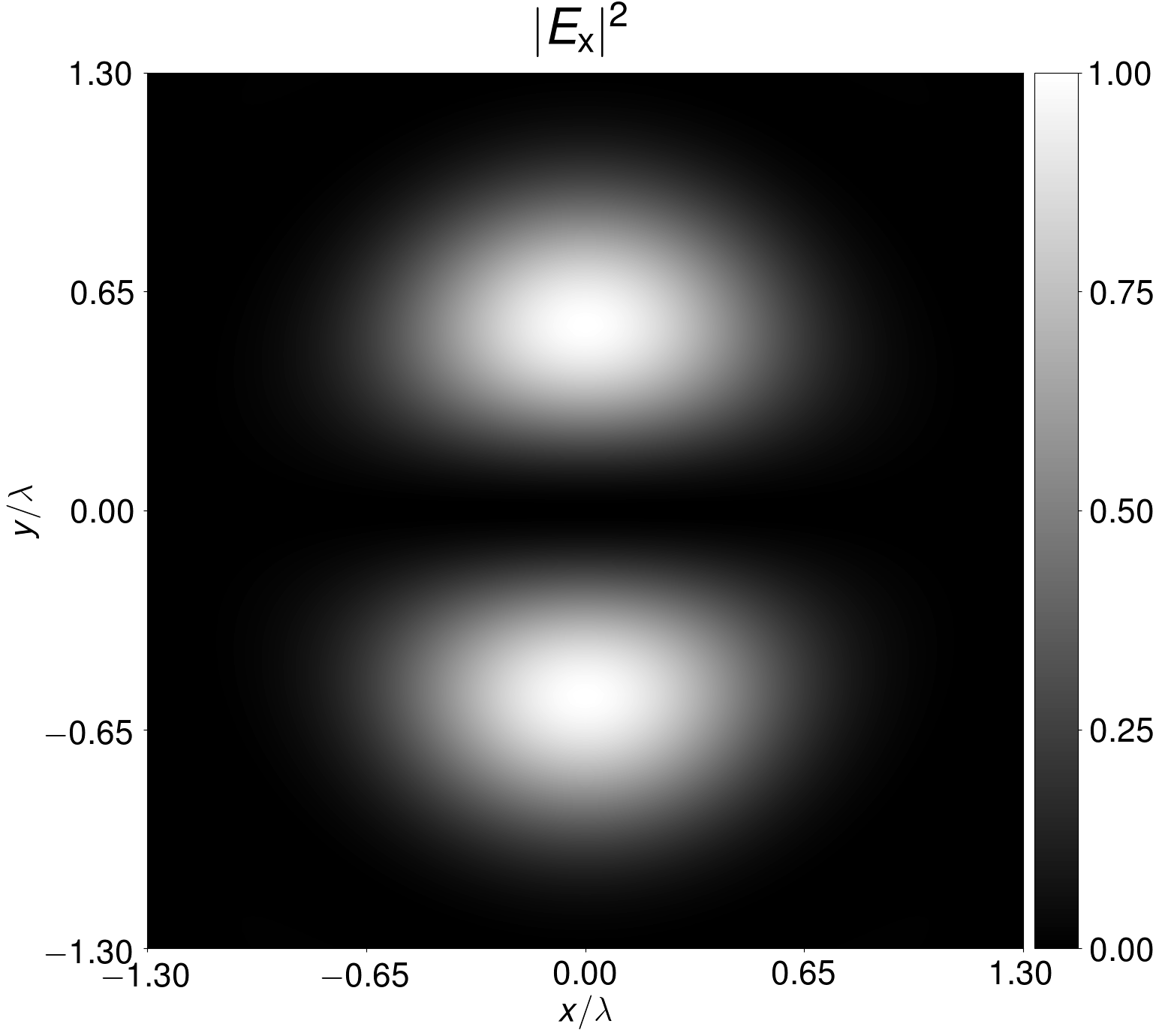}
     \includegraphics[width=0.4\textwidth]{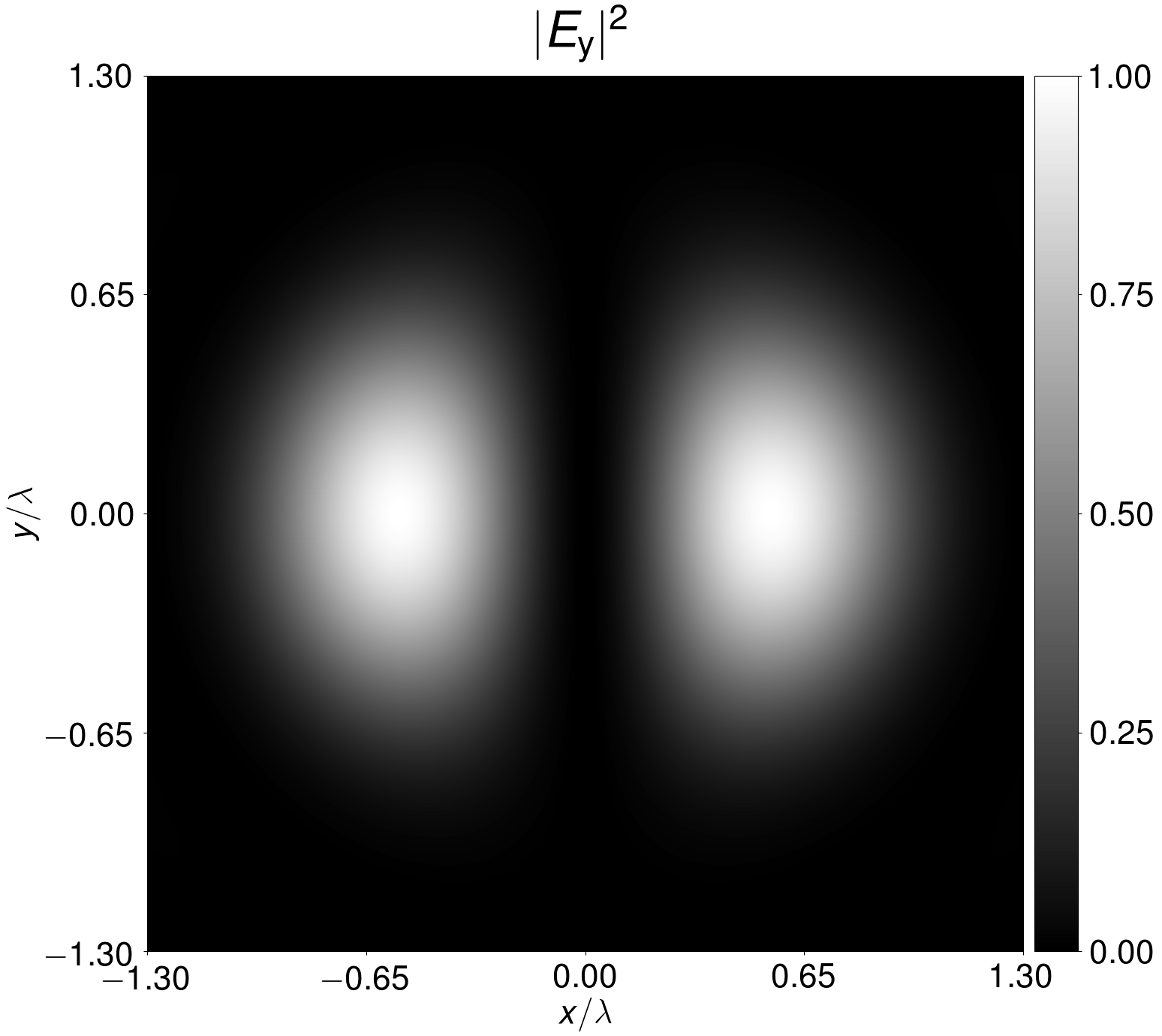}
     \\
   \includegraphics[width=0.4\textwidth]{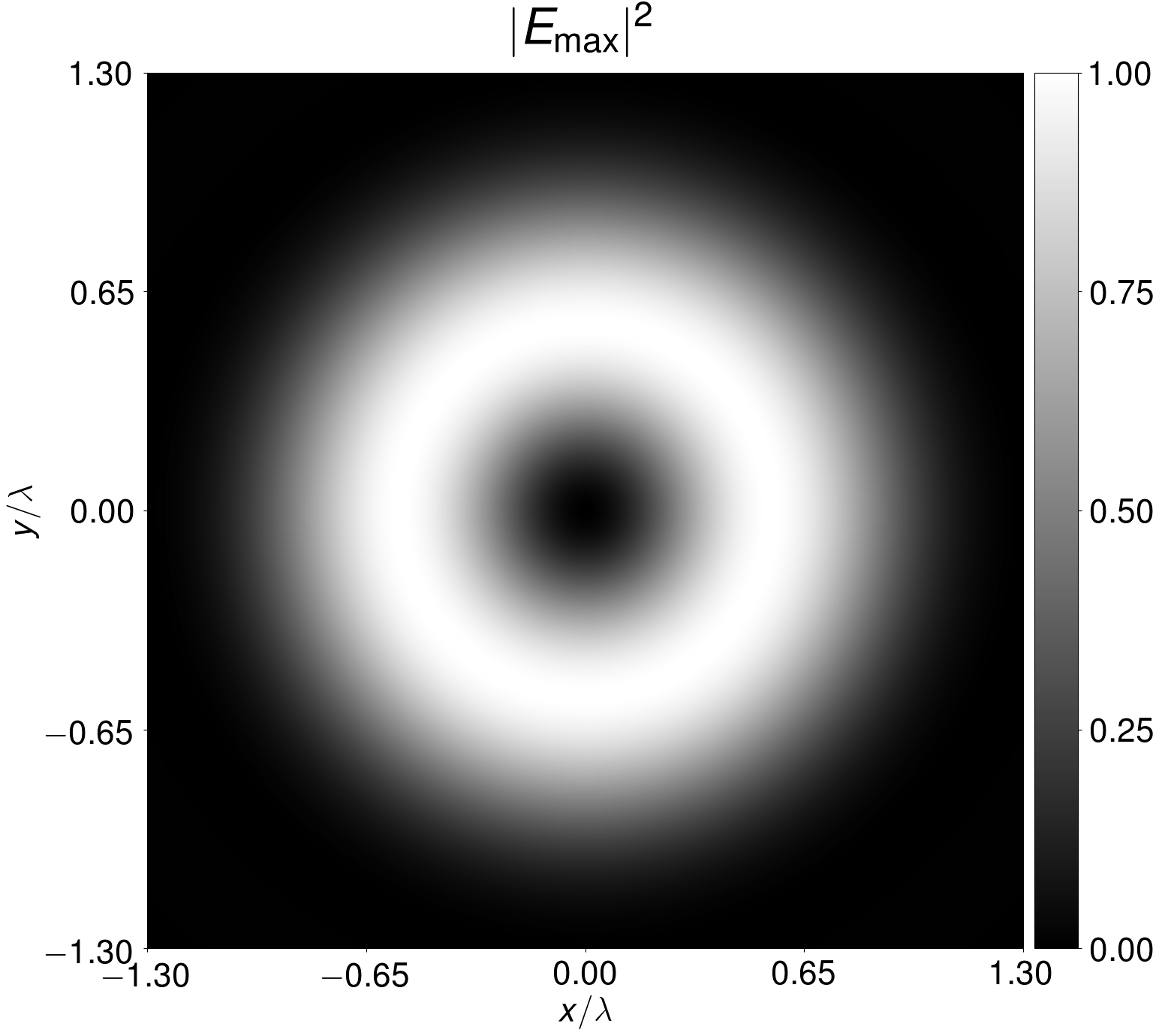}
   \caption{Optimum focused squared electric field amplitudes and electric energy density in the focal plane 
   for $R = 1.25\lambda$ and $\NA=0.75$. The solution has $\ell=0$. Top left: $|E_x|^2$, top right: $|E_y|^2$, bottom:
    $|\bm{E}|^2$. The longitudinal component $E_z$ vanishes. The amplitude and energy density are normalised such that the maximum of the energy density is unity.}
 \label{Fig.ExyzefocalR1.25NA75}
\end{figure}

In Fig.~\ref{Fig.pupilR1.25NA75} a snapshot of the optimum pupil field is shown for  $R=1.25 \lambda$ and $\NA=0.75$,  for which $\ell=0$. It is found that $ \widehat{a_p}(\alpha,0)=0$ and in fact this property holds for all solutions where  $\ell=0$. Then
  (\ref{eq.Epupil_opt2}) implies that the optimum pupil field is azimuthally polarised with amplitude that is rotational invariant and depends only on $\varrho_e$. The focal field is a superposition of S-polarised plane waves and hence the $E_z$ component of the field in the focal region vanishes.
 As is seen in Figs. \ref{Fig.ExyzefocalR1.25NA75} the transverse electric field amplitudes in the  focal point  vanish  and the electric energy density has a doughnut shape.  


\begin{figure}
  \centering
  \includegraphics[width=0.5\textwidth]{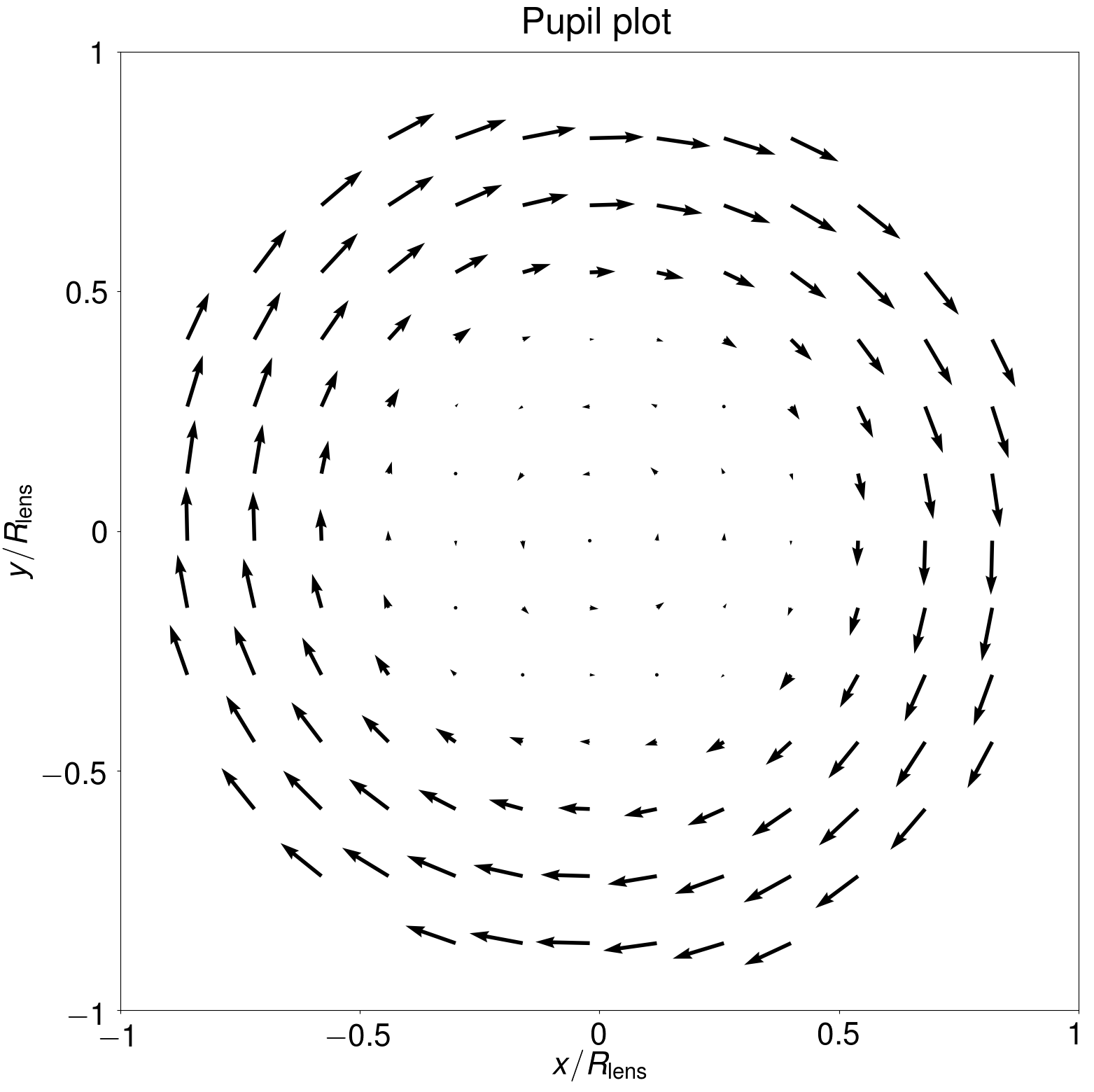}
   \caption{Pupil field for $R = 1.25\lambda$ and $\NA=0.95$. In this case  $\ell = 0$.}
  \label{Fig.pupilR1.25NA95}
\end{figure}

\begin{figure}
  \centering
   	 \includegraphics[width=0.4\textwidth]{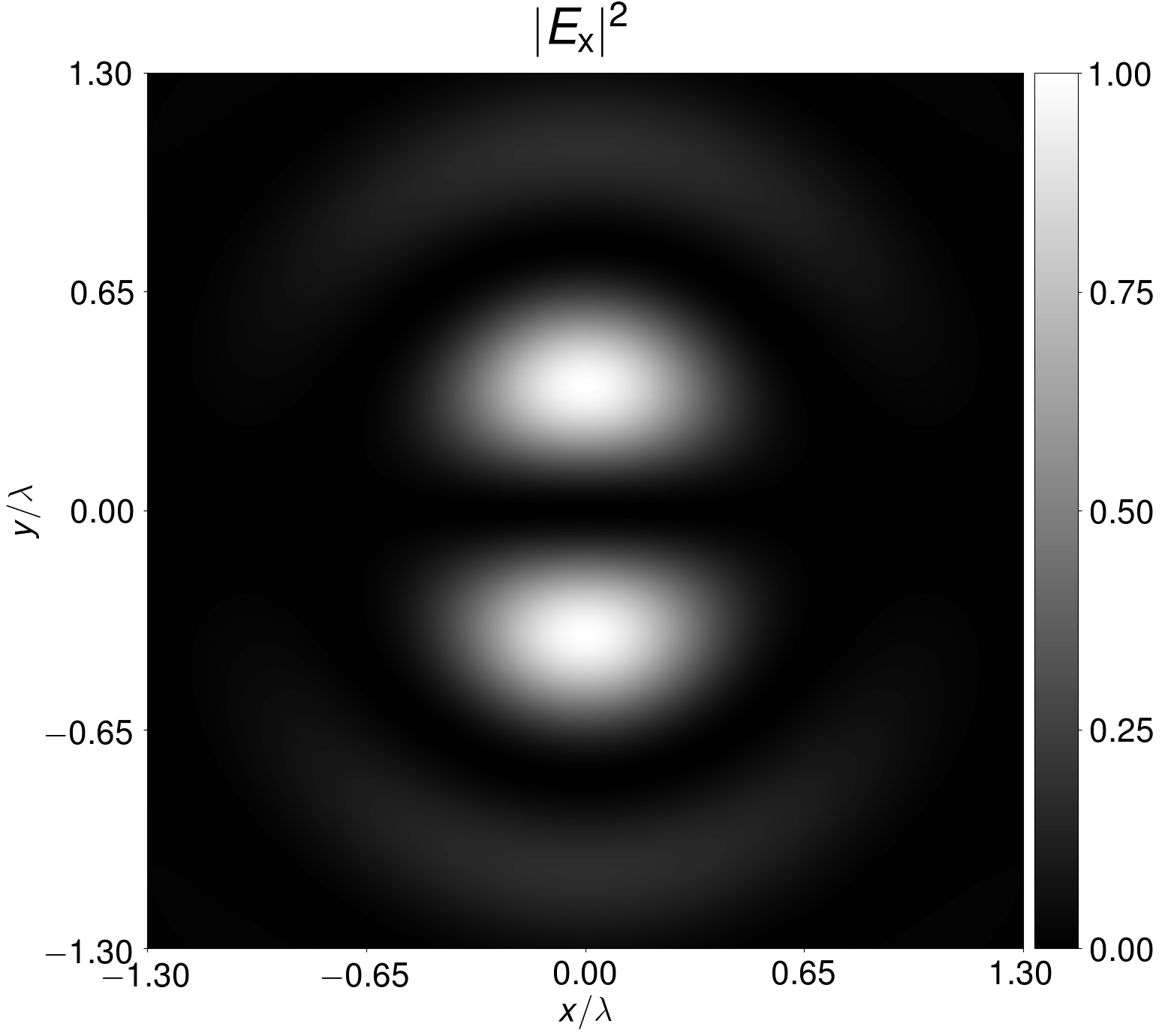}
   	  \includegraphics[width=0.4\textwidth]{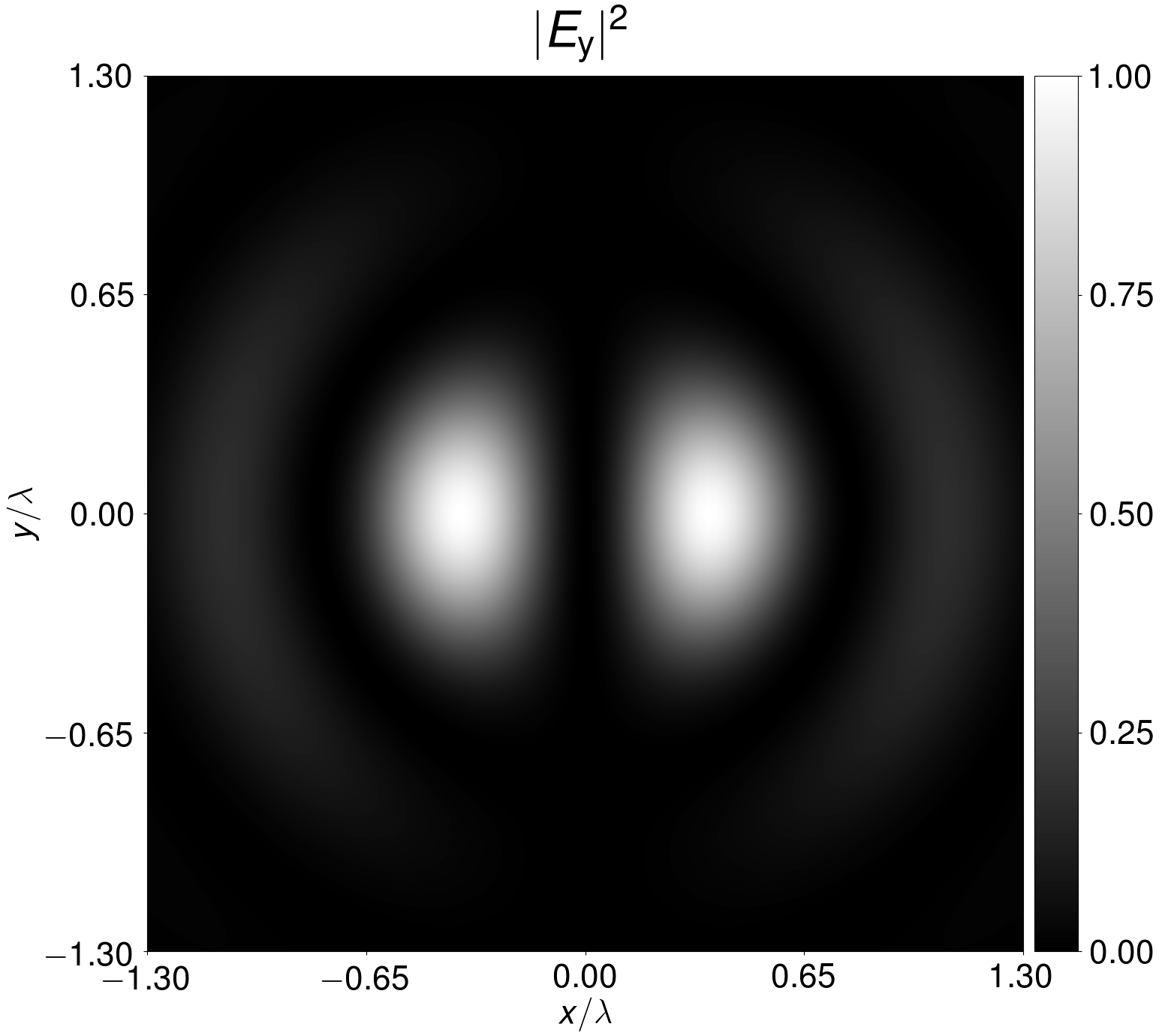}
   	\\
   \includegraphics[width=0.4\textwidth]{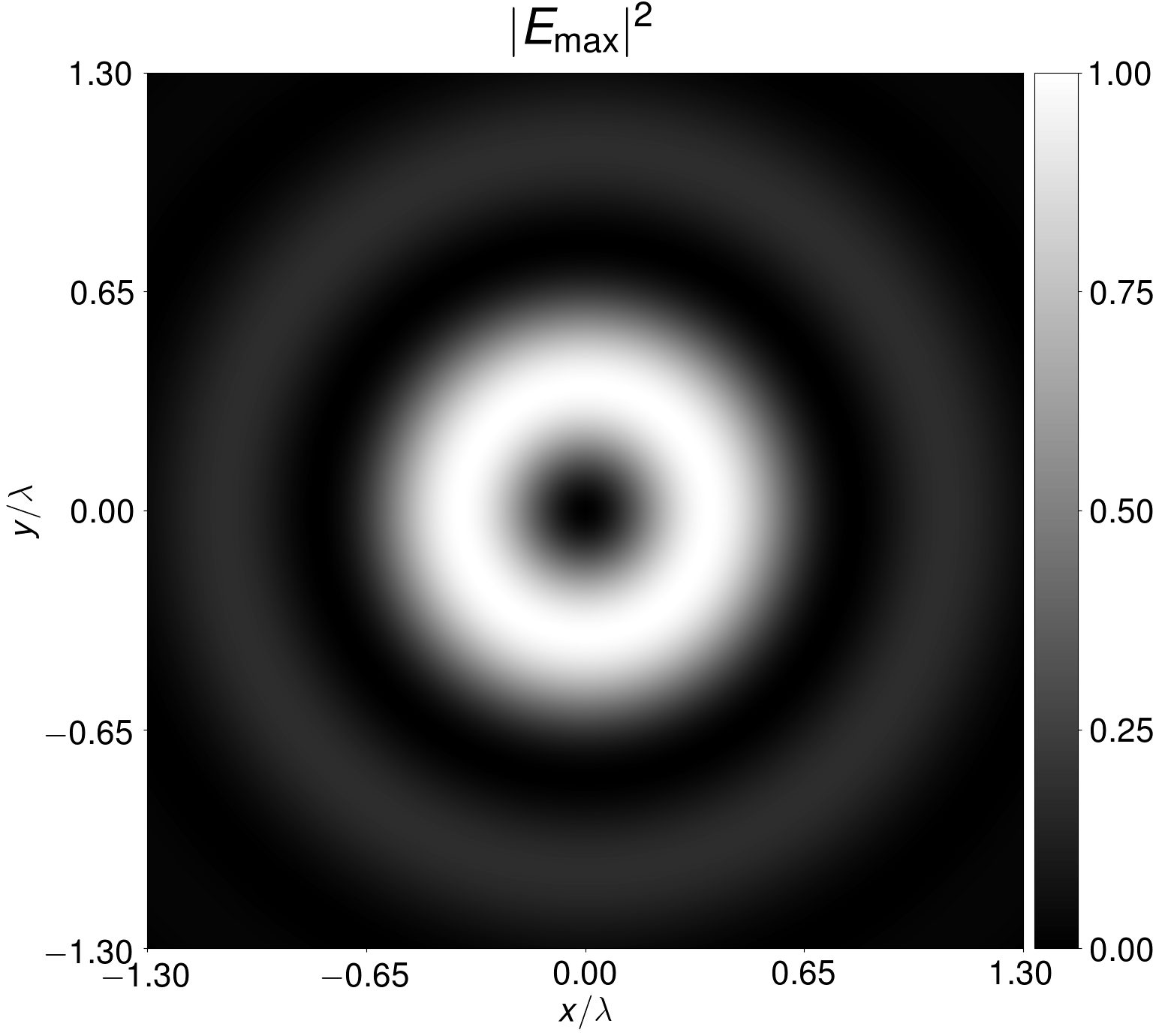}
   \caption{Optimum focused squared electric field amplitudes and electric energy density in the focal plane 
   for $R = 1.25\lambda$ and $\NA=0.95$. The solution has $\ell=0$. Top left: $|E_x|^2$, top right: $|E_y|^2$, bottom:
    $|\bm{E}|^2$. The longitudinal component $E_z$ vanishes. The amplitude and energy density are normalised such that the maximum of the energy density is unity. }
 \label{Fig.ExyzefocalR1.25NA95}
\end{figure}

When the $\NA$ of the lens is increased, the optimum pupil field for the same $R=1.25\lambda$ becomes more concentrated at the edge of the pupil. This is confirmed by Fig.~\ref{Fig.pupilR1.25NA95} where the results are shown for $\NA=0.95$. In this case the rotational symmetric solution: $\ell=0$ applies as for $\NA=0.75$,  but the ratios of the amplitudes in the centre to those at the edge are much smaller than in 
Fig.~\ref{Fig.pupilR1.25NA75}.  The optimum electric field components in the focal region for  $R=1.25\lambda$ and $NA=0.95$ are shown in Fig.~\ref{Fig.ExyzefocalR1.25NA95}. They are more narrow than in Fig.~\ref{Fig.ExyzefocalR1.25NA75} for $\NA=0.75$ (note the different scales of the figures for $\NA=0.75$ and $\NA=0.95$).


\begin{figure}
  \centering
   \includegraphics[width=0.5\textwidth]{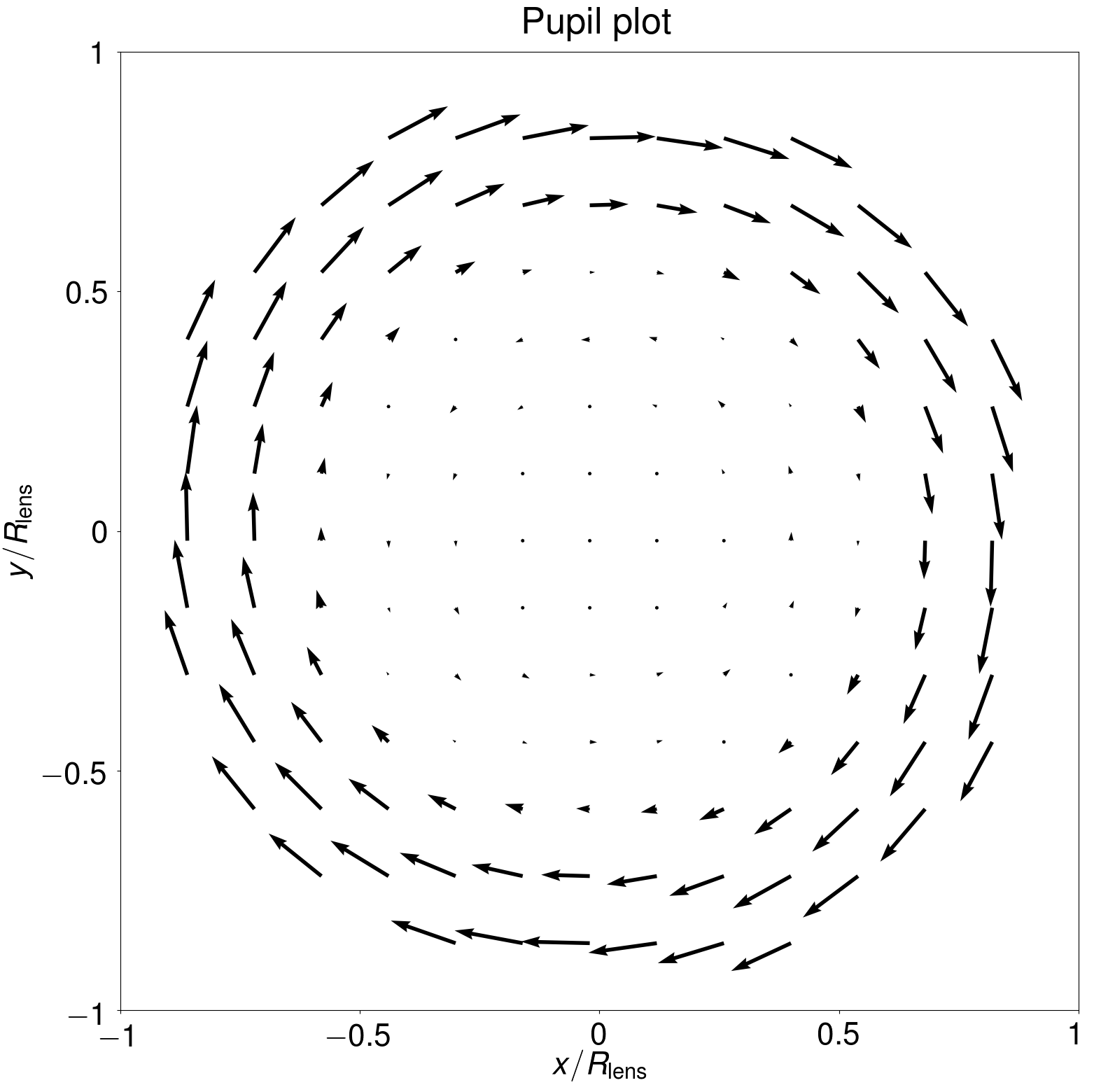}
     \caption{Pupil field for $R = 1.753656\lambda$ and $\NA=0.95$ for which two solutions exist. The solution shown here  has  $\ell = 0$.}
  \label{Fig.pupilR1_753656NA95ell0}
\end{figure}

\begin{figure}
  \centering
    \includegraphics[width=0.4\textwidth]{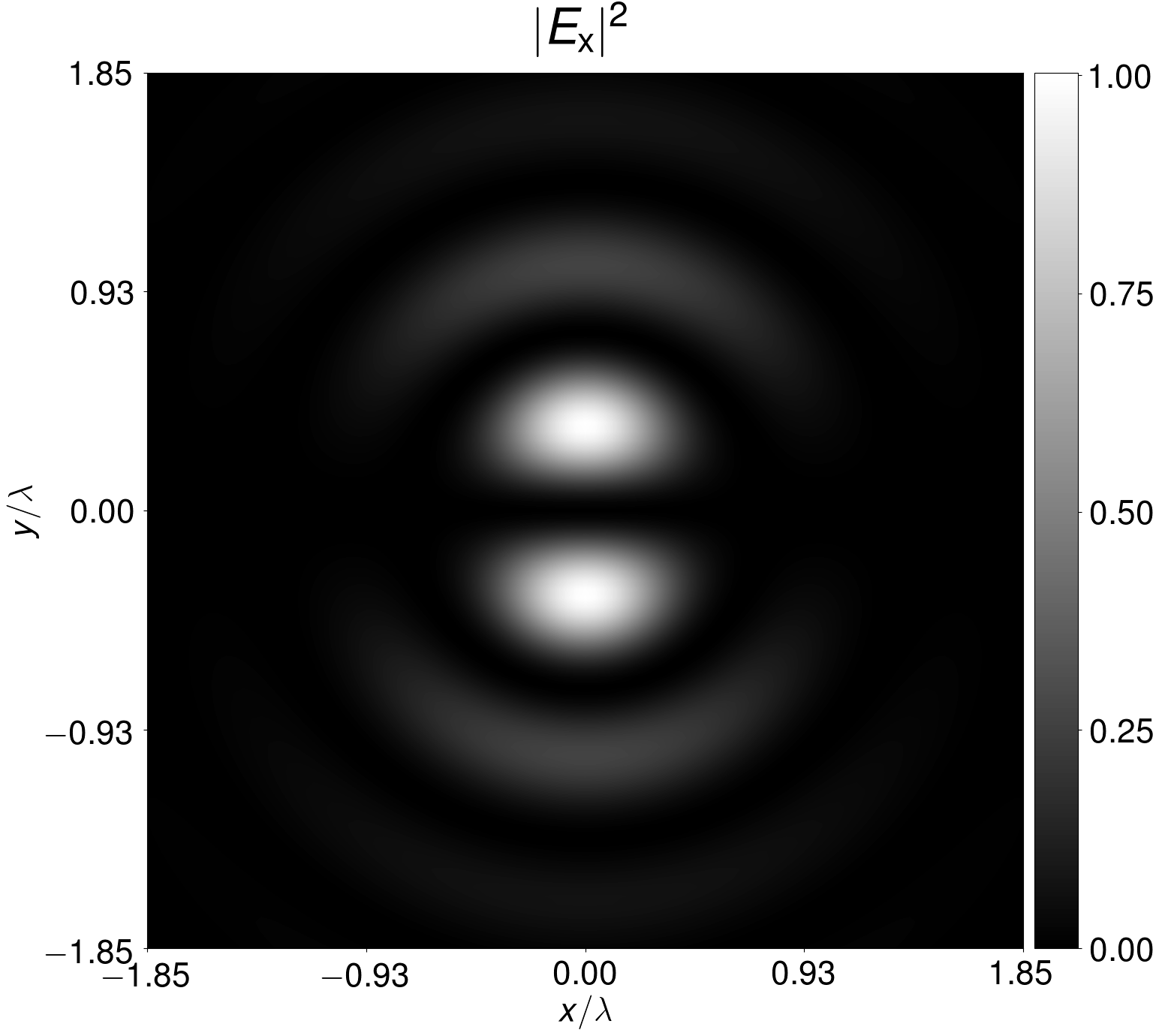}
    \includegraphics[width=0.4\textwidth]{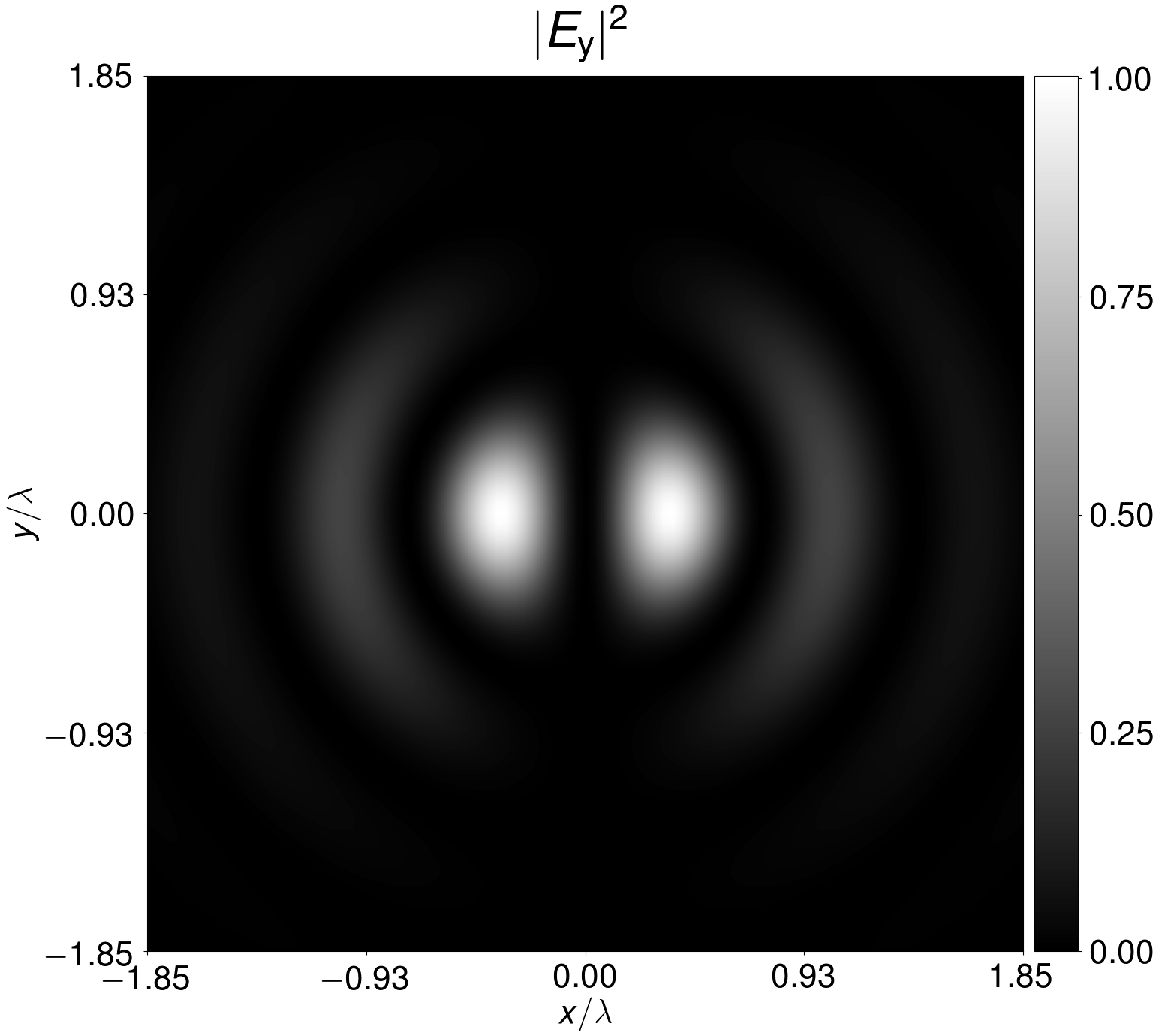}
 \\
     \includegraphics[width=0.4\textwidth]{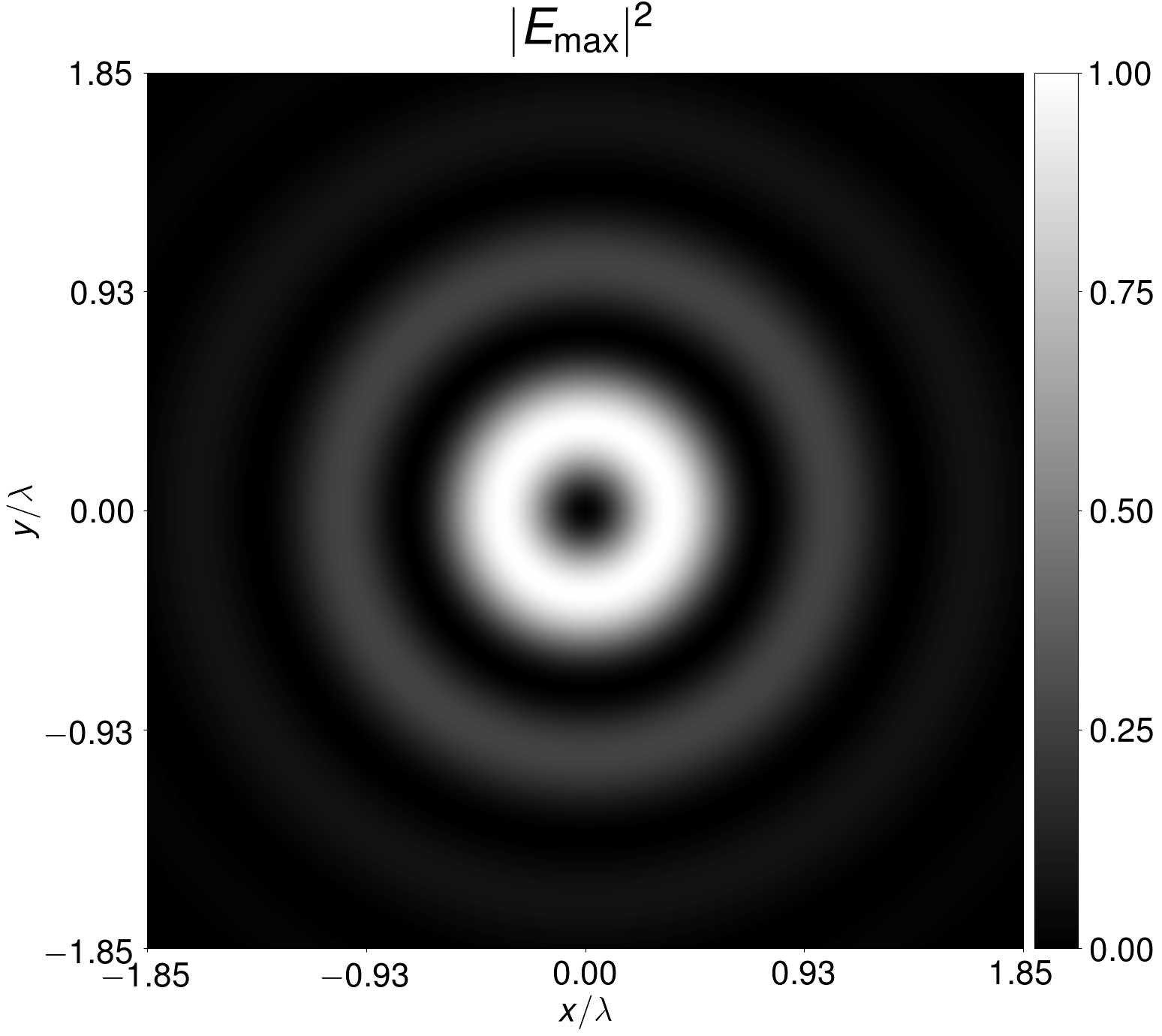}
      \caption{Optimum focused squared electric field amplitudes and electric energy density in the focal plane 
   or $R = 1.753656 \lambda$ and $\NA=0.95$. The solution has $\ell=0$. Top left: $|E_x|^2$, top right: $|E_y|^2$,  bottom:
    $|\bm{E}|^2$. The longitudinal component $E_z$ is everywhere zero. The amplitude and energy density are normalised such that the maximum of the energy density is unity.  }
   \label{Fig.ExyzefocalR1_753656NA95ell0}
\end{figure}

\begin{figure}
  \centering
   \includegraphics[width=0.5\textwidth]{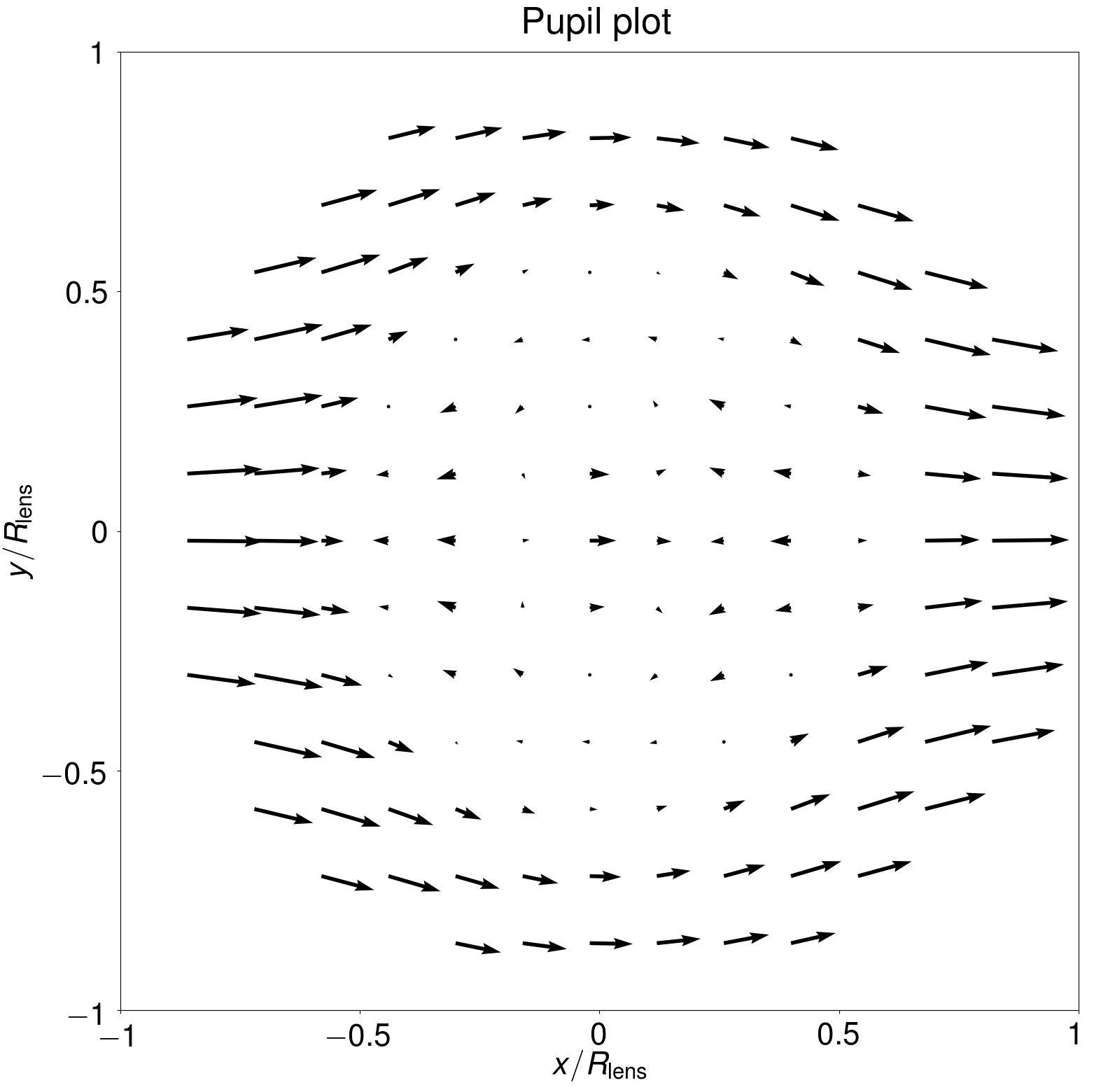}
    \caption{Pupil field for $R = 1.753656\lambda$ and $\NA=0.95$ for which two solutions exist. In the case  shown here $\ell = 1$.}
  \label{Fig.pupilR1.753656NA95ell1}
\end{figure}

\begin{figure}
  \centering
    \includegraphics[width=0.4\textwidth]{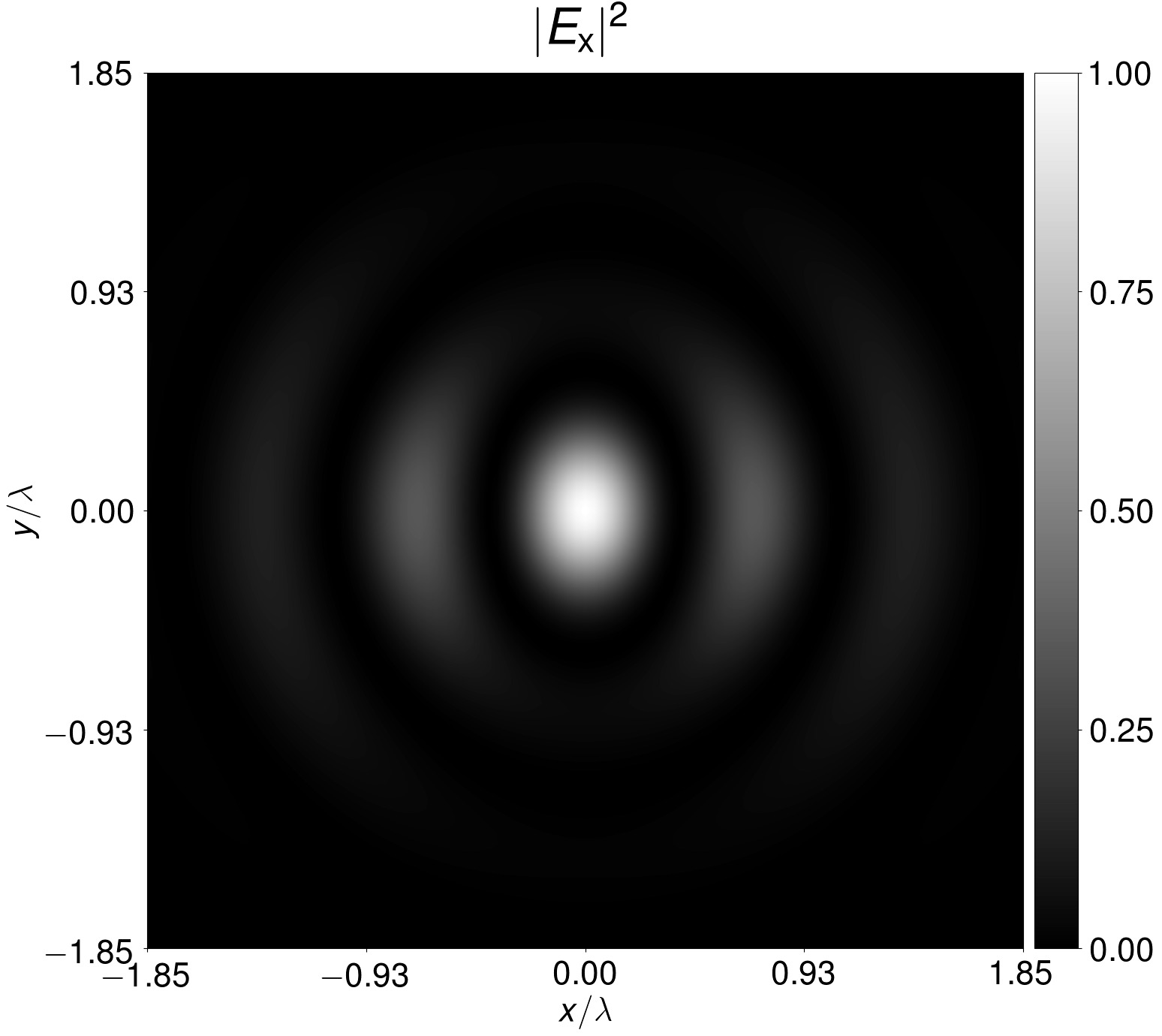}
     \includegraphics[width=0.4\textwidth]{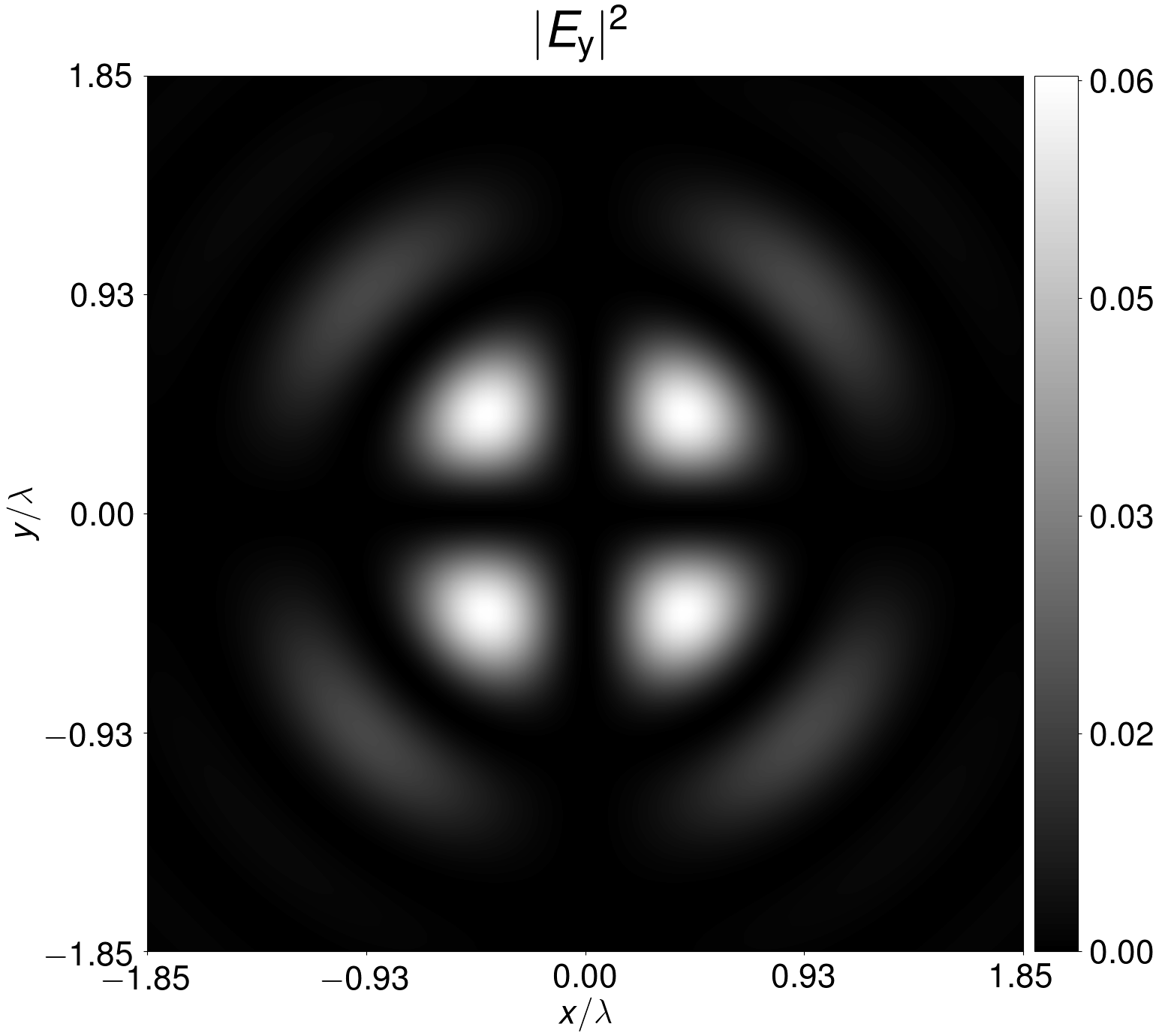}
     \\
    \includegraphics[width=0.4\textwidth]{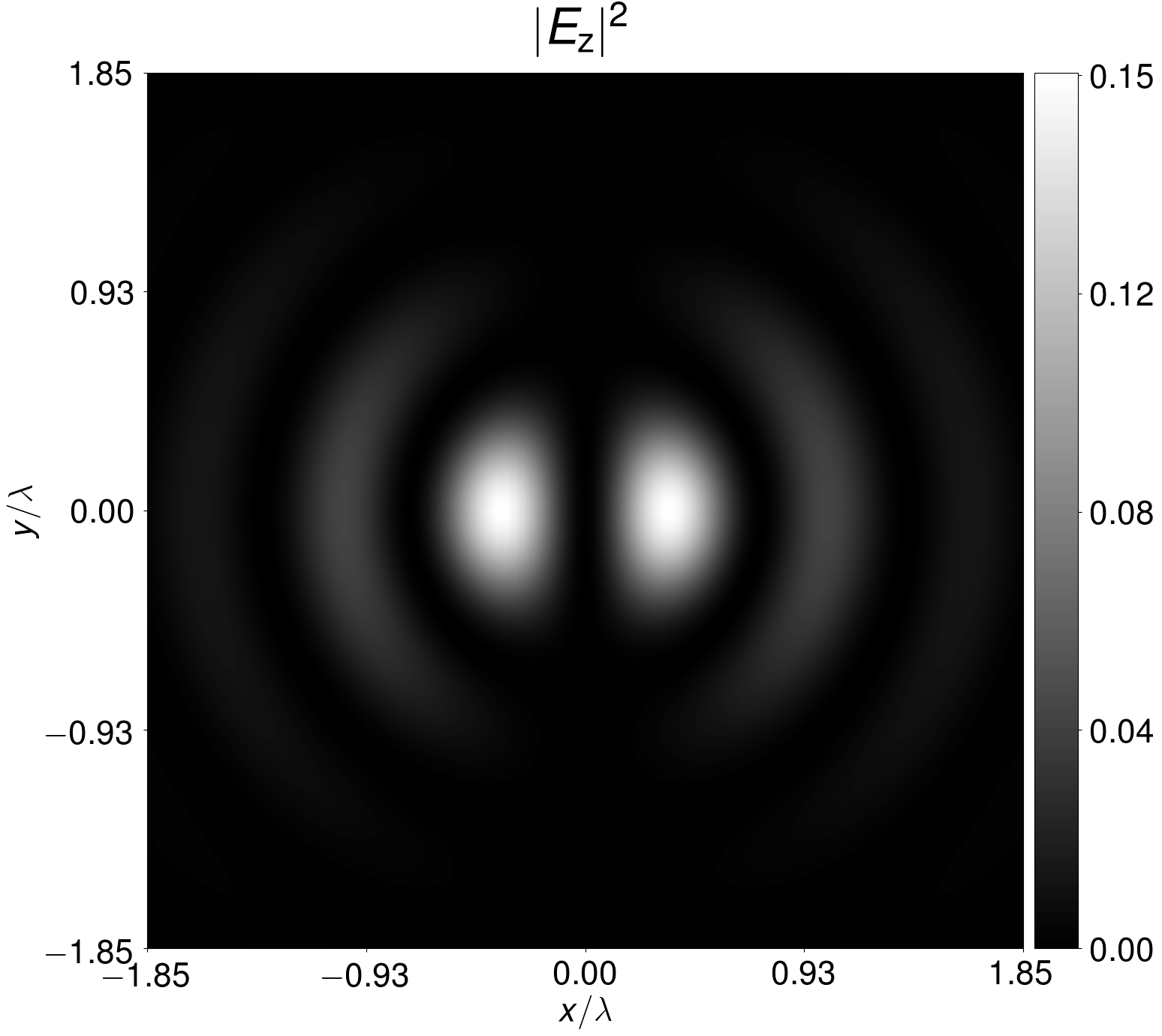}
        \includegraphics[width=0.4\textwidth]{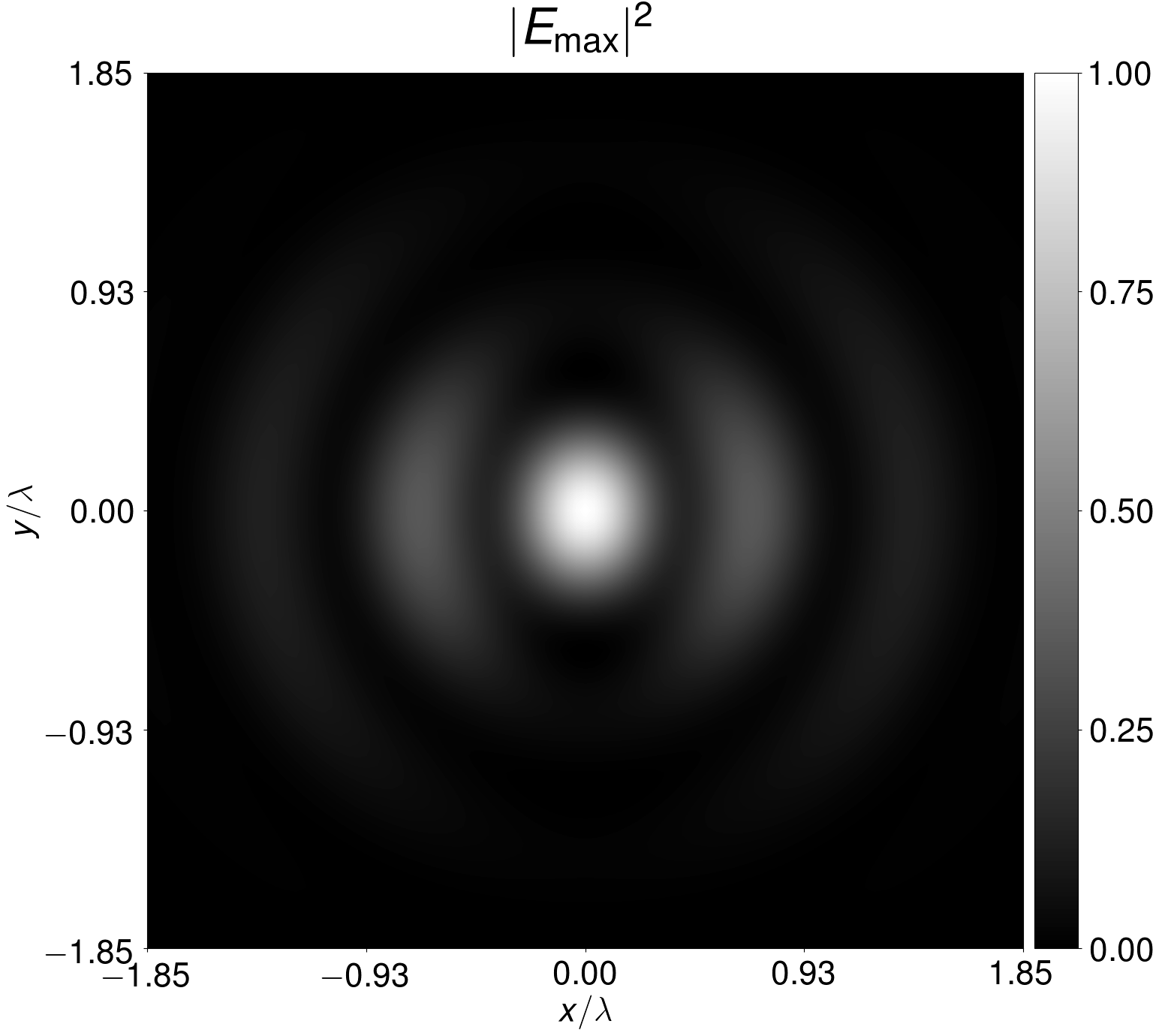}
         \caption{Optimum focused squared electric field amplitudes and electric energy density in the focal plane 
  for $R = 1.753656 \lambda$ and $\NA=0.95$. Top left: $|E_x|^2$, top right: $|E_y|^2$, bottom left: $|E_z|^2$ and bottom right:
    $|\bm{E}|^2$. The amplitude and energy density are normalised such that the maximum of the energy density is unity. The solution has $\ell=1$. }
   \label{Fig.ExyzefocalR1_753656NA95ell1}
\end{figure}

Next we consider the optimisation problem at the border between two regions where $\ell=0$ and $\ell=1$.
For $R=1.753656\lambda$ and $\NA=0.95$ two solutions are found. Fig.~\ref{Fig.pupilR1_753656NA95ell0} and
Fig.  \ref{Fig.ExyzefocalR1_753656NA95ell0} show the optimum pupil field and the optimum electric field components in the focal plane for $\ell=0$.
We have $\widehat{a_p}(\alpha,0)=0$  and hence the pupil field is azimuthally polarised.
  It is seen that the pupil field is strongly concentrated at the rim of the pupil similar to the case  of Fig.~\ref{Fig.pupilR1.25NA95}.
   In Fig.~\ref{Fig.pupilR1.753656NA95ell1} and 
 Fig.  \ref{Fig.ExyzefocalR1_753656NA95ell1}
  the optimum pupil field and the corresponding electric field components in the focal plane are shown  for the case $\ell=1$. It is seen that the pupil field amplitudes are largest at the rim. Furthermore it strongly deviates from that of a linearly polarised plane wave which is a general trend when $R$ is increased.  


\begin{figure}
  \centering
   \includegraphics[width=0.45\textwidth]{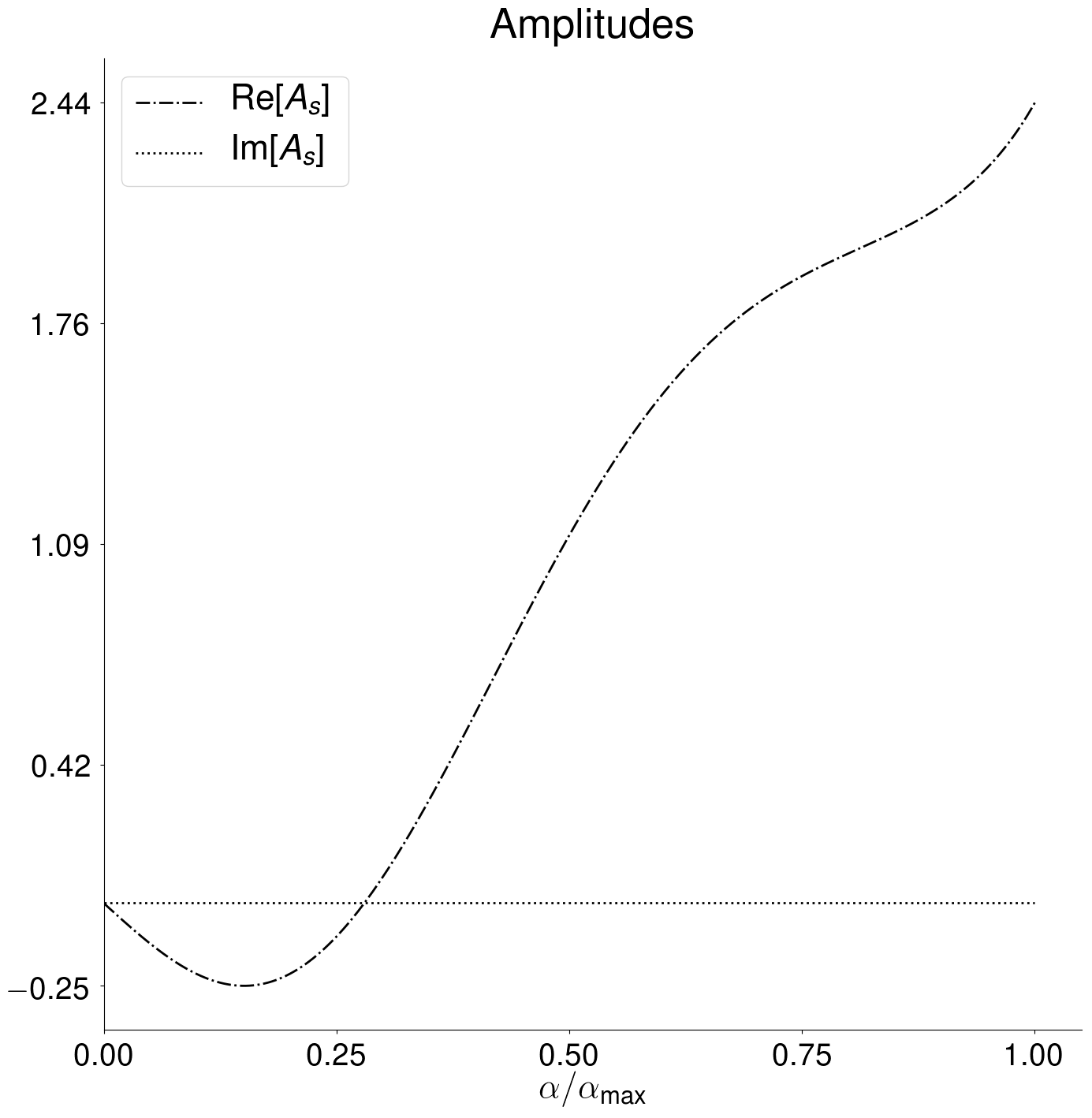}
   \includegraphics[width=0.45\textwidth]{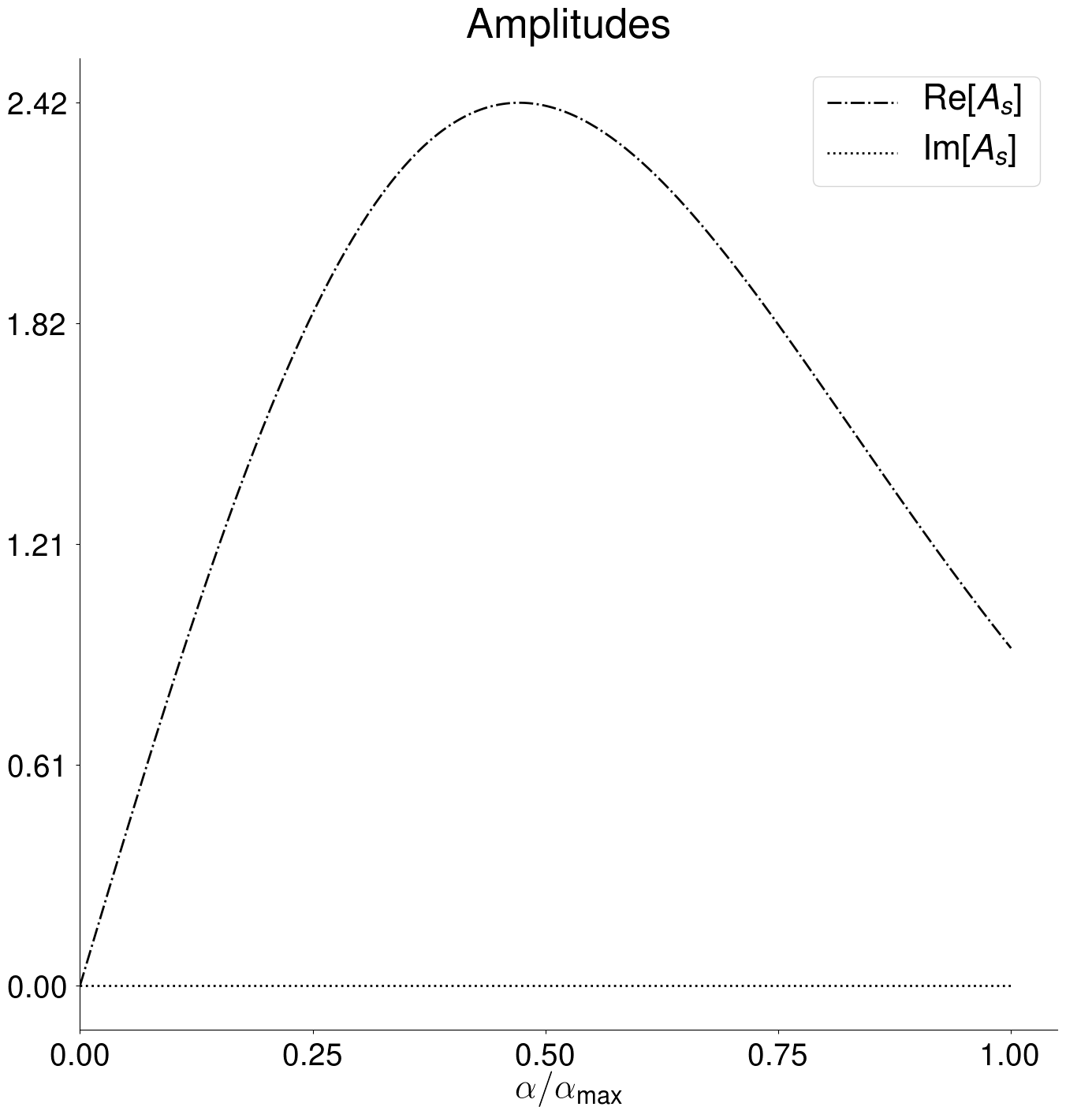}
   \\
   \includegraphics[width=0.45\textwidth]{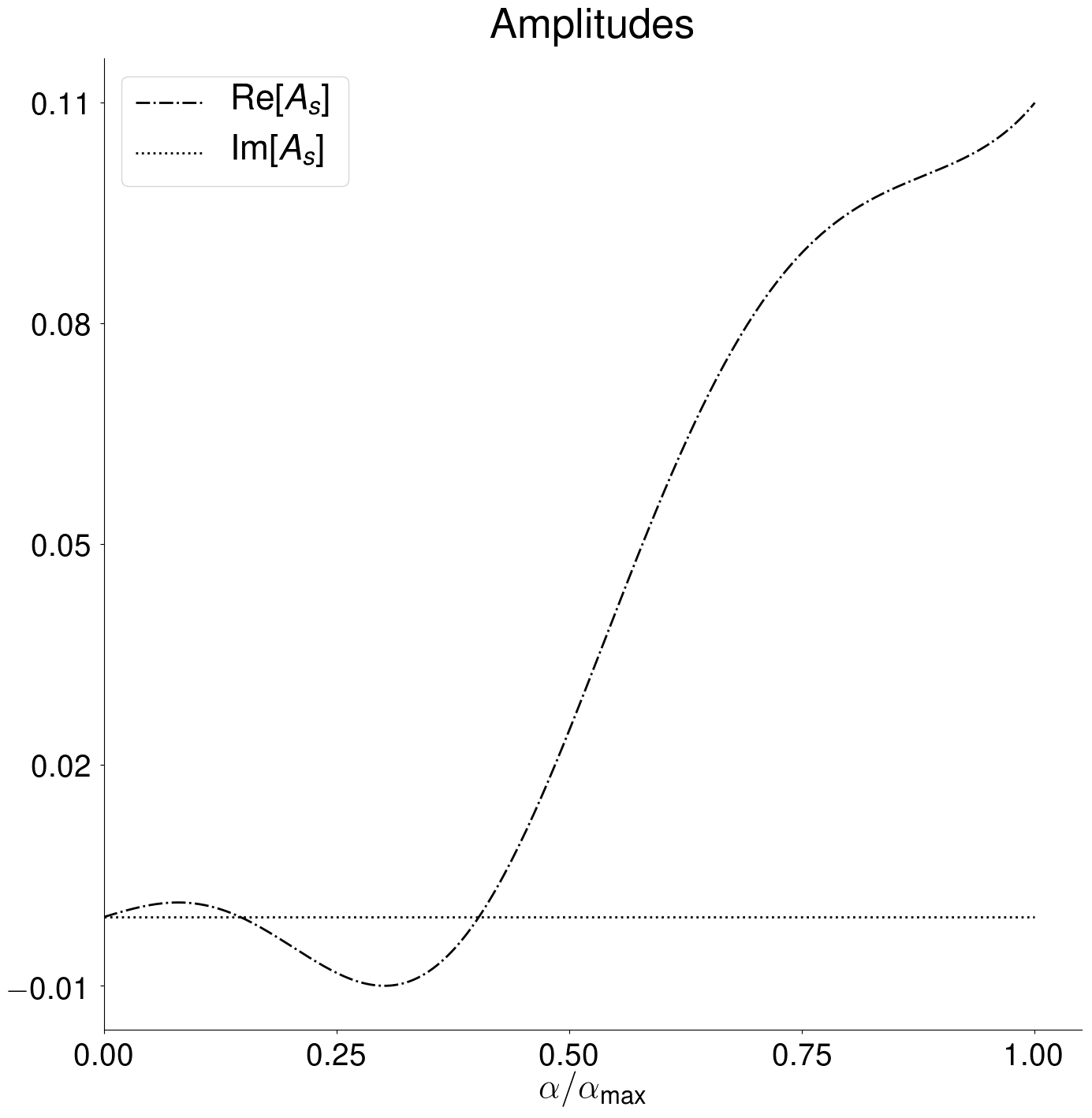}
    \caption{The  function $\alpha \mapsto \widehat{a}_s(\alpha,\ell=0)$ for the optimum pupil fields
    of Figs. \ref{Fig.pupilR1.25NA75} (top left),   \ref{Fig.pupilR1.25NA95} (top right),   \ref{Fig.pupilR1_753656NA95ell0}  (bottom).}
  \label{Fig.as_ell0}
\end{figure}

To better explain the optimum pupil fields, we show in Fig.~\ref{Fig.as_ell0} the corresponding $\widehat{a}_s(\alpha,\ell=0)$ as function of $\alpha$.

\begin{figure}
  \centering
  \includegraphics[width=0.45\textwidth]{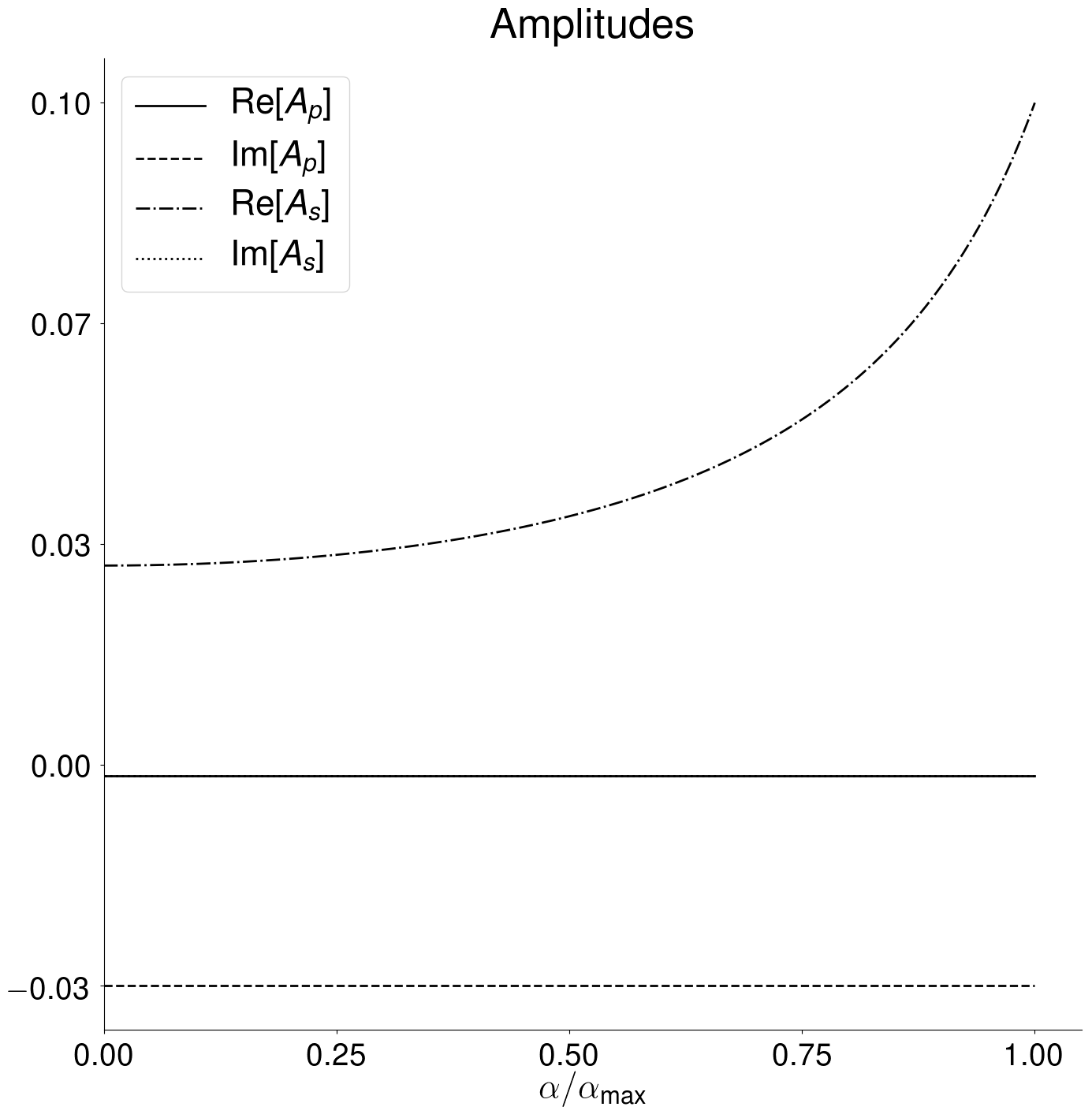}
   \includegraphics[width=0.45\textwidth]{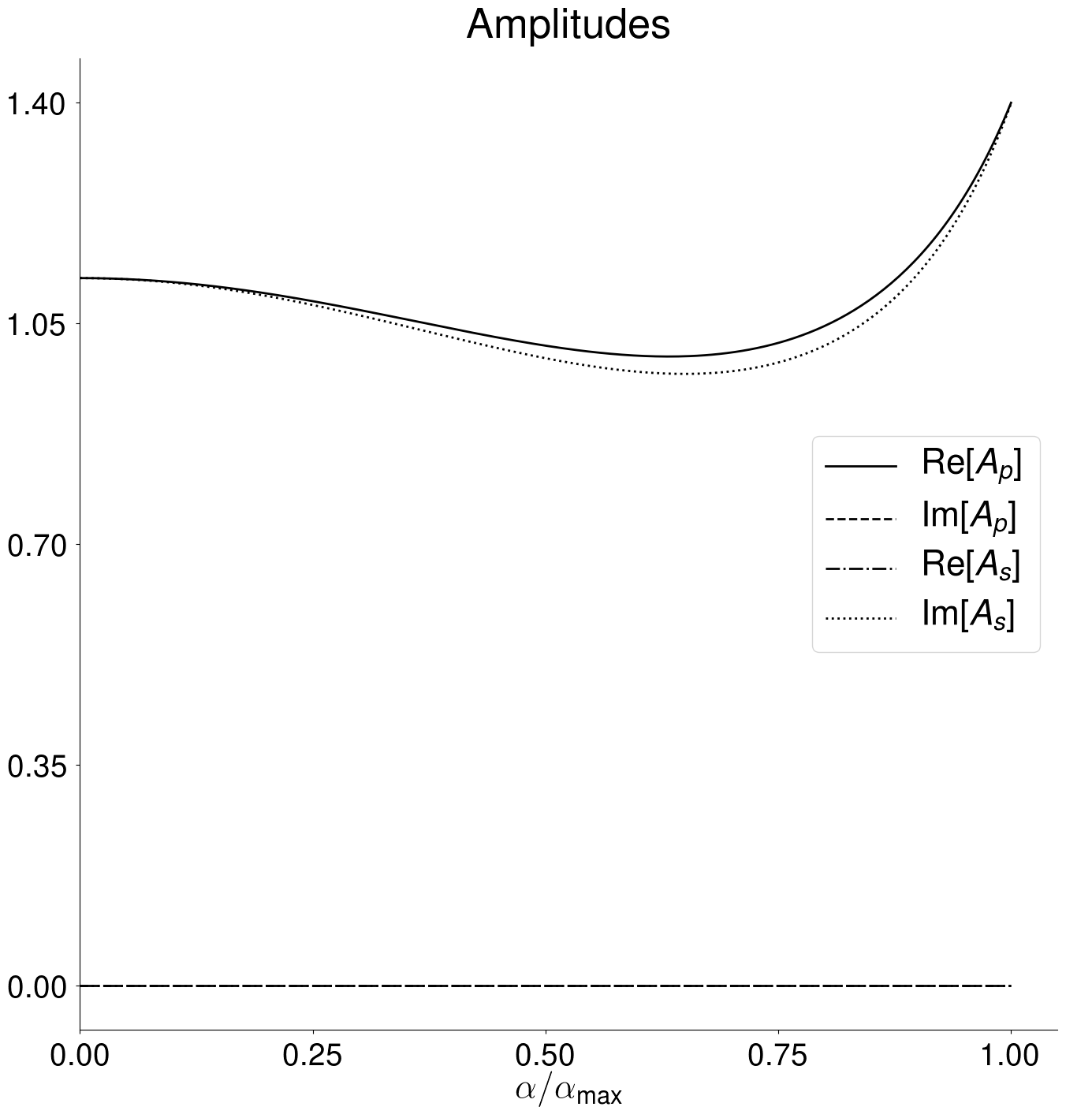}
   \\
    \includegraphics[width=0.45\textwidth]{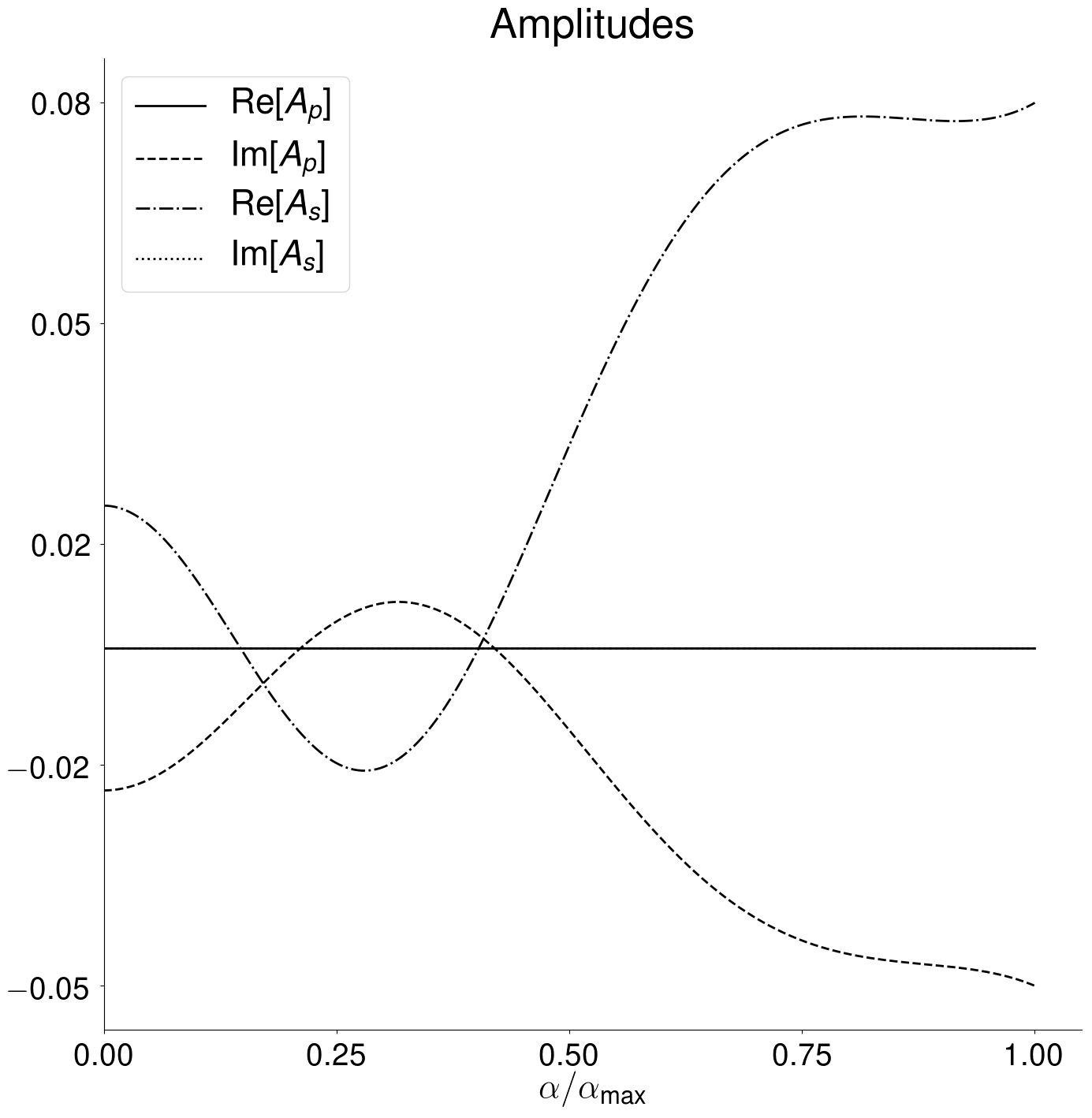}
    \caption{The  functions  $\alpha \mapsto \widehat{a}_p(\alpha,\ell=1)$ and
     $\alpha \mapsto \widehat{a}_s(\alpha,\ell=1)$ for the optimum pupil fields
    of Figs. \ref{Fig.pupilR0NA95} (top left),   \ref{Fig.pupilR0.5NA95} (top right),  
 \ref{Fig.pupilR1.753656NA95ell1} (bottom). }
  \label{Fig.apas_ell1}
\end{figure}


\section{Conclusion}
We have derived a general formulism for obtaining the electromagnetic field with given power  and given numerical aperture of which the electric energy averaged over a bounded set is maximum. The set can be chosen arbitrarily: it may consist of finitely many points,  it may be a curve, a (curved) surface or a three dimensional region. It has been shown that the Lagrange multiplier rule implies that the optimum field is eigenfield with maximum eigenvalue of an integral operator whose kernel is determined by the set. This integral operator is compact and also hermitian, provided the proper scalar product is chosen. Hence its spectrum is discrete and there is a maximum eigenvalue. It was shown that when  the set over which the electric energy is averaged is scaled by a parameter $\sigma>0$,  the optimum solution remains the same if the numerical aperture and the ratio of $\sigma$ and the wavelength are kept constant. We have studied in more detail the problem of maximizing the electric energy in a disc perpendicular to and symmetric with respect to the optical axis. If the radius of the disc vanishes, the energy in a single  point is maximized. In this case the optimum pupil field which after focusing gives maximum eleectric energy density in the focal point can be computed in closed form and is similar to that of a linear polarised plane wave. For general radii, the optimum solutions must be computed numerically. It is found that when the numerical aperture is fixed and the radius of the disc is increased, the optimum pupil fields alternate between a field that resembles more or less that of a polarised plane wave with constant direction of polarisation, and an azimuthally polarized pupil field . At values of $\NA$ and the radius over the wavelength where the transitions between the two types of solutions occurs, multiple optimum fields exist. 

\section*{Acknowledgements}
H.P.U acknowledges interesting discussions with Bogathi V. Reddy during the intial phase of the research. The authors thank Jan M.A.M.  van Neerven of the  Delft Institute of Applied Mathematics for mathematical advise.

\appendix  
\section{Derivation of Eq. \ref{eq.GSAv}}\label{appendix.A}
We express $G_{S,\Pi}$ in terms of the plane wave amplitudes $\Av$. The
following derivation is formal but can be mathematically justified.

First we remark that \eqref{eq:electric-field} implies for every $z$:
\begin{align}
  \label{eq.appendixA_FE}
  \mathcal{F}_2(\Pi(\Ev))(\kp, z) &= \Pi(\Av)(\kp) \e^{i k_z z}, \\
  \label{eq.appendixA_FH}
  \mathcal{F}_2(\Pi(\Ev)^\ast)(\kp, z) &= \Pi(\Av)(-\kp)^\ast \e^{-i k_z z}, 
\end{align}
where $k_z=k_z(\kp)$ given by \eqref{eq.defkzkp} and
$\mathcal{F}_2$ is the 2D  Fourier transform defined by
  (\ref{eq:def_Fourier-transform}) and its inverse by 
  (\ref{eq:inv-Fourier-transform})
We apply Plancherel's identity and the convolution theorem using the 3D Fourier transform:
\begin{equation}
  \begin{split}
    G_{S,\Pi}(\Ev) &= \langle T_S, |\Pi(\Ev)|^2 \rangle_{\R^3} 
    = \frac{1}{(2\pi)^3} \langle \FIII(T_S), \FIII[\Pi(\Ev) \Pi(\Ev)^*]^*\rangle\\
    &= \frac{1}{(2\pi)^6} \langle \FIII(T_S), \, \FIII(\Pi(\Ev))^\ast \ast
    \FIII(\Pi(\Ev^*))^\ast \rangle_{\R^3}.
    \label{eq.appendixA_GSP1}
  \end{split}
\end{equation}
Next we write the 3D Fourier transform of $\Pi(\Ev)$ as the  composition of the 2D
and the 1D Fourier transform:
\begin{equation}
  \FIII(\Pi(\Ev))(\xip,\xi_z) = \iiint_{\R^3}\Pi( \Ev)(\rp,z)
  \e^{-i(\xip\cdot \rp + \xi_z z)} \Dxip \D{z} = \int_\R \e^{-i \xi_z z} \FII(\Pi(\Ev))(\xip,z) \D{z}.
  \label{eq.appendixA_F3F2}
\end{equation}
Using \eqref{eq.appendixA_FE}, we find:
\begin{subequations}
  \label{eqFIIIEv}
  \begin{align}
    \FIII(\Pi(\Ev))(\xip,\xi_z) &= \int_\R \e^{-i(\xi_z - k_z(\xip))z} \D{z}
                                  \, \Pi(\Av)(\xip) = 2\pi \delta(\xi_z - k_z(\xip)) \cdot \Pi(\Av)(\xip),\\
    \FIII(\Pi(\Ev)^\ast)(\xip,\xi_z) &= 2\pi \delta(\xi_z + k_z(\xip)) \cdot \Pi(\Av)(-\xip)^\ast.
  \end{align}
\end{subequations}
Hence,
\begin{align*}
  [\FIII(\Pi(\Ev))^\ast \ast \FIII(\Pi(\Ev)^\ast)^\ast](\xip, \xi_z) = 4\pi^2
  \iint_{\R^2} \delta(\xi_z + k_z(\xip') - k_z(\xip - \xip'))
  \Pi(\Av)(-\xip') \cdot \Pi(\Av)(\xip - \xip')^\ast \D\xip',
\end{align*}
where we have used that
\begin{equation*}
  \int_\R \delta(\xi_z - \xi_z' - k_z(\xip - \xip'))
  \delta(\xi_z' + k_z(\xip - \xip')) \D\xi_z' = \delta(\xi_z + k_z(\xip') - k_z(\xip - \xip')),
\end{equation*}
which can be verified by integrating against a test function. Substitution into
(\ref{eq.appendixA_GSP1})
  yields after computing the integral
over the $\delta$-function:
\begin{equation*}
  G_{S,\Pi}(\Ev) = \frac1{(2\pi)^4} \iint_{\R^2} \iint_{\R^2}\FIII(T_S)(\xip, k_z(\xip -
  \xip') - k_z(\xip'))  \Pi(\Av)(-\xip') \cdot
  \Pi(\Av)(\xip - \xip')^\ast  \Dxip \Dxip'.
\end{equation*}
By a change of integration variables we get:
\begin{equation}
  \label{eq.appendixA_GSAv}
  G_{S,\Pi}(\Ev) = \frac1{(2\pi)^4} \iint_\Omega \iint_{\Omega} \FIII(T_S)(\xip - \xip', k_z(\xip) - k_z(\xip'))  \Pi(\Av)(\xip') \cdot
  \Pi( \Av)(\xip)^\ast  \Dxip \Dxip',
\end{equation}
Since the integral is over 2D Fourier variables $\xip$, we can switch back to the $\kp$-variables to 
finally get:
\begin{equation}
  \label{eq.appendixA_GSAv_2}
  G_{S,\Pi}(\Ev) = \frac1{(2\pi)^4} \iint_{\Omega} \iint_{\Omega} \FIII(T_S)(\kp - \kp', k_z(\kp) - k_z(\kp'))  \Pi(\Av)(\kp') \cdot
  \Pi( \Av)(\kp)^\ast  \Dkp \Dkp',
\end{equation}
This is the averaged energy density in $S$ expressed in terms of the plane wave amplitudes $\Av$.

\section{The Fourier coefficients of $C_R$}\label{sec:fourier-coefficients}
\label{appendix_B}
In this appendix we compute the Fourier coefficients of $C_R$. As a first step,
we expand the Bessel function $J_1$ in
\eqref{eq:Fourier-transform-disc-R} into its Taylor series
\begin{equation}
  \label{eq:Fourier-coeff-CR-1}
  \frac{J_1(R |\kp - \kp'|)}{R |\kp - \kp'|} =  \sum_{k = 0}^\infty \frac{(-1)^k}{k! (k + 1)!}
  \biggl(\frac{R|\kp - \kp'|}2 \biggr)^{2k}.
\end{equation}
Next, we recall \eqref{eq:dependence-kpmkpp} to which we apply the binomial theorem
after setting $\gamma=\beta - \beta'$, $\xi=\sin \alpha$ and $\xi'=\sin \alpha'$:
\begin{align*}
  (\xi^2 + &\xi'^2 - 2 \xi \xi'
             \cos\gamma)^k = \sum_{\ell = 0}^k \binom{k}{\ell} (-1)^\ell
             \cos^\ell\gamma \,  (2 \xi\xi')^\ell (\xi^2 + \xi'^2)^{k - \ell}.
\end{align*}
Combining with \eqref{eq:Fourier-coeff-CR-1}:
\begin{align*}
      \frac{J_1(R |\kp - \kp'|)}{R |\kp - \kp'|} =
      \sum_{\ell = 0}^\infty \sum_{m = \ell}^\infty  \frac{(-1)^{m + \ell}}{(m +
      1)! \ell! (m - \ell)!}
  \biggl(\frac{R k}2 \biggr)^{2m}
  \cos^\ell\gamma \, (2 \xi\xi')^\ell (\xi^2 + \xi'^2)^{m - \ell}.
\end{align*}
In the next step, we apply the binomial theorem to $2 \cos\gamma =
\e^{i \gamma} + \e^{-i \gamma}$. Combining and rearranging gives
\begin{align*}
  %
                        \frac{J_1(R |\kp - \kp'|)}{R |\kp - \kp'|} =
  &= \frac1{4\pi^2} \sum_{\ell = 0}^\infty \e^{i \ell \gamma} \sum_{s = 0}^\ell \binom{\ell}{s} \e^{-2 i s \gamma}
    (\xi\xi')^\ell \sum_{m = \ell}^\infty  \frac{(-1)^{m + \ell}}{(m +
    1)! \ell! (m - \ell)!} \biggl(\frac{R k}2 \biggr)^{2m}
    (\xi^2 + \xi'^2)^{m - \ell}.
\end{align*}
Rearranging the sums gives
\begin{equation*}
  \frac{J_1(R |\kp - \kp'|)}{R |\kp - \kp'|} 
  = \frac1{4\pi^2}  \sum_{\ell = -\infty}^\infty \e^{i \ell \gamma} \sum_{s = \max(0,
    -\ell)}^\infty \frac{(\xi \xi')^{\ell + 2s}}{s! (\ell
    + s)!} \biggl(\frac{R k}2 \biggr)^{2\ell + 4s} \sum_{m = 0}^\infty
  \frac{(-1)^{m} (\xi^2 + \xi'^2)^m}{m! (m + \ell + 2s + 1)!} \biggl(\frac{R k}2 \biggr)^{2m}.
\end{equation*}
The last sum over $m$  is the expansion of a Bessel function:
\begin{equation*}
  \sum_{m = 0}^\infty \frac{(-1)^{m} \sqrt{\xi^2 +
      \xi'^2}^{2m}}{m! (m + \ell + 2s + 1)!} \biggl(\frac{R k}2
  \biggr)^{2m} = 2^{\ell + 2s + 1} \frac{J_{\ell + 2s + 1}(Rk \sqrt{\xi^2 + \xi'^2})}{(Rk \sqrt{\xi^2 +
      \xi'^2})^{\ell + 2s + 1}}.
\end{equation*}
Hence,
\begin{equation*}
  \frac{J_1(R |\kp - \kp'|)}{R |\kp - \kp'|} 
  = \frac1{4\pi^2}  \sum_{\ell = -\infty}^\infty \e^{i \ell \gamma} \sum_{s = \max(0,
    -\ell)}^\infty \frac{(\xi \xi')^{\ell + 2s}}{s! (\ell
    + s)!} \biggl(\frac{R k}2 \biggr)^{2\ell + 4s} 
  2^{\ell + 2s + 1} \frac{J_{\ell + 2s + 1}(Rk \sqrt{\xi^2 + \xi'^2})}{(Rk \sqrt{\xi^2 +
      \xi'^2})^{\ell + 2s + 1}}.
\end{equation*}
Since  by \eqref{eq.defCR}
\begin{equation}
  C_R(\alpha,\alpha',\beta) = \frac{2 \cos \alpha' \sin \alpha'}{\cos\alpha} \,  \frac{ J_1(k R \sqrt{ \sin^2\alpha + \sin^2 \alpha' - 2 \sin\alpha \sin\alpha' \cos\beta})}{  R \sqrt{ \sin^2\alpha + \sin^2 \alpha' - 2 \sin\alpha \sin\alpha' \cos\beta}},
  \label{eq.appendix_defCR}
\end{equation}
it follows that the Fourier coefficients of $\beta \mapsto C_R(\alpha,\alpha',\beta)$ are: 
\begin{equation}
  \label{eq:CR-Fourier-coefficients}
  \widehat{C_R}(\alpha, \alpha',\ell) = \frac1{4\pi^2}  \frac{2 \cos \alpha' \sin \alpha'}{\cos\alpha} \, \sum_{s = \max(0,
    -\ell)}^\infty \frac{(\sin\alpha \sin\alpha')^{\ell + 2s}}{s! (\ell
    + s)!} \biggl(\frac{R k}{\sqrt{2}} \biggr)^{2\ell + 4s}
  \frac{J_{\ell + 2s + 1}(Rk \sqrt{\sin^2 \alpha + \sin^2\alpha'})}{(Rk \sqrt{\sin^2 \alpha +
      \sin^2\alpha'})^{\ell + 2s + 1}}.
\end{equation}
Recall that we have shown that is is sufficient to consider $\ell \geq
0$.  For these $\ell$ the expression can be simplified slightly. The partial sums of the series converge very fast.

\section{Analytical evaluation of the integrals with respect to polar angle of the focused field.}
\label{appendix_C}
We will use the following notations:
\begin{subequations}
  \begin{align}
    N_{x,\ell}(\phi,\gamma) &= \int_0^{2\pi} \e^{i \gamma
                              \cos(\beta - \phi)} \e^{i \ell \beta} \cos \beta  \D\beta,\\
    N_{y,\ell}(\phi,\gamma) &= \int_0^{2\pi} \e^{i \gamma \cos(\beta - \phi)} \e^{i
                              \ell \beta} \sin \beta \D\beta,\\
    N_{z,\ell}(\phi,\gamma) &= \int_0^{2\pi} \e^{i \gamma
                              \cos(\beta - \phi)} \e^{i \ell \beta} \D\beta.
  \end{align}
\end{subequations}
These integrals can be computed analytically, using the
integral representations of the Bessel functions \cite[Equation
4.7.6]{Andrews1999}:
\begin{subequations}
  \begin{align}
    N_{x,\ell}(\phi,\gamma) &= -\pi i^{\ell - 1} \e^{i \ell \phi}  [\e^{i \phi} J_{\ell + 1}(\gamma) - \e^{-i \phi} J_{\ell - 1}(\gamma)],\\
    N_{y,\ell}(\phi,\gamma) &= \pi i^\ell \e^{i \ell \phi} [\e^{i \phi} J_{\ell + 1}(\gamma) + \e^{-i \phi} J_{\ell - 1}(\gamma)],\\
    N_{z,\ell}(\phi,\gamma) &= 2\pi i^\ell \e^{i \ell \phi} J_\ell(\gamma).
  \end{align}
\end{subequations}
Using these expressions it follows that
\begin{equation}
  \int_0^{2\pi} \pu(\alpha, \beta) \e^{i \ell \beta} \e^{i \gamma \cos(\beta - \phi)} \D\beta = 
  \begin{pmatrix}
    - N_{x, \ell}(\phi, \gamma) \cos \alpha \\
    - N_{y, \ell}(\phi, \gamma) \cos \alpha,\\
    N_{z, \ell}(\phi, \gamma) \sin \alpha
  \end{pmatrix},
  \label{eq.intpu}
\end{equation}
and
\begin{equation}
  \int_0^{2\pi} \su(\beta) \e^{i \ell \beta} \e^{i \gamma \cos(\beta-\phi)} \D\beta =
  \begin{pmatrix}
    -N_{y, \ell}(\phi, \gamma)\\
    N_{x, \ell}(\phi, \gamma)\\
    0
  \end{pmatrix}.
  \label{eq.intsu}
\end{equation}
Note that 
\begin{align*}
  N_{x, -\ell}(\phi, \gamma) &= (-1)^{\ell - 1} N_{x, \ell}(\phi, \gamma)^\ast,\\
  N_{y, -\ell}(\phi, \gamma) &= (-1)^{\ell - 1} N_{y, \ell}(\phi, \gamma)^\ast,\\
  N_{z, -\ell}(\phi, \gamma) &= (-1)^{\ell} N_{z, \ell}(\phi, \gamma)^\ast.
\end{align*}
So that
\begin{equation}
  \int_0^{2\pi} \pu(\alpha, \beta) \e^{-i \ell \beta} \e^{i \gamma \cos(\beta - \phi)} \D\beta = (-1)^{\ell }
  \begin{pmatrix}
    N_{x, \ell}(\phi, \gamma)^\ast  \cos \alpha\\
    N_{y, \ell}(\phi, \gamma)^\ast  \cos\alpha ,\\
    N_{z, \ell}(\phi, \gamma)^\ast  \sin\alpha
  \end{pmatrix},
  \label{eq.intpu2}
\end{equation}
and
\begin{equation} 
  \int_0^{2\pi} \su(\beta) \e^{-i \ell \beta} \e^{i \gamma \cos(\beta - \phi)} \D\beta = (-1)^{\ell }
  \begin{pmatrix}
    N_{y, \ell}(\phi, \gamma)^\ast\\
    -N_{x, \ell}(\phi, \gamma)^\ast\\
    0
  \end{pmatrix}.
  \label{eq.intsu2}
\end{equation}
Then
\begin{align}
  \int_0^{2\pi} \Re[   \widehat{a_p}(\alpha,\ell) \e^{i \ell \beta}   \pu(\alpha, \beta)] \e^{i k ( \rho
  \sin\alpha \cos (\beta-\phi)} \D\beta 
  & =  \frac12 \int_0^{2\pi}   \widehat{a_p}(\alpha,\ell) \e^{i \ell \beta}   \pu(\alpha, \beta) \e^{i k ( \rho
        \sin\alpha \cos (\beta-\phi)} \D\beta  \nonumber \\
  &  + \frac12 \int_0^{2\pi}   \widehat{a_p}(\alpha,\ell)^* \e^{-i \ell \beta}   \pu(\alpha, \beta) \e^{i k ( \rho
      \sin\alpha \cos (\beta-\phi)} \D\beta \nonumber \\
  &= \frac12 \begin{pmatrix}
              - \Re[ \widehat{a_p}(\alpha,\ell) N_{x,\ell}(\phi,k\rho\sin \alpha)] \cos \alpha\\
              - \Re[ \widehat{a_p}(\alpha,\ell) N_{y,\ell}(\phi,k\rho\sin \alpha)] \cos \alpha\\
              i \Im [ \widehat{a_p}(\alpha,\ell) N_{z,\ell}(\phi,k\rho\sin \alpha)] \sin \alpha
            \end{pmatrix},  \mbox{ if } \ell \mbox{ is odd}, \label{eq.integralpoddell}\\
  & = \frac12 \begin{pmatrix}
                - i \Im[ \widehat{a_p}(\alpha,\ell) N_{x,\ell}(\phi,k\rho\sin \alpha)] \cos \alpha\\
                - i\Im[ \widehat{a_p}(\alpha,\ell) N_{y,\ell}(\phi,k\rho\sin \alpha) ]\cos \alpha\\
                \Re [ \widehat{a_p}(\alpha,\ell) N_{z,\ell}(\phi,k\rho\sin \alpha)] \sin \alpha
              \end{pmatrix},  \mbox{ if } \ell \mbox{ is even}, \label{eq.integralpevenell}
\end{align}
and
\begin{align}
  \int_0^{2\pi} \Re[   \widehat{a_s}(\alpha,\ell) \e^{i \ell \beta}   \su(\alpha, \beta)] \e^{i k ( \rho
  \sin\alpha \cos (\beta-\phi)} \D\beta 
  & = \frac12 \int_0^{2\pi}   \widehat{a_s}(\alpha,\ell) \e^{i \ell \beta}   \su(\alpha, \beta) \e^{i k ( \rho
        \sin\alpha \cos (\beta-\phi)} \D\beta  \nonumber \\
  & +\frac12 \int_0^{2\pi}   \widehat{a_s}(\alpha,\ell)^* \e^{-i \ell \beta}   \su(\alpha, \beta) \e^{i k ( \rho
      \sin\alpha \cos (\beta-\phi)} \D\beta \nonumber \\
  &= \frac12 \begin{pmatrix}
              - \Re[ \widehat{a_s}(\alpha,\ell) N_{y,\ell}(\phi,k\rho\sin \alpha)] \\
              \Re[ \widehat{a_s}(\alpha,\ell) N_{x,\ell}(\phi,k\rho\sin \alpha) ]\\
              0
            \end{pmatrix},  \mbox{ if } \ell \mbox{ is odd}, \label{eq.integralsoddell}\\
  &= \frac12 \begin{pmatrix}
                - i \Im[ \widehat{a_s}(\alpha,\ell) N_{y,\ell}(\phi,k\rho\sin \alpha)]\\
                i\Im[ \widehat{a_s}(\alpha,\ell) N_{,\ell}(\phi,k\rho\sin \alpha) ] \\
                0
              \end{pmatrix},  \mbox{ if } \ell \mbox{ is even}. \label{eq.integralsevenell}
\end{align}
Hence, with \eqref{eq.Evfocal}:
\begin{align}  
  \Ev(\rho,\phi,z) = \frac12 \frac{k^2}{2\pi^2} \int_0^{\amax} \left[ 
  \begin{pmatrix}
          - \Re[ \widehat{a_p}(\alpha,\ell) N_{x,\ell}(\phi,k\rho\sin \alpha)] \cos \alpha\\
          - \Re[ \widehat{a_p}(\alpha,\ell) N_{y,\ell}(\phi,k\rho\sin \alpha)] \cos \alpha\\
          i \Im [ \widehat{a_p}(\alpha,\ell) N_{z,\ell}(\phi,k\rho\sin \alpha)] \sin \alpha
        \end{pmatrix} + 
  \begin{pmatrix}
          - \Re[ \widehat{a_s}(\alpha,\ell) N_{y,\ell}(\phi,k\rho\sin \alpha)] \\
          \Re[ \widehat{a_s}(\alpha,\ell) N_{x,\ell}(\phi,k\rho\sin \alpha) ]\\
          0
        \end{pmatrix} \right] \nonumber \\
  \times \e^{i k z \cos\alpha} \cos\alpha \sin \alpha \D \alpha,
  \label{eq.Evoddell}
\end{align}
if $\ell $ is odd, and
\begin{align}  
  \Ev(\rho,\phi,z) = \frac12 \frac{k^2}{2\pi^2} \int_0^{\amax} \left[
  \begin{pmatrix}
          - i \Im[ \widehat{a_p}(\alpha,\ell) N_{x,\ell}(\phi,k\rho\sin \alpha)] \cos \alpha\\
          - i\Im[ \widehat{a_p}(\alpha,\ell) N_{y,\ell}(\phi,k\rho\sin \alpha) ]\cos \alpha\\
          \Re [ \widehat{a_p}(\alpha,\ell) N_{z,\ell}(\phi,k\rho\sin \alpha)] \sin \alpha
        \end{pmatrix} +
  \begin{pmatrix}
          - i \Im[ \widehat{a_s}(\alpha,\ell) N_{y,\ell}(\phi,k\rho\sin \alpha)]\\
          i\Im[ \widehat{a_s}(\alpha,\ell) N_{,\ell}(\phi,k\rho\sin \alpha) ] \\
          0
        \end{pmatrix} \right] \nonumber \\
  \times \e^{i k z \cos\alpha} \cos\alpha \sin \alpha \D \alpha, 
  \label{eq.Evevenell}
\end{align}
if $\ell$ is even.
Similarly, using  \eqref{eq.Hvfocal}:
\begin{align}
  \Hv(\rho,\phi,z)  = \frac12
  n \sqrt{\frac{\epsilon_0 }{\mu_0}}  \frac{k^2}{2\pi^2}\int_0^{\amax}  \Re \left[
  \begin{pmatrix}
          - \Re[ \widehat{a_p}(\alpha,\ell) N_{y,\ell}(\phi,k\rho\sin \alpha)] \\
          \Re[ \widehat{a_p}(\alpha,\ell) N_{x,\ell}(\phi,k\rho\sin \alpha) ]\\
          0
        \end{pmatrix}
  - 
  \begin{pmatrix}
          - \Re[ \widehat{a_s}(\alpha,\ell) N_{x,\ell}(\phi,k\rho\sin \alpha)] \cos \alpha\\
          - \Re[ \widehat{a_s}(\alpha,\ell) N_{y,\ell}(\phi,k\rho\sin \alpha)] \cos \alpha\\
          i \Im [ \widehat{a_s}(\alpha,\ell) N_{z,\ell}(\phi,k\rho\sin \alpha)] \sin \alpha
        \end{pmatrix}
  \right]
  \nonumber \\
  \times \e^{i k z \cos\alpha} \cos\alpha \sin \alpha \D \alpha, \nonumber \\
  \label{eq.Hvoddell}
\end{align}
for $\ell$ odd, and
\begin{align}
  \Hv(\rho,\phi,z)  = \frac12
  n \sqrt{\frac{\epsilon_0 }{\mu_0}}  \frac{k^2}{2\pi^2}\int_0^{\amax}  \Re \left[
  \begin{pmatrix}
    - i \Im[ \widehat{a_p}(\alpha,\ell) N_{y,\ell}(\phi,k\rho\sin \alpha)]\\
    i\Im[ \widehat{a_p}(\alpha,\ell) N_{,\ell}(\phi,k\rho\sin \alpha) ] \\
    0
  \end{pmatrix}
  - 
  \begin{pmatrix}
    - i \Im[ \widehat{a_s}(\alpha,\ell) N_{x,\ell}(\phi,k\rho\sin \alpha)] \cos \alpha\\
    - i\Im[ \widehat{a_s}(\alpha,\ell) N_{y,\ell}(\phi,k\rho\sin \alpha) ]\cos \alpha\\
    \Re [ \widehat{a_s}(\alpha,\ell) N_{z,\ell}(\phi,k\rho\sin \alpha)] \sin \alpha
  \end{pmatrix}
  \right]
  \nonumber \\
  \times \e^{i k z \cos\alpha} \cos\alpha \sin \alpha \D \alpha, \nonumber \\
  \label{eq.Hevenell}
\end{align}
for $\ell$ even.
The  integrals over azimuthal angle $\alpha$ have to be computed numerically.

\section{Discretization of the integral equation}
\label{appendix:discretization}
In this Appendix we will discretize
\eqref{eq.eigenell}. This means that for each $\ell \in
\mathbb{Z}$ we discretize $\alpha$ and $\alpha'$ and approximate
\begin{equation}
  \label{eq:discretized-problem-start}
  \Lambda \Ahv{\alpha}{\ell}  =  \int_0^{\amax}
  \hat{\Mv}_{R, \ell}(\alpha, \alpha') \Ahv{\alpha'}{\ell}  \D{\alpha'}
\end{equation}
by a matrix equation. First, we subtitute $s:\alpha \to (\alpha + 1) {\amax}/2$ to obtain
\begin{equation}
  \label{eq:discretized-problem-start-with-subs}
  \Lambda \Ahv{s(\alpha)}{\ell}  = \frac{\amax}2\int_{-1}^{1}
  \hat{\Mv}_{R, \ell}'(s(\alpha), s(\alpha')) \Ahv{s(\alpha')}{\ell} \D{\alpha'}.
\end{equation}
We discretize the integral with the Gaussian quadrature rule on
the interval $-1 < s(\alpha) < 1$, which will, given the number of data points $N$ return nodal points $-1
= \alpha'_1 < \alpha'_2 < \ldots < \alpha'_N = 1$ and weights $(w_n)_{n =
  1}^N$ so that we can write
\begin{equation*}
  \Lambda \Ahv{s(\alpha)}{\ell}  = \frac{\amax}2 \sum_{n = 1}^N w_n
  \hat{\Mv}_{R, \ell}'(s(\alpha), s(\alpha'_n)) \Ahv{s(\alpha'_n)}{\ell}.
\end{equation*}
If we discretize $\alpha$ on the integration nodal points, we get $N$ equations, that is for each $m = 1,\ldots,N$ we have
\begin{equation*}
  \Lambda \Ahv{s(\alpha_m)}{\ell}  = \frac{\amax}2 \sum_{n = 1}^N w_n
  \hat{\Mv}_{R, \ell}'(s(\alpha_m), s(\alpha'_n)) \Ahv{s(\alpha'_n)}{\ell},
\end{equation*}
for $m = 1, \ldots, N$.
We rewrite this as  a matrix eigenvalue problem. Let
$(\alpha_i)_{i = 1}^N$ be the integration nodal points with corresponding weights $(w_i)_{i
  = 1}^N$ and set
\[
\vec{\alpha} = \begin{pmatrix} \alpha_1\\ \vdots\\
  \alpha_N \end{pmatrix}, \, \vec{w} = \operatorname{diag}\{w_1,
\ldots, w_N\} \text{ and } \vec{W} = \operatorname{diag}\{\vec{w}, \vec{w}\}.
\]
Next we define the block matrix $\vec{M}_D$:
\begin{equation}
  \label{eq:discretized-matrix}
  \vec{M}_D = \begin{pmatrix}
    \vec{M}_D^{11} &  \vec{M}_D^{12}\\
    \vec{M}_D^{21} &  \vec{M}_D^{22}
  \end{pmatrix},
\end{equation}
where the matrices $\vec{M}_D^{mn}$ are defined as
\begin{equation*}
  \vec{M}_D^{mn} = \hat{M}_{R,mn}'(s(\vec{\alpha}), s(\vec{\alpha}')).
\end{equation*}
Using this, we can write the discretized equation as an eigenvalue problem
\begin{equation}\label{eq:discretized-problem-final}
  \frac{2\Lambda}{\amax} \begin{pmatrix}
    \hat{a}_{P, \ell}(s(\vec{\alpha}))\\ 
    \hat{a}_{S, \ell}(s(\vec{\alpha}))\\
  \end{pmatrix}  =\vec{M}_D \vec{W} \begin{pmatrix}
    \hat{a}_{P,\ell}(s(\vec{\alpha}))\\ 
    \hat{a}_{S,\ell}(s(\vec{\alpha}))
  \end{pmatrix}.
\end{equation}
The method we have used above is the so-called Nystr\"om
method. For the discretized problem to be a good approximant to the
solution of integral equation \eqref{eq:discretized-problem-start}
the solution of \eqref{eq:discretized-problem-final} should converge to it
as $N \to \infty$.

As we have seen in Section~\ref{subsection.mathprop}
the integral operator is compact. Applying \cite[Theorem 3]{aS75} gives us the numerical
stability for problem \eqref{eq:discretized-problem-final}.


\bibliography{bibliography}

\end{document}